\documentclass[11pt]{article}
\usepackage[margin=1in]{geometry}

\usepackage{inputenc}
\usepackage{etoolbox}
\usepackage{etoc}
\usepackage{float}
\inputencoding{utf8}
\usepackage{amssymb}
\usepackage[hidelinks]{hyperref}
\usepackage[round]{natbib}
\usepackage{amsthm}

\usepackage{tabularx}
\usepackage{xcolor}
\usepackage{subcaption}
\usepackage{placeins}
\usepackage{amsmath}
\usepackage{thm-restate}
\usepackage{float}
\usepackage{enumitem}
\usepackage{xr}
\usepackage{graphicx}
\usepackage{thmtools}
\usepackage{url}

\newcommand\bbR{\ensuremath{\mathbb{R}}}

\newcommand\bbE{\ensuremath{\mathbb{E}}}

\newcommand{\bbrp}{\ensuremath{\bbR_+}}

\usepackage[group-separator={,}]{siunitx}
\newtheorem{lemma}{Lemma}
\newtheorem{remark}{Remark}

\newcommand{\parbold}[1]{~\\\noindent\textbf{#1}. }
\newlength\epirule%
\newcommand\epiline{%
	\noalign{\global\epirule\arrayrulewidth\global\arrayrulewidth 1pt}\hline%
	\noalign{\global\arrayrulewidth\epirule}%
}
\newcommand\epigraph[2]{%
	\hfill\begin{tabular}{@{}r@{}}
		{\small\textit{#1}}\\[.5em] \epiline%
		{\small\textsc{#2}}
	\end{tabular}
}

\newcommand\lij{\lambda_{i\to j}}

\newcommand\lonetwo{\lambda_{1\to 2}}
\newcommand\ltwoone{\lambda_{2\to 1}}

\newcommand\paru{\ensuremath{\frac{\partial}{\partial u}}}
\newcommand\parl{\ensuremath{\frac{\partial}{\partial \ell}}}
\newcommand\partildeR[1]{\ensuremath{{\frac{\partial}{\partial {#1}}} {\hat{R}(\sigma)}}}

\newcommand\argmax{\arg\max}

\providetoggle{tableofcontentsapp}
\toggletrue{tableofcontentsapp}

\begin{document}

\title{Driver Surge Pricing}

\newcommand\blfootnote[1]{%
	\begingroup
	\renewcommand\thefootnote{}\footnote{#1}%
	\addtocounter{footnote}{-1}%
	\endgroup
}

\author{
	Nikhil Garg\\
	Stanford University\\
	\texttt{nkgar6@gmail.com} \\
	\and
	Hamid Nazerzadeh\\
	USC Marshall \& Uber Technologies \\
	\texttt{nazerzad@usc.edu} \\
}

\maketitle

\begin{abstract}
Ride-hailing marketplaces like Uber and Lyft use dynamic pricing, often called surge, to balance the supply of available drivers with the demand for rides.
We study driver-side payment mechanisms for such marketplaces, presenting the theoretical foundation that has informed the design of Uber's new additive driver surge mechanism.
We present a dynamic stochastic model to capture the impact of surge pricing on driver earnings and their strategies to maximize such earnings. In this setting, some time periods (surge) are more valuable than others (non-surge), and so trips of different time lengths vary in the induced driver opportunity cost.
First, we show that multiplicative surge, historically the standard on ride-hailing platforms, is not incentive compatible in a dynamic setting. We then propose a structured, incentive-compatible pricing mechanism. This closed-form mechanism has a simple form and is well-approximated by Uber's new additive surge mechanism. Finally, through both numerical analysis and real data from a ride-hailing marketplace, we show that additive surge is more incentive compatible in practice than is multiplicative surge.%

\blfootnote{
	We would like to thank Uber's driver pricing data science team, in particular Carter Mundell, Jake Edison, Alice Lu, Michael Sheldon, Margaret Tian, Qitang Wang, Peter Cohen, Kane Sweeney, and Jonathan Hall for their support and suggestions without which this work would have not been possible.
		We also thank Leighton Barnes, Ashish Goel, Ramesh Johari, Vijay Kamble, Anurag Komanduri, Hannah Li, Virag Shah, and anonymous reviewers.
		This work was funded in part by the Stanford Cyber Initiative, the Office of Naval Research grant N00014-15-1-2786, and National Science Foundation grants 1544548 and 1839229.
	}
\end{abstract}

\section{Introduction}
Ride-hailing marketplaces like Uber, Lyft, and Didi match millions of riders and drivers every day. A key component of these marketplaces is a {\em surge} (dynamic) pricing mechanism. On the rider side of the market, surge pricing reduces the demand to match the level of available drivers and maintains the reliability of the marketplace, cf.,~\citet{hall2015effects}, and so allocates the rides to the riders with the highest valuations. On the driver side, surge encourages drivers to drive during certain hours and locations, as drivers earn more during surge~\citep{lu_surge_2018,hall_labor_2017,chen_dynamic_2015}.
\citet{castillo_surge_2017} show that surge balances both sides of this spatial market by moderating the demand and the density of available drivers, hence avoiding so called ``Wild Goose Chase'' equilibria in which drivers spend much of their time on long distance pick ups.
Surge pricing -- along with centralized matching technologies -- is often considered the primary reason that ride-hailing marketplaces outperform traditional taxi services on metrics such as driver utilization and overall welfare~\citep{cramer_disruptive_2016,buchholz_spatial_2017,ata_spatial_2019}.

However, variable pricing (across space and time) must be carefully designed, since it can create incentives for ``cherry-picking'' and rejecting certain trip requests. Such behavior increases earnings of strategic drivers at the expense of other drivers, who may then disproportionately receive such trip requests after they are rejected by others, cf.,~\citet{cook_gender_2018}. It also reduces overall platform reliability, inconveniencing riders who may have to wait longer before receiving a ride.

\begin{figure}[t!]
	\centering
	\begin{subfigure}[t]{0.48\textwidth}
		\centering
		\includegraphics[scale=1.08]{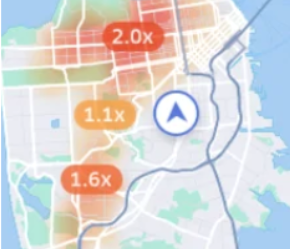}
		\caption{{\em Multiplicative surge heatmap.} ``1.6x'' on the map means that the standard fares for trips from the corresponding area are increased by $60\%$.}
		\label{fig:multi_heatmap}
	\end{subfigure}%
	~ \hfill
	\begin{subfigure}[t]{0.48\textwidth}
		\centering
		\includegraphics{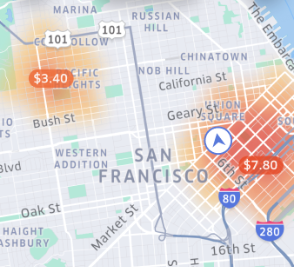}
		\caption{{\em Additive surge heatmap.}  ``\$7.8'' on the map means that $\$7.8$ is added to each trip's standard fare from the corresponding area.}
		\label{fig:add_heatmap}
	\end{subfigure}
	\caption{Driver surge heatmaps with multiplicative and additive surge. On Uber, drivers see a heatmap of surge when they are logged in but not on a trip, guiding them to higher earning opportunities by signaling each location's value~\citep{lu_surge_2018}. Structural simplicity is essential to clearly communicate payments to drivers, and additive and multiplicative surge represent the two simplest options. }
	\label{fig:heatmaps}
\end{figure}

Uber recently revamped its driver surge mechanism, to improve the driver experience and make earnings more dependable~\citep{uber_dependable_earnings}. The main change is making surge  ``additive'' instead of ``multiplicative.'' Under {\bf multiplicative surge}, the driver payout from a surged trip scales with the length of the trip.
In contrast, under {\bf additive surge}, the payout surge component is constant (independent of trip length), with some adjustment for very long trips~\citep{uber_driver_surge}. Figure~\ref{fig:heatmaps} depicts the driver app surge heat-map for each type of surge. %
We show that the change directly addresses a primary reason that drivers who strategically reject trip requests may earn more than others, even as total payments remain the same.

\subsection{Contributions}
We consider the design of \textit{incentive compatible} (IC) pricing mechanisms in the presence of surge. Trips differ by their length $\tau \in \bbrp$, and the platform sets the payout $w(\tau)$ for each trip in each world state (i.e., surge vs non-surge). Drivers decide which trip requests $\sigma \subseteq \bbrp$ to accept in each world state, in response to the payout function $w$.\footnote{Drivers' level of sophistication and experience varies, cf.~\cite{cook_gender_2018}. An IC mechanism aligns the incentives of drivers to accept all trips, for any level of strategic response to pricing strategies.} The technical challenge is to design an IC pricing mechanism $w$, for which accepting all trips is an earning maximizing strategy for drivers over a long horizon, i.e., where $\sigma = (0, \infty)$ in each world state maximizes driver earnings.

We first study a continuous-time, infinite horizon
{\em single-state model}, where trip requests arrive over time according to a stationary Poisson process. %
We show that in this model, multiplicative pricing -- where the payout of a trip is proportional to the length of that trip -- is incentive compatible.
To obtain this result, we show in Theorem~\ref{thm:basicmodeloptimal} that the best response strategy of a driver to function $w$, to maximize earnings, is a threshold strategy where the driver accepts all trips with payout rate $\frac{w(\tau)}{\tau}$ above some threshold. Hence, a mechanism that equalizes the payout rate of all trips is incentive compatible.

We then present a model where the world state stochastically transitions over time between surge and non-surge states, with trip payments, distributions, and intensity varying between states. %
In such a \textit{dynamic} system, completing a given trip affects a driver's earnings beyond just the length of the trip, i.e., it imposes a future-time externality on the driver that is a function of the trip length. The driver's trip opportunity cost thus includes both what occurs during a trip, and a continuation value. This externality causes {\em multiplicative pricing to not be incentive compatible in the presence of surge} (Theorem~\ref{prop:multnotoptimal}), in contrast to the single-state model. Namely, drivers can benefit from rejecting long trips in a non-surge state, and short trips in the surge state.

In Theorem~\ref{thm:ICpolicy}, our main result, we propose a class of incentive compatible pricing functions described in closed form of the model primitives. The prices incorporate driver temporal externalities: during surge, short trips pay more per unit time than do long trips.

Next, we study surge pricing in our model numerically, showing that additive surge is incentive compatible in more regimes of interest than is multiplicative surge. %
Finally, using RideAustin data, we show that our theoretical insights extend to practice: additive surge correctly values trips amid temporal externalities, unlike multiplicative surge. %

To our knowledge, ours is the first ride-hailing pricing work to incorporate dynamic (non-constant), stochastic demand and pricing.
This component is essential to uncover how a particular trip imposes substantial temporal externalities on a driver's future earnings.%

\subsection{Related Work}
We discussed some of the related work on surge pricing above. Here, we briefly review the lines of research closest to ours. We refer the reader to a recent survey by~\citet{korolko_dynamic_2018} for a broader overview of the growing literature on ride-hailing markets.

\textbf{Driver spatio-temporal strategic behavior.} Several works model strategic driver behavior in a spatial network structure, and across time in a single-state.~\citet{ma_spatio-temporal_2018} develop spatially and temporally smooth prices that are welfare-optimal and incentive compatible in a deterministic model. Their prices form a competitive equilibrium and are the output of a linear program with integer solutions.
We similarly seek to develop incentive compatible pricing schemes, and both works broadly construct VCG-like prices that account for driver opportunity costs. Our focus is on structural aspects (e.g., multiplicative in trip length) in a non-deterministic model.

\citet{bimpikis_spatial_2016} show how the platform would price trips between locations, taking into account strategic driver re-location decisions, in a stationary model with discrete locations. They show that pricing trips based on the origin location substantially improves surplus, as well as the benefits of ``balanced'' demand patterns.
~\citet{besbes_surge_2018} consider a continuous state space setting and show how a platform may optimally set prices across the space in reaction to a localized demand shock to encourage drivers to relocate; their model has driver cost to re-locate, but no explicit time dimension. They find that localized prices have a global impact, and, e.g., the optimal pricing solution incentivizes some drivers to move away from a demand shock.
 ~\citet{afeche_ride-hailing_2018} consider a two state model with demand imbalances and compare platform levers such as limiting ride requests and directing drivers to relocate, in a two-state fluid model with strategic drivers. They upper-bound performance under these policies, and find that it may be optimal for the platform to reject rider demand even in over-supplied areas, to encourage driver movement.
 A similar insight is developed by~\citet{guda_your_2019} who explicitly model market response to surge pricing. Finally,~\citet{yang2018mean} analyze a mean-field system in which agents compete for a location-dependent, time-varying resource, and decide when to leave a given location. They leverage structural results---agents' equilibrium strategies depend just on the current resource level and number of agents---to numerically study driver relocation decisions as a function of the platform commission structure.

\textbf{{Pricing in ride-sharing and service systems}.}
There is a growing literature on queuing and service systems motivated in part by the ride-sharing market.
For example,~\citet{besbes_spatial_2018} revisit the classic square root safety staffing rule in spatial settings, cf.,~\cite{bertsimas_stochastic_1991,bertsimas_stochastic_1993}.
Much of the focus of this line of work is how pricing affects the arrival rate of (potentially heterogeneous) customers, and thus the trade-off between the price and rate of customers served in maximizing revenue.

\citet{banerjee_pricing_2015} consider a network of queues in which long-lived drivers enter the system based on their expected earnings but cannot reject specific trip requests. Under their model, dynamic pricing \textit{cannot} outperform the optimal static policy in terms of throughput and revenue, but is more robust.~\citet{cachon_role_2017} argue in contrast that surge pricing and payments are welfare increasing for all market participants when drivers decide when to work.
~\citet{chen_pricing_2018} consider a marketplace with forward-looking buyers and sellers who arrive sequentially and can wait for better prices in the future. They develop strategy-proof prices whose variation over time matches the participants' expected utility loss incurred by waiting.
~\citet{lei2016real} consider a model where customers arrive over time and utilize a capacity constrained resource for a certain amount of time. They develop an asymptotically revenue-maximizing, dynamic, customer-side pricing policy, even when service times may be heterogeneous.
~\cite{glazer1983economics} consider taxi-driver strategic responses to multiplicative and affine pricing, as we do, focusing on deviations in which a driver can take a circuitous route in order to increase the length of a trip.

One of the most related to our work in modeling approach,~\citet{kamble_revenue_2018} studies how a freelancer can maximize long-term earnings with job-length-specific prices, balancing on-job payments and utilization time. In his model, a freelancer sets their own prices for a discrete number of jobs of different lengths and, with assumptions similar to our single-state model, it is optimal for the freelancer to set the same price per hour for all jobs. We further discuss the relationship of this work to our single-state model below.

\paragraph{\bf Organization.} The rest of the paper is organized as follows. Section~\ref{sec:surgemodel} contains our model; we derive driver earnings as it depends on their strategy, and formalize the platform objective. In Section~\ref{sec:rewardaffinesec}, we formulate a driver's best response strategy to affine pricing functions in each model. In Section~\ref{sec:IC}, we present incentive compatible pricing functions for our surge model. In Section~\ref{sec:additivesurgesims}, we numerically compare the IC properties of additive and multiplicative surge. Finally, in Section~\ref{sec:empiricalevidence}, we empirically compare additive and multiplicative surge using data from the RideAustin marketplace.
\section{Model, driver earnings, and platform objective}

\label{sec:surgemodel}
We consider a large ride-hailing market with decoupled pricing, from the perspective of a \textit{single driver}.
This driver receives trip requests of various lengths. The trips' rate, distribution, and payment are known to the driver and determined exogeneously to decisions to accept or decline requests.
We do not consider spatial heterogeneity, to focus on the temporal opportunity cost and continuation value based on a length of the trip.\footnote{We believe our insight can be extended to a spatial setting where the price can be decomposed to a time-based component, based on the length of the trip, and a spatial component based on the destination of the trip. However, this would be beyond the scope of this work, cf.,~\cite{bimpikis_spatial_2016}.}

In this section, we first in Section~\ref{sec:modelprimitives} present the primitives of our two models, a single-state model and a dynamic model with surge pricing. Then in Section~\ref{sec:driverstratmodel} we describe the driver's strategy space and derive the driver reward in each model. Next, in Section~\ref{sec:platformobjective}, we formalize the platform objective and technical challenge solved in this work. We conclude with a short discussion on our model's relationship to practice in Section~\ref{sec:limitationsshort}.

\subsection{Model primitives}
\label{sec:modelprimitives}
We start with the model primitives in each model.
\subsubsection{Single-state model}
\label{sec:stationarymodel}
We start with a
model where there is a single world state, i.e., all model components are constant over time. Time is continuous and indexed by $t$.
At each time $t$, the driver is either \textit{open}, or \textit{busy}. While open, the driver receives job (trip) requests from riders according to a Poisson process at rate $\lambda$, i.e., the time between requests is exponential with mean $\frac{1}{\lambda}$. Job lengths, denoted by $\tau$, are drawn independently and identically from a continuous distribution $F$.

If the driver accepts a job request of length $\tau$ at time $t$ (as discussed below), they receive a payout of $w(\tau)$ at time $t + \tau$, at which time they become open again. Otherwise, the driver remains open.
Except where specified, the only assumptions on $w$ are that it is continuous and asymptotically (sub)-linear: $\exists c: \lim \inf_{\tau \to \infty} \frac{w(\tau)}{\tau} \leq c$, which ensures that the driver reward is also bounded.\footnote{These restrictions are innocuous. With continuity, similar trips pay similarly. Asymptotic sub-linearity means that the marginal value of additional length of a trip remains bounded; it trivially holds if the domain of $F$ is bounded.}

\subsubsection{Dynamic model with surge pricing}
\label{sec:surgepricingmodel}
 A model with fixed pricing and arrival rates of jobs is not a realistic representation of %
ride-hailing platforms. In particular, rider demand (both in intensity and in distribution) may vary substantially over time, even within a day (cf. Appendix Figure~\ref{fig:surgeover3days}). %
To study how this {\bf dynamic} nature affects driver decisions, we consider a model with %
two states, $i \in \{1, 2\}$, where $i = 2$ denotes the \textit{surge} state.
(At a high level, the surge state provides a higher earnings rate to the driver. The precise definition is in Section~\ref{sec:statrewarddynamic}, after we formulate the driver's earnings rate in each state). %

The world evolves stochastically between the two states, as a Continuous Time Markov Chain (CTMC). When the world is in state $i$, the state changes to $j$ according to a fixed exponential clock that ticks at rate $\lij$, independently of other randomness.

When open in state $i$, the driver receives job requests at rate $\lambda_i$ with lengths $\tau\sim F_i$, and collects payout according to payment function $w_i$, which is presumed to have the same properties as $w$ in the single-state model. The state of the world may change while a driver is on trip. Crucially, the driver receives payments according to the state of the world $i$ when the trip \textit{begins}. %
We will use $w = \{w_1, w_2 \}$ to denote the overall pricing mechanism. %

\subsection{Driver strategies and earnings}
\label{sec:driverstratmodel}
In our model, the driver can decide whether to accept the trip request, with no penalty.\footnote{This assumption follows Uber's current practice. We further discuss the driver's information set in Section~\ref{sec:limitationsshort}.}

In the single-state model, let $\sigma\subseteq \bbrp \triangleq (0, \infty)$ denote the driver's (fixed) strategy, where $\tau\in\sigma$ implies that a driver accepts job requests of length $\tau$.
In the dynamic model, the driver follows policy $\sigma = \{\sigma_1, \sigma_2\}$, where $\sigma_i \subseteq \bbrp$ indicates the jobs accepted in state $i$. We assume that driver policies are measurable with respect to $F$ (corresponding $F_i$ in dynamic model); for technical reasons, in the dynamic model we also assume that $\sigma_i$ consist of a union of open intervals, i.e., are open subsets of $\bbrp$. When we write equalities with policies $\sigma$, we mean equality up to changes of measure 0.

The driver is long-lived and aims to maximize their own lifetime average hourly earnings on the platform, including both open and busy times.
Let $R(w, \sigma, t)$ denote the (random) total earnings from jobs accepted from time $0$ up to time $t$ if the driver follows policy $\sigma$ and the payout function is $w$. Then, the driver's lifetime \textit{earnings rate} is
\[
R(w,\sigma) \triangleq {\lim\inf}_{t\to\infty}\frac{R(w,\sigma, t)}{t}.%
\]%
This earnings rate is a deterministic (non-random) quantity, and is a function of the driver policy $\sigma$, pricing function $w$, and the primitives.

A driver policy $\sigma^*$ is \textbf{optimal} (best-response) with respect to pricing function $w$ if it maximizes the lifetime earnings rate of the driver among all policies: $R(w,\sigma^*) \geq R(w,\sigma)$, \text{for all valid policies} $\sigma$ (i.e., measurable with respect to $F$ or $F_i$, with $\sigma_i$ open sets). Then, pricing function $w$ is \textbf{incentive compatible (IC)} if accepting all job requests is optimal with respect to $w$, i.e., $\sigma = (0, \infty)$ in the single-state model or $\sigma = \{(0, \infty),(0, \infty)\}$ in the dynamic model is optimal with respect to $w$. In other words, payment function $w$ is incentive compatible if an earnings-maximizing driver (who knows all the primitives, $w$, and the trip length $\tau$ at request time) accepts every trip request. %

We now analyze the driver's lifetime earnings rate $R(w, \sigma)$ for each model.

\subsubsection{Driver earnings in the single state model} \label{sec:statrewardsingle}
In the single-state model, the primitives directly induce a renewal reward process, where a given renewal cycle is the time a driver is newly open to the time they are open again after completing a job. Let $W(\sigma)$ be the mean earnings on trips $\tau \in \sigma$, i.e., the expected earning in a renewal cycle; let $T(\sigma)$ be the sum of the expected wait time to an accepted trip and the expected length of a trip, and thus the expected renewal cycle length; let $F(\sigma)$ be the probability the driver receives a request in $\sigma$. Then, the lifetime driver mean hourly earnings (earnings rate) is
\begin{align*}
R(w,\sigma) = \frac{W(\sigma)}{T(\sigma)} =  \frac{\frac{1}{F(\sigma)}\int_{\tau \in \sigma} w(\tau) dF(\tau)}{\frac{1}{F(\sigma)\lambda} + \frac{1}{F(\sigma)}\int_{\tau \in \sigma} \tau dF(\tau)}
\end{align*}
The first equality follows from the renewal reward theorem, and holds with probability 1.%

\subsubsection{Driver earnings in the dynamic model} \label{sec:statrewarddynamic}

For the dynamic model, on the other hand, %
we cannot directly use the renewal reward theorem with a renewal cycle containing just a single trip. The driver's earning on a given trip is no longer independent of earnings on other trips: given a job that starts in the surge state, the driver's next job is more likely to also start in surge. Given whether each job started in the surge state, however, job earnings are independent. We can use this property to prove our next lemma, which gives the driver earnings rate in the dynamic model. Let $\mu_i(\sigma)$ be the fraction of time the driver spends either open state $i$ or on a trip that \textit{starts} in state $i$.
\begin{restatable}{lemma}{lemstatingreward}
	\label{lem:statingreward}
	In the dynamic model, the earnings rate can be decomposed into each state $i$ earnings rate $R_i(w_i, \sigma_i)$ and fraction of time $\mu_i(\sigma)$ spent in state $i$:
	\begin{align*}
	R(w, \sigma) %
	&= \mu_1 (\sigma) R_1(w_1, \sigma_1) + \mu_2 (\sigma) R_2(w_2, \sigma_2) & \text{with probability } 1.
	\end{align*}
	As in the single-state model, $R_i(w_i, \sigma_i) = \frac{W_i(\sigma_i)}{T_i(\sigma_i)},$
	where
	\begin{align*}
	W_i(\sigma_i) &= \frac{1}{F_i(\sigma_i)}{\int_{\tau \in \sigma_i} w_i(\tau) dF_i(\tau)}, \text{\ \ \ \ \ \ \ \ \ \ \ \ }
	T_i(\sigma_i) = {\frac{1}{\lambda_iF_i(\sigma_i)} + \frac{1}{F_i(\sigma_i)}\int_{\tau \in \sigma_i} \tau dF_i(\tau)}
	\end{align*}
\end{restatable}
We prove the result by defining a new renewal process, in which a single reward renewal cycle is: the time between the driver is open in state 1 to the next time the driver is open in state 1 \textit{after} being open in state 2 at least once. In other words, \textit{each} renewal cycle is composed of some number (potentially zero) of sub-cycles in which the driver is open in state 1 and then is open in state 1 again after a completed trip; one sub-cycle starting with the driver open in state 1 and ending with being open in state 2 (either after a completed trip or a state transition while open); some number (potentially zero) of sub-cycles in which the driver is open in state 2 and then is open in state 2 again after a completed trip; and finally one sub-cycle starting in state 2 and ending with the driver open in state $1$.

Given the number of such renewal reward cycles completed up to time $t$, the total earnings on trips starting in each state (earnings in each sub-cycle) are independent of each other, and then we use Wald's identity~\citep{wald1973sequential} to separate $\mu_i(\sigma)$ and $R_i(\sigma_i)$.

Note that $T_i(\sigma_i)$ is not exactly the expected length of time in a single sub-cycle in a state given $\sigma_i$, but rather is proportional to it; the multiplicative constant $\frac{1}{\lambda_i F_i(\sigma_i) + \lambda_{i\to j}}$ cancels out with the same constant in the expected earnings in a single sub-cycle in a state given $\sigma_i$. This constant emerges from the primitives: when the driver is open in state $i$, there are two competing exponential clocks (with rates $\lambda_i F_i(\sigma_i)$ and $\lij$, respectively) that determine whether the driver will accept a request before the world state changes. %

What does $\mu_i(\sigma)$ look like? We defer showing the exact form to Section~\ref{sec:expectedtimeeachstate} in advance of developing incentive compatible pricing. Here, we provide some intuition: the trips that a driver accepts in each state determines the portion of their time spent on trips started in each state. If a driver never accepts trips in the non-surge state, they will be open and thus available for a trip as soon as surge begins. Inversely, if a driver accepts a long surge trip immediately before surge ends, they will be paid according to the surge payment function $w_2$ even though surge has ended. Surprisingly, given the complex formulation of the reward $R(w, \sigma)$ as it depends on $\sigma = \{\sigma_1, \sigma_2\}$, we find the structure of optimal policies as they depend on the pricing $w_i$, as well as incentive compatible pricing functions. %

Finally, we can now precisely define what it means for $i=2$ to be the surge state: it has a higher potential earning rate than state $1$. There exists some policy $\sigma_2$ such that $R_2(w_2, \sigma_2) > R_1(w_1,\sigma_1)$, for all $\sigma_1 \subseteq \bbrp$. In other words, suppose that instead we were in the single-state setting, where the primitives were set as either $(\lambda_1,F_1,w_1)$ or $(\lambda_2,F_2,w_2)$. Then the latter set of primitives would yield a higher maximum earnings.\footnote{		This assumption is different than the statement that each surge trip pays more than an equivalent non-surge trip, $w_2(\tau) \geq w_1(\tau), \forall \tau$, and neither statement implies the other. Under this definition, surge may be characterized as higher per-trip payments. Alternatively, if request arrival rate is high due to a demand shock, $\lambda_2 \gg \lambda_1$, then the driver waits less time between trip requests and so has a higher earnings rate -- even without higher per-trip payments. While a less common scenario in practice, our model further allows surge to be characterized by a more lucrative distribution of trips $F_2$ compared to $F_1$, even if the intensity of trips and on-trip payments conditional on trip length are identical. More generally, surge may be characterized by a combination of such scenarios.  }%

\subsection{Platform objective and constraints}
\label{sec:platformobjective}
Having derived the driver reward, we now describe the platform objective, setting up the technical challenge we solve in the rest of the work. Recall that {our model is {\bf decoupled}: rider and driver prices are determined separately. Under decoupled pricing, the platform has under its control both the price $p_i(\tau)$ charged to the rider and the payment $w_i(\tau)$ paid to the driver for a trip of length $\tau$---and the proportion of these two values may vary across trips. This modeling assumption follows the current practice~\citep{uber_service_fee} and allows us to focus on the drivers' perspective, without further complicating the analysis.\footnote{Coupled pricing imposes more constraints.~\citet{bai_coordinating_2018} and~\citet{bikhchandani2020intermediated} both find that the platform should adjust its payout ratio with demand---an example of decoupling---to maximize profit or overall welfare. }}

What should be the role of driver payments with decoupled pricing? In practice, the platform quotes the rider a price and `guarantees' fulfillment if a ride is requested; driver payments should thus primarily ensure that all requested rides are fulfilled, motivating our goal of designing incentive compatible prices.
In Appendix Section~\ref{sec:platformobjdetail}, we formalize this intuition by considering driver payments $w$ as a sub-problem of the comprehensive platform challenge, involving jointly setting both rider prices and driver payments to maximize an objective (e.g., profit or welfare).
We establish that -- with decoupled pricing and an earnings-maximizing driver within our model -- this joint problem can be decomposed into one in which the rider pricing (not considered in this work) determines the objective value, subject to finding a driver payment policy $w$ that satisfies incentive compatibility and a driver \textit{participation constraint}: that the driver earnings rate is higher than an outside option earnings rate (denoted $R$), i.e., $\max_\sigma \  R(w, \sigma) \geq R$.

In the dynamic model, we additionally consider \textit{per-state driver earnings constraints}, $R_i(w_i, (0, \infty)) = R_i$, for some exogenous $R_2 > R_1$. This constraint comes from practice, via features not directly captured in our model. As detailed in Appendix Section~\ref{sec:eachstateearningsdetail}, following the current practice, in our model platforms impose a business constraint to approximately pass on rider revenue in each world state to the driver, i.e., the constraints $R_i$ are determined by per-state revenue, a function of latent demand and rider prices.%

	If the platform has more flexibility, $R_i$ may also be optimized, for example to induce drivers to position themselves in areas with more frequent surges. \citet{lu_surge_2018} find empirically that drivers do re-position to higher surge areas. \citet{freund_escrow_lyft2020} describe how Lyft manages an incentive budget over time and space to incentivize driver re-positioning, and in a coupled pricing setting~\citet{besbes_surge_2018} show theoretically how to set prices to induce driver movement.
 More broadly, the revenue during one spatio-temporal period may be used to smooth out driver payments in another period, cf.~\citet{asadpour_escrow_2019,bai_coordinating_2018}. In this work, we do not directly consider how the platform should set $R_i$ (or $R$); how to do so over space and time is an interesting avenue for future work.
Instead, we establish our results for a range of $R_i$ for which incentive compatible prices can be constructed.
This decomposition reflects how decoupled surge pricing is set in practice, and for the rest of this work we seek a payment policy that satisfies these conditions.

\subsection{{Practical considerations}}
\label{sec:limitationsshort}
Our model is stylized in several important respects, and ride-hailing practice is not consistent across marketplaces, time, or geography. Our theoretical model reflects our view on the most relevant components from practice.

{\bf Driver heat-maps and affine pricing} We are especially interested in \textit{affine} pricing schemes, where $w_i(\tau) = m_i\tau + a_i$, with $m_i\geq 0$ (in the single-state model: $w(\tau) = m\tau + a$, with $m\geq 0$; we refer to the case with $a_i > 0$ ($a_i < 0)$ as positive (negative) affine pricing). Such pricing functions can be communicated as time and distance rates (see, e.g.,~\citet{uber_fares_calculated}), and the surge component displayed on a heat-map. This simplicity is an important desiderata from practice, where payments should be clear to drivers.

{\bf Driver information structure: trip time and time to the rider.}
We assume that the platform reveals the total trip length to the driver at the time of request, and that the driver can freely reject it without penalty.
Drivers often cannot see the rider's destination or the trip length until they pick up the rider (but they can reject a request based on the pick-up time to the rider, without penalty).\footnote{{This practice is not consistent across marketplaces and locations. For example, in California as of January 2020, Uber shows the driver the destination and payment estimate at request time. Incentive compatible pricing is an important stepping stone to showing this information.}} Some drivers call ahead to find out the rider's destination or even cancel the trip at the pick-up location, creating negative experiences for both the rider and the driver.\footnote{We note that destination discrimination is against Uber's guidelines and could lead to deactivation~\citep{uber_community_guidelines}.}
Our notion of incentive compatibility is {\em ex-post}, in which drivers would accept all trips even knowing the trip length. This notion is stronger than an \textit{ex-ante} setting in which the trip length is not revealed to drivers.
Furthermore, in practice, jobs have two components: the time it takes to pick up the rider, and the time while the rider is in the driver's vehicle -- and the former component is typically unpaid.\footnote{Lyft has recently experimented with paying drivers for the time it takes to pick up the rider~\citep{auerbach_paying_2019}.} Our model combines these two components into an overall trip length, which determines payments.

{\bf Markovian surge and model limitations.} In practice, surge has strong intra-day patterns -- for example, rush hours have higher average surge values, cf. Appendix Figure~\ref{fig:surgetripbyhour}. However, evolution of surge on finer time scales, on the level of drivers' individual trip decisions, is more volatile and believably Markovian, cf. Appendix Figure~\ref{fig:surgeover3days}. Our theoretical model assumes that surge is Markovian and binary and the response of a single driver, and further ignores spatial effects. We discuss such issues in Sections~\ref{sec:limitations} and~\ref{sec:moreempiricalfacts}, and our empirical analysis in Section~\ref{sec:empiricalevidence} provides evidence that our insights extend to practice despite these theoretical limitations.

\section{Incentive compatibility with affine pricing}
\label{sec:rewardaffinesec}

In this section, we study the incentive compatibility of affine pricing. In Section~\ref{sec:singlestateaffine},
we first characterize the driver's best-response strategy with respect to any pricing function $w$ in the single-state model. We then observe that multiplicative pricing,
a special case of affine pricing where $w(\tau) = m\tau$,
is incentive compatible. %
In contrast, in Section~\ref{sec:dynamicmodelaffine}, we show that in the dynamic model, multiplicative pricing may no longer be incentive compatible. We further derive the structure of optimal driver policies in each state with respect to affine or multiplicative pricing, which will enable numerical study of the incentive compatibility properties of additive and multiplicative surge in Section~\ref{sec:additivesurgesims}. Section~\ref{sec:discussionmult} discusses the key differences in the two models, setting up Section~\ref{sec:IC} where we derive incentive compatible pricing functions for the dynamic model.

\subsection{Single-state model: multiplicative pricing is incentive compatible}
\label{sec:singlestateaffine}
Our first result is a simple optimal driver policy in the single-state model.

\begin{restatable}{theorem}{thmbasicmodeloptimal}
	\label{thm:basicmodeloptimal}
	With a single state, for each $w$ there exists a constant $c_w \in \bbrp$ such that the policy ${\sigma}^* = \left\{\tau : \frac{w(\tau)}{\tau} \geq c_w \right\}$ is optimal for the driver with respect to $w$.
\end{restatable}

Theorem~\ref{thm:basicmodeloptimal} establishes that, in a single-state model with Poisson job arrivals, the \textit{length} of the job is not important, only the hourly rate while busy on the job. The optimal $c_w$ in the policy is not necessarily $c_w = \sup \frac{w(\tau)}{\tau}$: drivers must trade off the earnings rate while on a trip with their utilization rate; the more trips that a driver rejects, the longer the wait for an acceptable trip. In the appendix we prove the result by, starting at an arbitrary policy $\sigma$, making changes to the policy that increase the earnings rate while on a job \textit{without decreasing the utilization rate}. Thus, each such change improves the reward $R(w, \sigma)$, and the sequence of changes results in a policy of the above form, for some threshold $c'$. Then, this threshold $c'$ can be optimized, leading to an optimal policy of this form.

An immediate corollary of Theorem~\ref{thm:basicmodeloptimal} is that $w(\tau) = m\tau$, for $m>0$, is IC. In other words, if the platform pays a constant rate $\frac{w(\tau)}{\tau} = m$ to busy drivers, then in the single-state model it is in the driver's best interest to accept every trip. This result is driven by the following insight for Poisson arrivals: while \textit{receiving} long trip requests is more beneficial to drivers in the single-state setting as they increase one's utilization rate (the driver is busy for a longer time until the next open period), \textit{rejecting} short trips to cherry-pick long trips decreases utilization by the same amount.\footnote{This insight is similar to a result of~\citet{kamble_revenue_2018}; however, in our setting the driver's strategy $\sigma$ is a \textit{subset} of $\bbrp$ denoting the job requests accepted, as opposed to a discrete set of prices charged. Further, in our settings the driver responds to the platform's prices instead of setting prices, enabling a wider range of IC pricing mechanisms.} Further note that, given an earnings rate target $R$, calculating the multiplier $m$ and thus an IC pricing policy is trivial.

On the other hand, affine pricing may not be incentive compatible because short trips are worth more per unit time than are long trips: $\frac{w(\tau)}{\tau} = m + \frac{a}{\tau}$. The optimal policy may thus be to accept trips in $\sigma^* = (0, T)$ for some $T$. However, our next proposition establishes that affine pricing is incentive compatible if the additive component stays small enough as a function of the request arrival rate:

\begin{restatable}{proposition}{lemaffineICstationary}
	\label{lem:affineICstationary}
	With a single state, $w(\tau) = m\tau + a$ is incentive compatible if $0 \leq a \leq \frac{m}{\lambda}$.
\end{restatable}

The sufficient condition has a simple intuition: when open, the expected amount of time the driver must wait for the next request is $\frac{1}{\lambda}$; if on-trip time is valued at $m$ per unit-time, then with $a = \frac{m}{\lambda}$ the additive component can be interpreted as paying for the driver's expected waiting time. Thus, while a driver may earn more per hour for a short trip than a long trip with affine pricing, such a short trip is not worth the time the driver must wait for the next trip request. We further note that the condition in the proposition is not a necessary one; however, deriving necessary and sufficient conditions in closed form requires specifying the trip distribution $F$.

As we'll see in the next sub-section, the structure of optimal driver policies in reaction to affine pricing differs sharply in the dynamic model.

\subsection{Dynamic model: multiplicative pricing is not incentive compatible}
\label{sec:dynamicmodelaffine}
 In the single-state model, multiplicative pricing is incentive compatible; a driver cannot benefit in the future by rejecting certain trips if all trips have the same on-trip earning rate. In contrast, we now show that the same insight does not hold for the dynamic model, as a driver can influence future trips through the decision to accept or reject certain trips. %

\begin{restatable}{theorem}{propmultnotoptimal}
	\label{prop:multnotoptimal}
	If $w = \{w_1, w_2\}$, there exists an optimal policy $\sigma = \{\sigma_1, \sigma_2\}$ $($i.e., that maximizes $R(w, \sigma))$,
	defined with parameters $t_1, t_2, t_3, t_4, t_5,t_6 \in [0, \infty) \cup \{\infty\}$,
	such that

	\begin{itemize}
		\item Non-surge state driver optimal policy $\sigma_1$: %
		\begin{itemize}
			\item If $w_1$ is multiplicative or positive affine, $\sigma_1$ rejects long trips, i.e., $\sigma_1 = (0, t_1)$.
			\item If $w_1$ is negative affine, $\sigma_1$ rejects short and long trips, i.e., $\sigma_1 = (t_2, t_3)$.
		\end{itemize}

		\item Surge state driver optimal policy $\sigma_2$: %

		\begin{itemize}
			\item If $w_2$ is multiplicative or negative affine, $\sigma_2$ rejects short trips, i.e., $\sigma_2 = (t_4, \infty)$.
\item If $w_2$ is positive affine, $\sigma_2$ rejects medium length trips, i.e., $\sigma_2 =(0,t_5)\cup (t_6, \infty)$.
		\end{itemize}
	\end{itemize}

	Furthermore, there exist settings where $t_i$'s take positive finite values, and in which multiplicative pricing is not incentive compatible in either state. Finally, only policies of the appropriate form as indicated (up to differences of measure 0) can be optimal.
\end{restatable}
We discuss the intuition in the next section. %
In the appendix, we prove the result for each case as follows: fixing $\sigma_j$ for $j \neq i$, we start with an arbitrary open set $\sigma_i = \cup_{k}^\infty (\ell_k, u_k)$, recalling that open sets can be written as a countable union of such disjoint intervals. Then, we find $\frac{\partial} {\partial u_k} R(w, \sigma)$, the derivative of the set function $R(w, \sigma)$ with respect to one of the interval upper end-points of $\sigma_i$, i.e., $u_k$. This derivative is the infinitesimal change in the overall reward if $\sigma_i$ is expanded by increasing $u_k$, and it has useful properties. In the surge state with multiplicative pricing, for example, $\frac{\partial} {\partial u} R(w, \sigma)$ has the same sign as a function that is \textit{increasing} in $u$, for each fixed $\sigma$. With affine pricing, it has the same sign as a \textit{quasi-convex} (positive affine in the surge state) or \textit{quasi-concave} (negative affine in the non-surge state) function in $u$, for a fixed $\sigma$. Such properties enable constructing a sequence of changes to $\sigma_i$ that each do not decrease the reward $R(w, \sigma)$, with the limit being a policy of the appropriate form. In particular, we can show that any policy that is not of the appropriate form above has $\frac{\partial} {\partial u_k} R(w, \sigma) > 0$ for some $u_k$, allowing local improvements until adjacent intervals $(\ell_k, u_k), (\ell_{k+1}, u_{k+1})$ can be combined or expanded to infinity. The numerics in Section~\ref{sec:additivesurgesims} provide examples in which multiplicative pricing is not incentive compatible, i.e., where policies of the form above with positive finite constants strictly increase driver earnings over the driver policy that accepts all trip requests.

The results of rejecting long trips in non-surge (and short trips in surge) extend to arbitrary functions where $\frac{w_1(\tau)}{\tau}$ is non-increasing (respectively, $\frac{w_2(\tau)}{\tau}$ is non-decreasing). The other two results do not hold with such generality, as the behavior of the derivative may be arbitrarily complex.

\subsection{Why is multiplicative surge pricing not incentive compatible? } %
\label{sec:discussionmult}
~
\epigraph{``I thoroughly dislike short trips ESPECIALLY when I'm picking up in a waning surge zone''}{Anonymous driver}%
~\\ \\ What explains the difference between multiplicative pricing being incentive compatible in the single-state model but not in the dynamic model? In the latter, a driver's policy affects not just their earnings while they are busy, but also the fraction of time during which they are busy during the lucrative surge state. In particular, it turns out, accepting short trips during surge may \textit{reduce} the amount of time that a driver is on a surge trip! Appendix Figure~\ref{fig:percenttimespentinstate2} shows in an example how the fraction of time in the surge state $\mu_2(\sigma)$ changes as a function of how many short trips the driver rejects.%

The anonymous driver we quote above identifies the key effect: when surge is short-lived, a driver may only have the chance to complete one surge trip before it ends. Thus, the driver may be better off waiting to receive a longer trip request, as with multiplicative surge they are paid a higher rate for the full duration of the longer trip. (Of course, there is a trade-off as rejecting too many trip requests risks not receiving any acceptable request before surge ends). In the surge state, then, multiplicative pricing does not compensate drivers enough to accept short trips that may reduce their future surge earnings. In the non-surge state, analogously, multiplicative pricing under-values long trips that may prevent taking advantage of a future surge.

Affine pricing is a first, reasonable attempt at fixing these issues. In the surge state, the additive value makes the previously under-valued short trips comparatively more valuable, as the earnings per unit time $\frac{w_2(\tau)}{\tau} = m_2+ \frac{a_2}{\tau}$ (with $a_2> 0$) are now higher for short trips. Unfortunately, with such pricing the structure for the surge optimal policy becomes $\sigma_2 = (0,t_5)\cup (t_6, \infty)$ -- if the values $m_2, a_2$ are not balanced correctly, the additive value is enough to make accepting extremely short trips $(0, t_5)$ profitable; for medium-length trips $\tau \in (t_5, t_6)$, however, the additive value is not large enough to make up for the fact that accepting the trip prevents accepting another surged trip before surge ends. Similarly, negative affine pricing in the non-surge state, $w_1(\tau) = m_1\tau + a_1$, (with $a_1 < 0$) is now too harsh on very short trips but potentially not enticing enough for long trips.

Next, we fix these issues and construct incentive compatible pricing schemes for our dynamic model. Then, in Section~\ref{sec:additivesurgesims} we leverage structural results derived here to numerically compare the incentive compatibility of additive and multiplicative surge.

\section{Incentive Compatible Surge Pricing}
\label{sec:IC}
We now present our main result, regarding the structure of incentive compatible pricing in the dynamic model.
To this aim, in Section~\ref{sec:expectedtimeeachstate}, we characterize $\mu_i(\sigma)$, how much time the driver spends in each state.
In Section~\ref{sec:ic_pricing}, we present incentive compatible prices, under a condition on the ratio of per-state earning rate constraints, $\frac{R_1}{R_2}$. Section~\ref{sec:icintuition} discusses an intuition of the IC pricing structure in terms of the driver's opportunity cost.%

\subsection{Transition probabilities and expected time spent in each state}
\label{sec:expectedtimeeachstate}
The expected fraction of time spent in each state, $\mu_i(\sigma)$, depends both on the evolution of the world state and the trips a driver accepts. To quantify the effects previewed in Section~\ref{sec:discussionmult}, we first analyze the evolution of the world state CTMC.

\begin{restatable}{lemma}{lemfindingq}
	\label{lem:findingq}
	Suppose the world is in state $i$ at time $t$. Let $q_{i\to j}(s)$ denote the probability that the world will be in state $j \neq i$ at time $t + s$. Then,
\begin{align*}
	q_{i\to j}(s) &= \frac{\lambda_{i\to j}}{\lambda_{i\to j}+\lambda_{j\to i}}\left[1 - e^{-(\lambda_{i\to j} + \lambda_{j\to i})s}\right]\nonumber
	\end{align*}
\end{restatable}
Note that $q_{i\to j}(s)$ is not just the probability that the world state transitions once during time $(t, t + s)$, but the probability that it transitions an odd number of times. This formulation emerges through a standard analysis of two-state CTMCs, in which this probability can be found through the inverse of the Laplace transform of the inverse of the resolvent of the Q-matrix for the system. Incorporating this value in closed form is the main hurdle in extending our results to general systems with more than two states. Using this formulation, the following lemma shows $\mu_i(\sigma)$.

\begin{restatable}{lemma}{lembreakingdownmui}
	\label{lem:breakingdownmui}
	Let $T_i(\sigma_i)$ be as defined in Lemma~\ref{lem:statingreward}. The fraction of time a driver following strategy $\sigma = \{\sigma_1, \sigma_2\}$ spends either open in state $i$ or on a trip started in state $i$ is
	\begin{align*}
	\mu_i(\sigma) &= \frac{	\lambda_i F_i(\sigma_i)T_i(\sigma_i)Q_j(\sigma_j)}
	{\lambda_j F_j(\sigma_j)T_j(\sigma_j)Q_i(\sigma_i)	+ 	\lambda_i F_i(\sigma_i)T_i(\sigma_i)Q_j(\sigma_j)}\\
	\text{where\,\,\,\,\,\,}
	Q_i(\sigma_i) &= \lambda_{i\to j} + \lambda_i\int_{\tau \in \sigma_i} q_{i\to j}(\tau) dF_i(\tau)
	\end{align*}
\end{restatable}

We prove this lemma by finding the expected number of sub-cycles in each state $i$, i.e., within a larger renewal reward cycle as defined, the expected number of sub-cycles that start with the driver being open in state $i$. This expectation is the mean of a geometric random variable parameterized by the probability that the driver will next be open in state $j$, given the driver is currently open in state $i$.
$Q_i(\sigma_i)$ is proportional to this probability. (As with $T_i(\sigma_i)$, there is a normalizing constant $\frac{1}{\lambda_i F_i(\sigma_i) + \lambda_{i\to j}}$); the larger it is, the fewer sub-cycles spent in state $i$. It has two components: the first is the probability that the state changes before the driver accepts a trip request; the second is the probability that the world state is $j$ when the driver completes a trip. Thus, the numerator in $\mu_i(\sigma)$ is proportional to the length of a sub-cycle in state $i$, times the fraction of sub-cycles that are started in state $i$. The larger $Q_j(\sigma_j)$ or $T_i(\sigma_i)$, the more time the driver spends in state~$i$.

\subsection{Incentive Compatible pricing in the dynamic model} \label{sec:ic_pricing}

How can the platform create incentive compatible pricing given the previously described effects? Our main result establishes when such IC prices exist, and reveals their form.

\begin{restatable}{theorem}{thmicpolicy}
	\label{thm:ICpolicy} %
Let $R_1 < R_2$ be target earning rates during non-surged and surge states, respectively.
There exist %
prices $w = \{w_1, w_2\}$ of the form \[w_i(\tau) = m_i\tau + z_i q_{i \to j}(\tau),\]
where $m_1,m_2,z_2 \geq 0$ $($but $z_1$ may be either positive or negative$)$, such that the optimal driver policy is to accept every trip in the surge state and all trips up to a certain length in the non-surge state. Furthermore, for $\frac{R_1}{R_2} \in (C,1)$, there exist fully incentive compatible prices of this form, where %
\begin{align*}
C&=	1 - \frac{1}{T_1}\frac{Q_2 (\lambda_{12}T_1 - Q_1) + Q_1(T_2\lambda_{1\to 2} + Q_2) }{Q_2 (\lambda_{1\to 2}T_1 - Q_1) + \lambda_{1\to 2}(T_2\lambda_{1\to 2} + Q_2)} \in [0,1),
\end{align*}
and $T_i = \lambda_iF_i(\sigma_i)T_i((0, \infty))$, and $Q_i = Q_i((0, \infty))$. For such prices, the driver policy to accept all requests is the unique optimal driver policy (up to differences of measure 0).
\end{restatable}

\begin{figure*}[t!]
	\centering
	\begin{subfigure}[t]{0.48\textwidth}
		\centering
		\includegraphics[width=\linewidth]{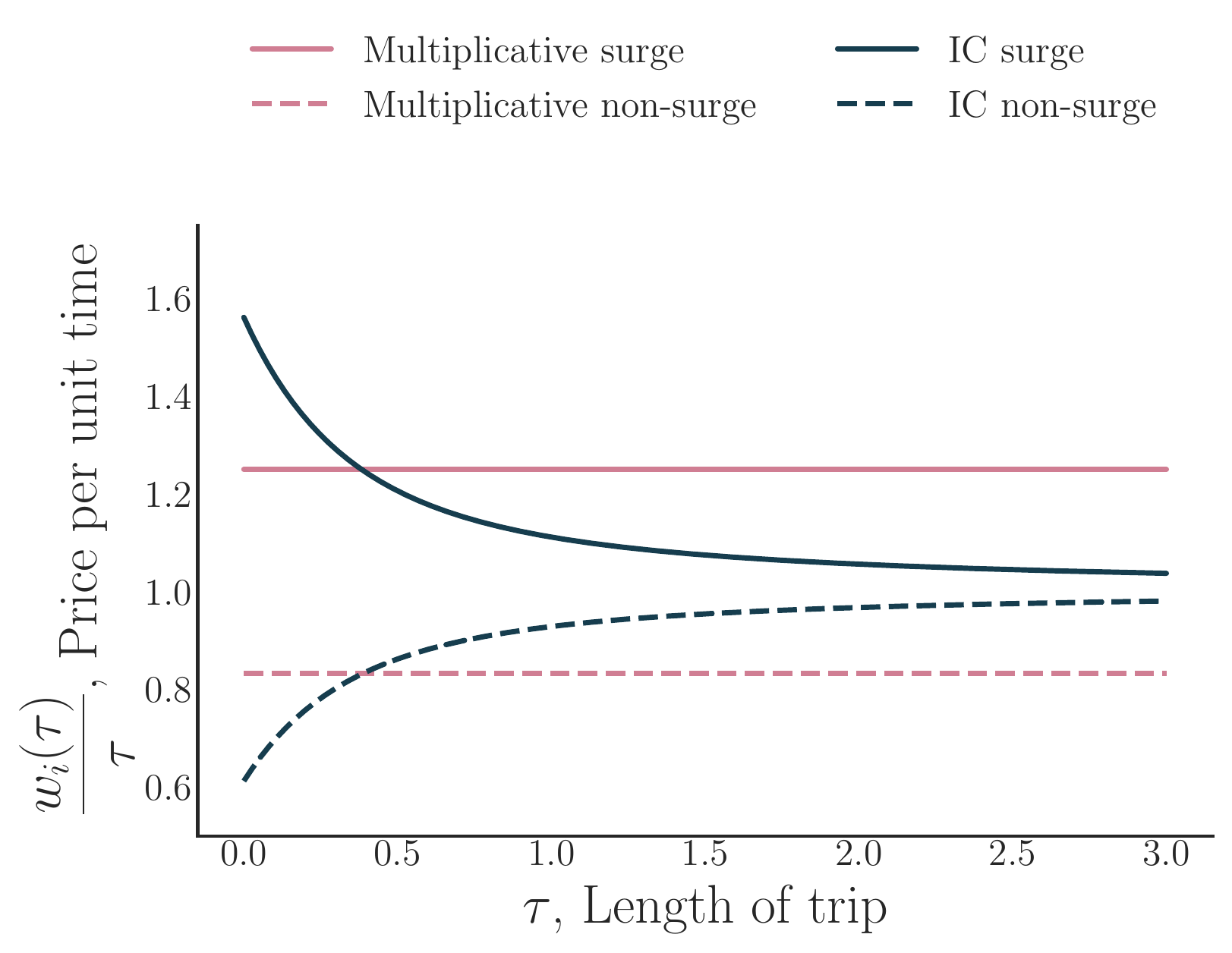}
		\caption{Price per unit time $\frac{w_i(\tau)}{\tau}$ for trips of different lengths $\tau$ in the each state for Incentive Compatible and multiplicative pricing when $R_2 = 1$ and $R_1 = \frac{2}{3}$.}
		\label{fig:surgeplotpricingcomparative}
	\end{subfigure}
	~ \hfill
	\begin{subfigure}[t]{0.48\textwidth}
	\centering
	\includegraphics[width=\linewidth]{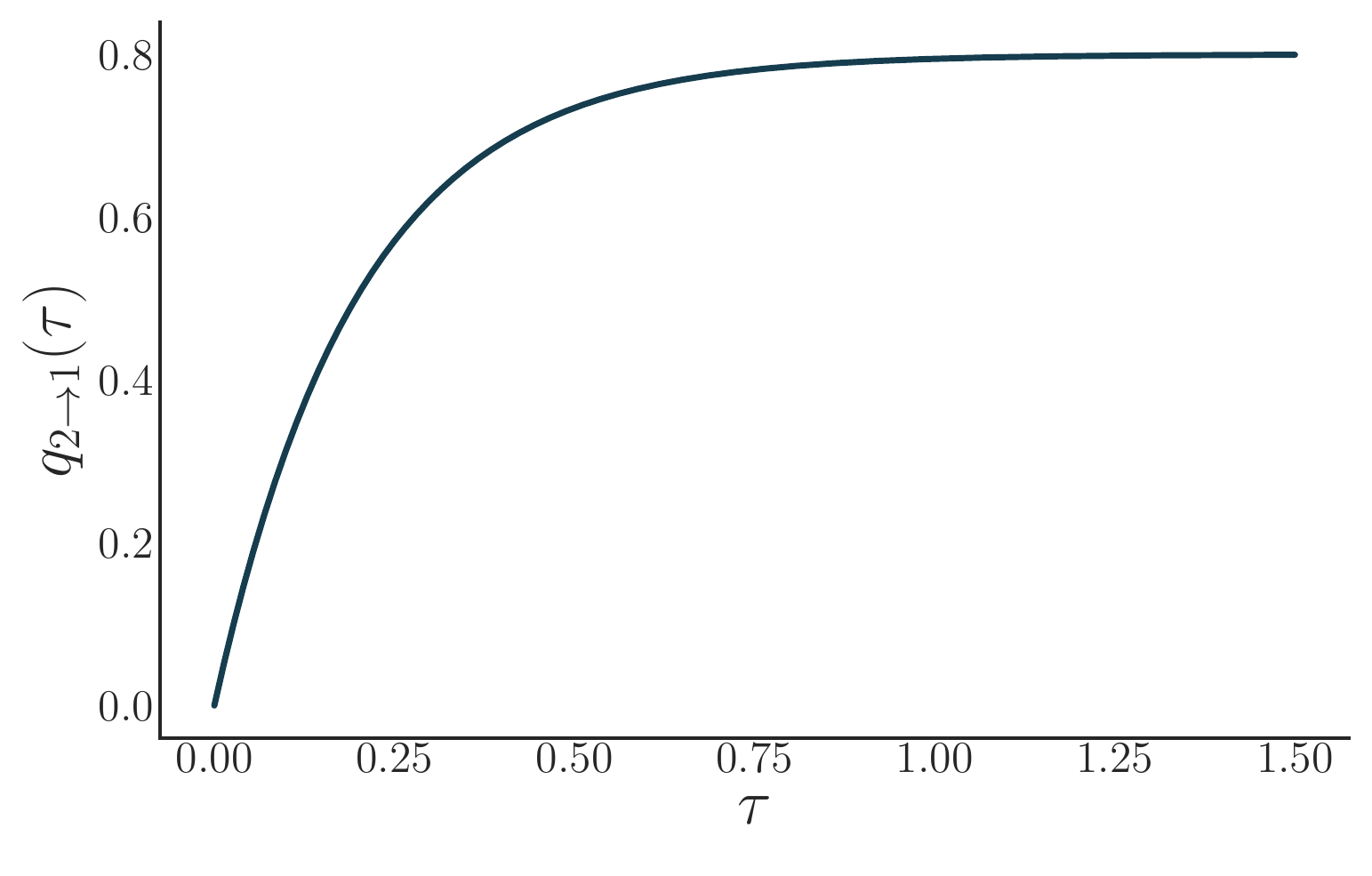}
	\caption{$q_{2\to 1}(\tau)$ when $\lonetwo = 1, \ltwoone = 4$. IC surge pricing is well-approximated by an affine function: $z_2 q_{2\to 1}(\tau)$ is approximately constant for longer trips.}%
	\label{fig:q21}
\end{subfigure}
\caption{The primitives are as follows: $\lambda_1 = \lambda_2 = 12, \lonetwo = 1,\ltwoone = 4$; in both states, trip lengths are distributed according to a Weibull distribution with shape $2$ and mean $\frac{1}{3}$. These parameters reflect realistic average trip to wait time values, and that surge tends to be short-lived compared to non-surge times.}
	\label{fig:surgepolicyexample}
\end{figure*}

Section~\ref{sec:ic_theorem_proof_sketch} contains a proof sketch.
To convey intuition, Figure~\ref{fig:surgeplotpricingcomparative} shows pricing functions in each state, plotting $\frac{w_i(\tau)}{\tau}$ against $\tau$.
Compared to multiplicative pricing with constant $\frac{w_i(\tau)}{\tau}$, IC surge pricing pays more for short trips and less for long trips. Inversely, IC non-surge pricing pays more for long trips than it does for short trips. Further, as $\tau$ increases, $w_1(\tau)$ approaches $w_2(\tau)$, reflecting the fact that the opportunity cost for long trips does not depend as strongly on the state in which it started (as discussed in Section~\ref{sec:icintuition}). Next, observe that IC surge pricing $w_2(\tau) = m_2\tau + z_2q_{2 \to 1}(\tau)$ is approximately affine, as $q_{2 \to 1}(\tau)$ (plotted in Figure~\ref{fig:q21}) is upper bounded by $\frac{\ltwoone}{\lonetwo + \ltwoone}$. The two components of pricing, $m_i$ and $z_i$, thus balance the comparative benefit of long and short trips.  We give further intuition for the form of payment scheme $w_i$ and the range $[C,1]$ in Section~\ref{sec:icintuition}, showing how they emerge from the driver's opportunity cost.

\begin{figure*}[t!]
	\centering
	\begin{subfigure}[t]{0.48\textwidth}
		\centering
		\includegraphics[width=\linewidth]{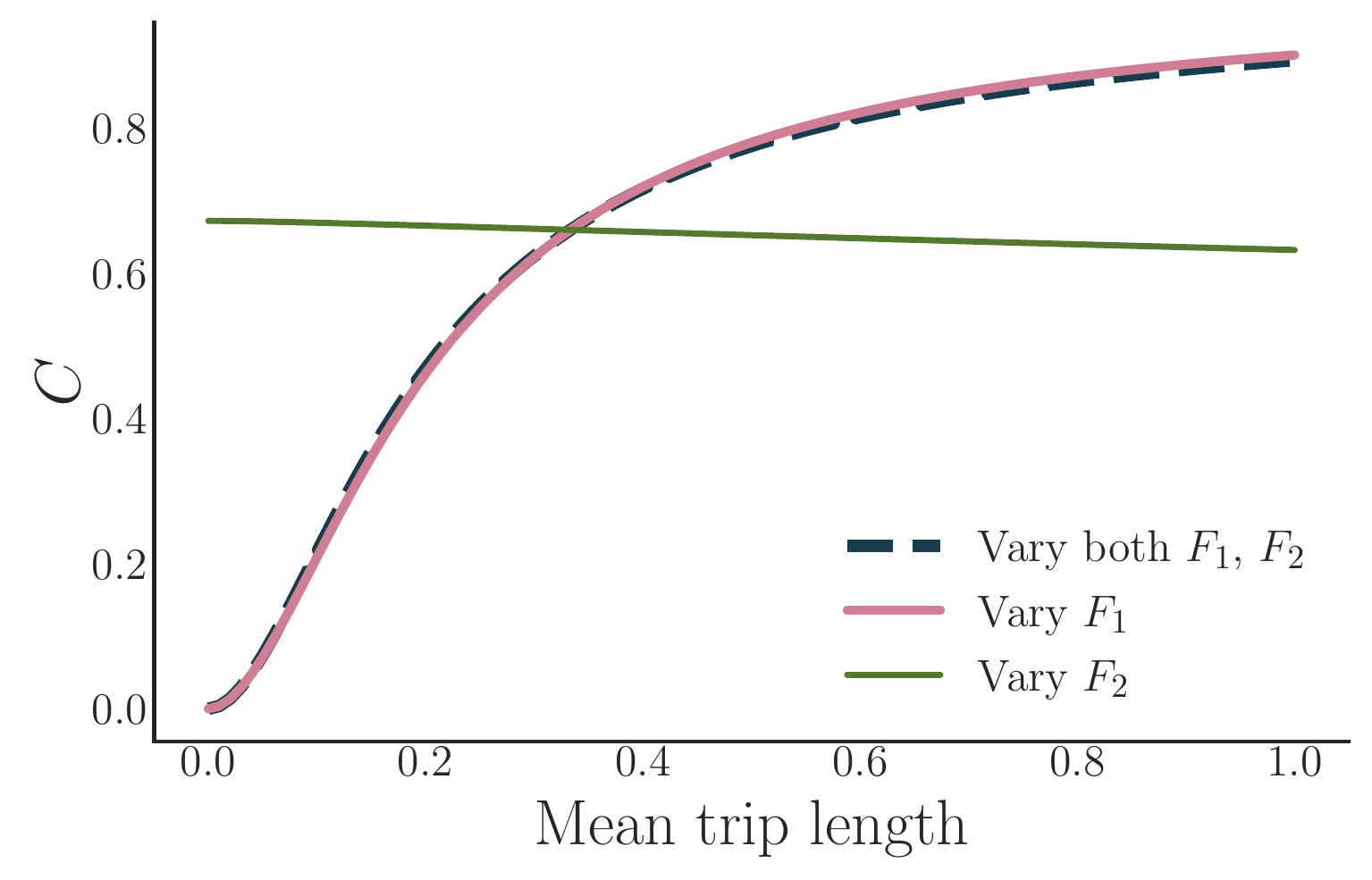}
		\caption{$C$ as the mean trip length changes. }
		\label{fig:C_change_with_triplength}
	\end{subfigure}
	~ \hfill
	\begin{subfigure}[t]{0.48\textwidth}
		\centering
		\includegraphics[width=\linewidth]{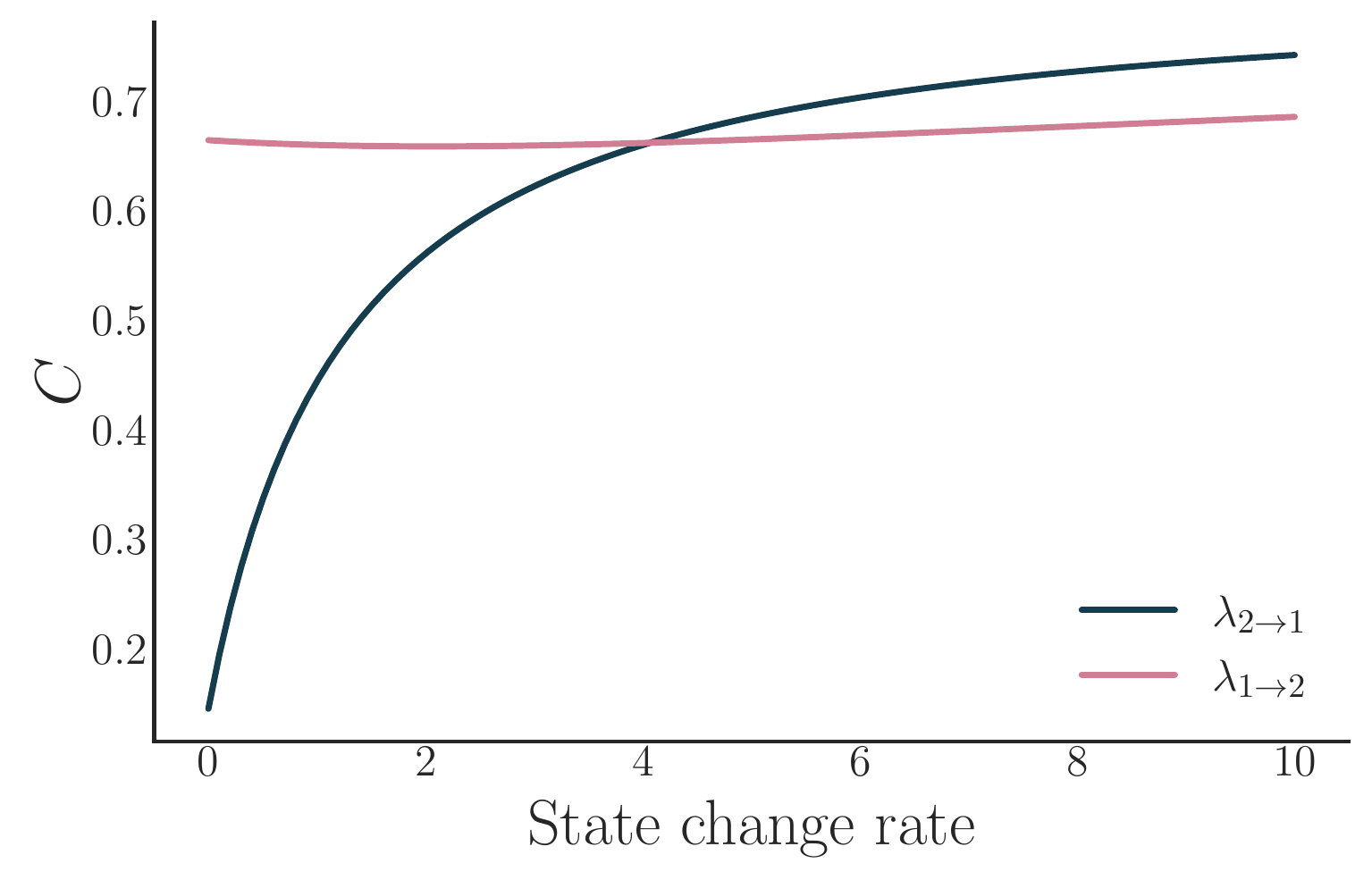}
		\caption{$C$ as $\lambda_{i \to j}$ change.}
		\label{fig:C_change_with_allrates}
	\end{subfigure}%
	\vspace{0.3em}
	\caption{How $C$, the ratio $R_1/R_2$ at which IC pricing is feasible from Theorem~\ref{thm:ICpolicy}, changes (1) with respect to the mean trip length, and (2) with respect to $\lambda_{i \to j}$. Except for those that are varied in each plot, the primitives are fixed to those used in Figure~\ref{fig:surgepolicyexample}: $\lambda_1 = \lambda_2 = 12, \lonetwo = 1,\ltwoone = 4$ and, in both states, trip lengths are distributed according to a Weibull distribution with shape $2$ and mean $\frac{1}{3}$.
}
	\label{fig:Cchanges}
\end{figure*}

Rather surprisingly and contrary to platform design focus, the \textit{non-surge} state is difficult to make incentive compatible. Our result establishes that there always exist payments, for any target driver earning rates $R_1 < R_2$, such that accepting every trip in the surge state is driver optimal; the same is not true for the non-surge state.\footnote{For $\frac{R_1}{R_2}$ small enough, no pricing function $w_1$ can be incentive compatible in non-surge periods. A driver would rather wait for the far more lucrative surge state.} Figure~\ref{fig:Cchanges} shows how $C$ changes with the primitives.

 Finally, for a given feasible $R_1, R_2$, there is a range of $m_i,z_i$ that form an incentive compatible pricing scheme. Why? A driver who rejects a trip request waits to receive another request, during which time they do not earn money. This wait time tilts the driver toward accepting any trip request to maximize earnings. Thus, there is flexibility in the balance between short and long trip earnings. The same insight drives Proposition~\ref{lem:affineICstationary}; even in the single-state model, trips do not have to have the same earnings per unit time, $\frac{w(\tau)}{\tau}$, as long as they meet some minimum threshold, $ \frac{w(\tau)}{\tau} \geq c_w$.

\subsection{Opportunity cost intuition for incentive compatible pricing}
\label{sec:icintuition}
We now present some intuition to understand Theorem~\ref{thm:ICpolicy} and our incentive compatible pricing scheme.
The payment $w_i(\tau)$ must account for the driver's opportunity cost (in a VCG-like manner), i.e., how much the driver can expect to earn if they instead reject the trip request. Of course, this opportunity cost itself depends on the pricing scheme $w$. We now break down parts of this opportunity cost.

\paragraph{On-trip opportunity cost.} While the driver is on-trip, the world state continues to evolve: surge might end or start, affecting the opportunity cost.%

 Let $\phi^k_i(\tau)$ be the expected amount of time that the world is in state $k$ during time $(t, t + \tau)$, given that it is in state $i$ at time $t$. Then, by integrating $q_{i\to j}(s)$ from $0$ to $\tau$: %
\begin{align*}
\phi^i_i(\tau) &= \left[\frac{\lambda_{j\to i}}{\lambda_{i\to j}+\lambda_{j\to i}}\right]\tau + \left[\frac{1}{\lambda_{i\to j}+\lambda_{j\to i}}\right]q_{i\to j}(\tau)\\
\phi^j_i(\tau) &= \left[\frac{\lambda_{i\to j}}{\lambda_{i\to j}+\lambda_{j\to i}}\right]\tau - \left[\frac{1}{\lambda_{i\to j}+\lambda_{j\to i}}\right]q_{i\to j}(\tau) = \tau - \phi^i_i(\tau)%
\end{align*}
Several insights emerge:

	One. As trip length $\tau \to \infty$, the first summand of each of $\phi^i_i(\tau), \phi^j_i(\tau)$ dominates, and this component does not depend on starting state $i$. As $\tau \to \infty$, we have $\phi^i_i(\tau) = \phi^i_j(\tau)$, $\phi^j_i(\tau) = \phi^j_j(\tau)$.
	 The stationary distribution of a positive recurrent CTMC does not depend on the starting state. We cannot always construct incentive compatible prices, for any $R_1, R_2$: as $\tau \to \infty$, the opportunity cost does not depend on the starting state $i$, and so payments must be similar, $w_1(\tau) \approx w_2 (\tau)$. When all non-surge trips are long, i.e., $F_1$ is concentrated around large values, the earnings rate in each state must be similar, $R_1 \approx R_2$.

$C$ encodes such constraints, as shown in Figure~\ref{fig:Cchanges}. As the mean of $\tau \sim F_1$ goes to $0$, then $\lambda_{12}T_1 - Q_1 \to 0$ and so $C\to 0$, and so the range of feasible $\frac{R_1}{R_2}$ expands. Similarly, $\lambda_{2 \to 1}$ also plays an important role. When small, the surge state is long. Thus, a driver will receive many trips during surge regardless of how long their last non-surge trip is---and so long trips during non-surge are no longer constrained to be highly paid compared to short trips.

	Two. The expected time spent in each state has the form, $m'_i \tau + z'_i q_{i\to j}(\tau)$, matching the form of our IC scheme. Thus, we can expect the ``network minutes'' on-trip opportunity cost -- the expected earnings during the time the driver would otherwise be on the given trip -- to have the same form as well.

\paragraph{Continuation value opportunity cost} It is not sufficient to consider just the opportunity cost for the duration of the trip: the driver's counter-factual earnings by rejecting the trip depends on future trips accepted. Such counter-factual trips both (1) pay the driver according to their starting state even after a world state transition, i.e., the difference between $R_i$ and $\tilde{R}_i$ above; and (2) potentially are still in progress past time $t + \tau$, when the current trip ends. This second complication is illustrated in Figure~\ref{fig:percenttimespentinstate2}, where a driver can extend the time spent on trips starting in the surge state by rejecting short surge trips. The effect depends on the lengths of future potential trips, i.e., $T_i(\sigma_i)$, and state transitions during those trips, $Q_i(\sigma_i)$, and is incorporated in both $C$ and the pricing scheme. %

\subsection{Proof sketch of Theorem~\ref{thm:ICpolicy}} \label{sec:ic_theorem_proof_sketch}
 The result is shown in the appendix %
 by manipulating the derivative of the reward function with respect to the policy $\sigma$. In particular, when the pricing function is of the given form with the appropriate constants $m_i,z_i$, then \textit{any} policy $\sigma = \{\sigma_1, \sigma_2\}$ can be locally improved by adding more trips to it, i.e., the overall reward is increasing as the driver accepts more trips: $R(w, \sigma') > R(w, \sigma), \forall \sigma \subsetneq \sigma'$. This result follows from $\paru R(w, \sigma) > 0$, for all $u, \sigma$, given the constraints, where $u$ is an upper endpoint of the policy in a state, $\sigma_i = \cup_k (\ell_k, u_k)$.

 The key step is finding sufficient constraints for this derivative to be positive with a pricing function of the given form, given any $\sigma_i$, as opposed to just $\sigma_i = (0, \infty)$. This difficulty emerges because incentive compatibility is a global condition on the set function $R(w, \sigma)$. In particular, we need to express these constraints simply---e.g., as a function of just $T_i((0, \infty)), Q_i((0, \infty))$, instead of the values $T_i(\sigma_i), Q_i(\sigma_i), \forall \sigma_i \subseteq \bbrp$. The $C$ presented in the theorem statement results from such a set of constraints on $m_i,z_i$.

\section{Numerics: Incentive Compatibility with Additive Surge}
\label{sec:additivesurgesims}

 We now analyze surge policies that reflect practice at ride-hailing platforms today. Non-surge pricing is typically approximately \textit{multiplicative}, i.e., $w_1(\tau) = m_1 \tau$, where $m_1$ is the base time (and distance) rate for a ride. We consider two types of affine \textit{surge} pricing $w_2$, which differ in their relationship to $w_1$ through a single parameter:
\begin{align*}
\text{\textbf{Multiplicative surge:}}\ \ \ \ \ \ \ \ \ \ \ \  & w_2(\tau) = m_2\tau &  m_2 \geq m_1\\
\text{\textbf{Additive surge:}}\ \ \ \ \ \ \ \ \ \ \ \  & w_2(\tau) = m_1\tau + a_2 &  a_2 \geq 0
\end{align*}
Multiplicative surge uses a multiplier $m_2$ larger than the base fare $m_1$, and $\frac{m_2}{m_1}$ is reported on the heat-map as in Figure~\ref{fig:multi_heatmap}; additive surge uses the {same} base fare multiplier $m_1$ but adds a factor $a_2$ that is  reported on the heat-map as in Figure~\ref{fig:add_heatmap}. These functions are trivial to calculate, given fixed primitives and target earnings rate $R_2$ in the surge state.%

Figure~\ref{fig:surgepolicyexample_affine} in the Appendix shows these types of pricing, compared to the incentive compatible pricing function. Multiplicative surge has constant $\frac{w_2(\tau)}{\tau}$ and so under-pays short trips and over-pays long-trips compared to IC pricing. Additive surge asymptotically (for large $\tau$) pays the same as multiplicative non-surge pricing, i.e. $\lim_{\tau\to\infty} \frac{w_2(\tau)}{\tau} = \lim_{\tau\to\infty} \frac{w_1(\tau)}{\tau} = m_1$. As a result, it over-pays short trips and under-pays long trips compared to IC surge pricing.

Uber has recently started a transition from multiplicative to additive surge. In this section, we argue that the additive component is more important than the multiplicative component for incentive compatibility in parameter regimes of interest.%

\subsection{Computing optimal driver policies }
Theorem~\ref{prop:multnotoptimal} establishes that multiplicative pricing (and, more generally, affine pricing) may not be incentive compatible in general. However, we still wish to compare the various types of surge pricing, and to analyze the regimes under which each is incentive compatible.

However, to do this comparison, one needs to calculate optimal driver policies with respect to a pricing function. Recall that the optimal driver policy in each state $\sigma_i$ is some subset of $\bbrp$. Finding such optimal subsets for general pricing functions $w$ is intractable, and so Theorem~\ref{prop:multnotoptimal} is particularly important for {computational} reasons. It establishes that, for any affine pricing structure in the surge state, all driver optimal policies are of the form $(0, t_1) \cup (t_2, \infty)$, for some $t_1, t_2$. We only need to find the values for these parameters that maximize the driver reward among sets of this form, and the resulting policy is optimal; this search is tractable with grid search and numeric integration.
Note that the proposition does not establish uniqueness of driver optimal policies; we thus choose the policy that maximizes the fraction of trips accepted in our computations.

\subsection{Results}

\begin{figure*}[t!]
	\centering
	\begin{subfigure}[t]{0.48\textwidth}
		\centering
		\includegraphics[width=\linewidth]{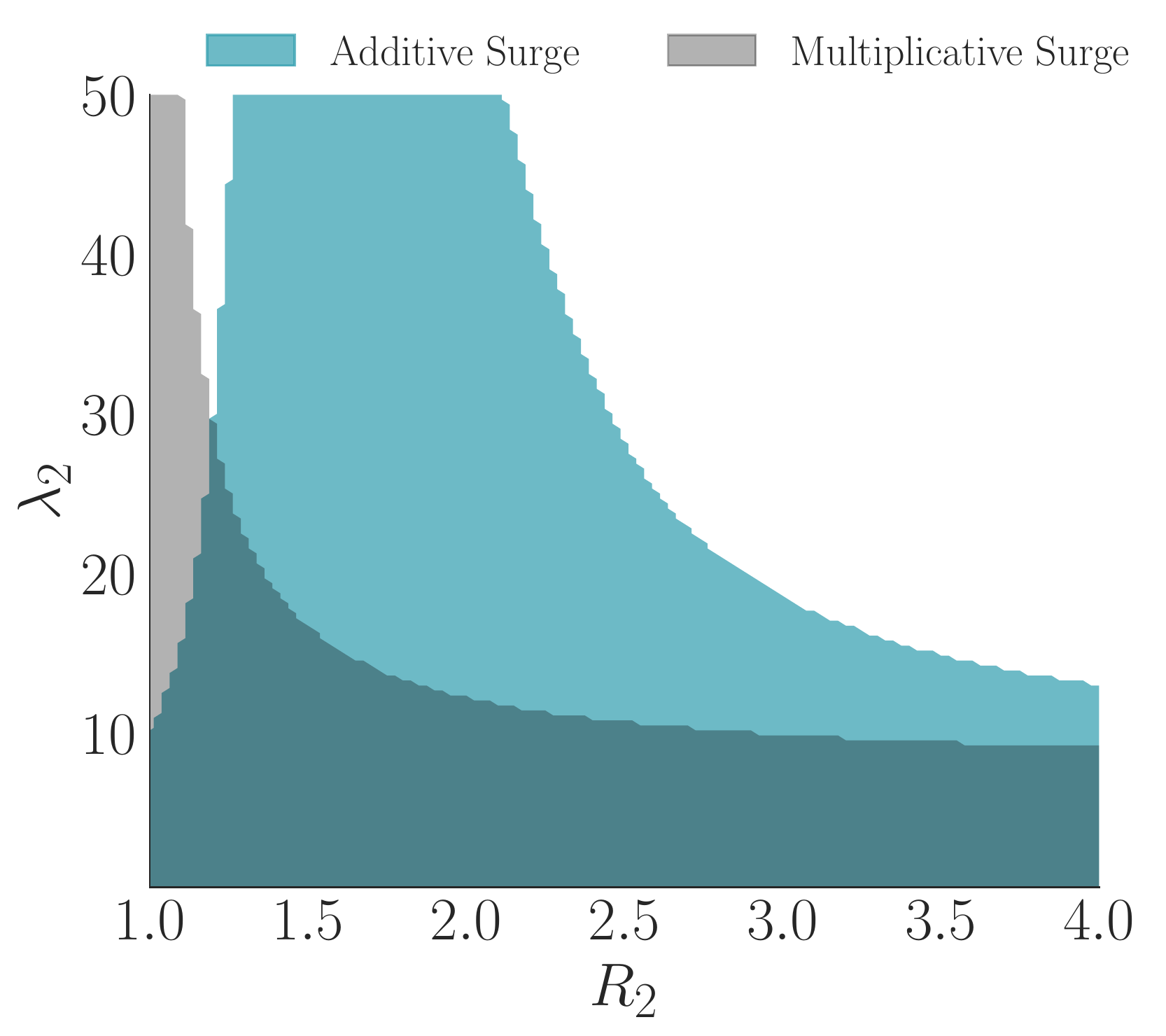}
		\caption{With $R_2$, surge state earnings rate, and $\lambda_2$, surge state job arrival rate. $R_2 \in [1.1, 3]$ is common in practice.}
		\label{fig:fracacceptedchangewithR1}
	\end{subfigure}
	~ \hfill
	\begin{subfigure}[t]{0.48\textwidth}
		\centering
		\includegraphics[width=\linewidth]{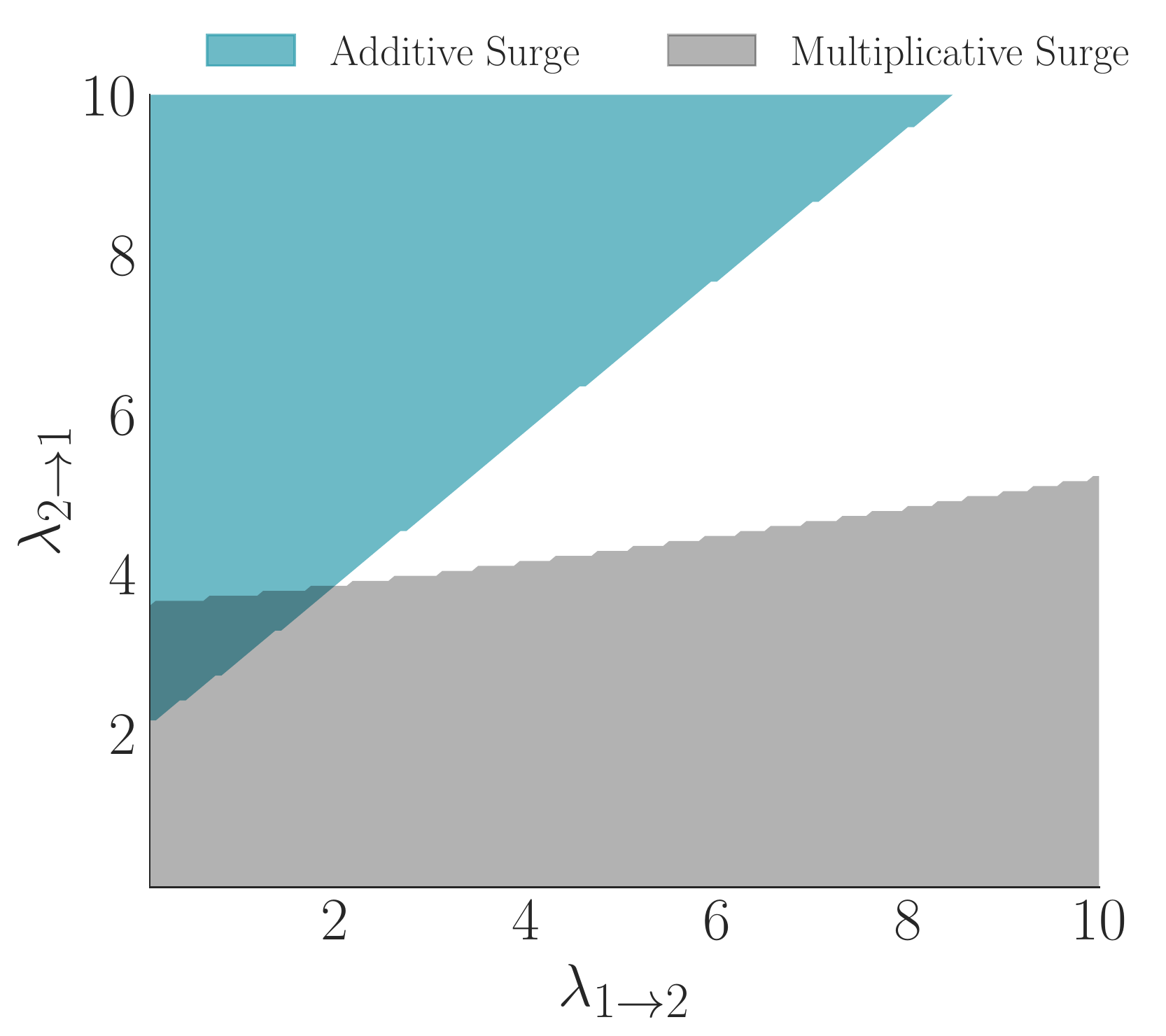}
		\caption{With $\ltwoone,\lonetwo$, rates for world state changing. $\ltwoone \gg \lonetwo$ is common in practice.}
		\label{fig:switchltwoone}
		\end{subfigure}
	\caption{Incentive compatibility for each type of surge. The shaded regions are where the respective scheme is incentive compatible in the surge state ($\sigma_2 = (0, \infty)$ is optimal). When not varied, $\lambda_1 = \lambda_2 = 10, \lonetwo = 1, \ltwoone = 4, R_2 = {3.33}, R_1 = 1$, and trip lengths in both states are distributed according to a Weibull distribution with shape $2$ and mean $0.3$. We assume every trip is accepted in the non-surge state.}
	\label{fig:affineapprox}
\end{figure*}

We now study the regimes in which each surge mechanism is incentive compatible. The shaded regions in Figure~\ref{fig:affineapprox} correspond to areas where the surge pricing function is fully incentive compatible in the surge state ($\sigma_2 = (0, \infty)$ is optimal). For example, when $R_2 = 2, \lambda_2 = 30$, additive surge is incentive compatible, but multiplicative surge is not.

As illustrated in Appendix~\ref{appsec:empiricalevidence} with data from the RideAustin marketplace, ride-hailing platforms most often operate in the following parameter regimes: (1) surge is between $1.1$ and $3$ times more valuable than non-surge; (2) surge is short-lived compared to non-surge periods ($\ltwoone \gg \lonetwo$); (3) and in a typical surge the driver receives several trip requests ($\frac{\lambda_2}{\ltwoone}> 1$, but small) but only completes one or two such trips ($\frac{1}{\ltwoone} \approx$ mean trip length).
Additive surge is incentive compatible in much more of this regime than is multiplicative surge, supporting Uber's recent shift from multiplicative to additive surge.  %

We can also draw qualitative insights in terms of sensitivity to the primitives, similar in spirit to effects in the form of $C$ in Theorem~\ref{thm:ICpolicy}.
Figure~\ref{fig:fracacceptedchangewithR1} shows the sensitivity with respect to $\lambda_2$ and $R_2$. As the arrival rate of jobs in the surge state, $\lambda_2$, increases, it becomes optimal for the driver to reject some trips:  ``cherry-picking'' becomes easier, as the driver is likely to receive many more trip requests before surge ends.
Similarly, as surge becomes increasingly more valuable compared to non-surge ($R_2$ increases), the incentive to reject non-valuable trips in the surge state increases.%

Additive surge contains an interesting non-monotonicity: when $R_2 \gg R_1$, the effect above dominates, and long trips are rejected.  When the surge state is moderately more valuable than non-surge, additive surge effectively balances the payments for different trip lengths and so is incentive compatible. When the two states are nearly equally valuable, again the optimal driver policy rejects long trips: our single-state model approximates the system, and so additive surge may not be incentive compatible, cf. Theorem~\ref{thm:basicmodeloptimal}.

Figure~\ref{fig:switchltwoone} shows the effects of the relative lengths of surge and non-surge. Here, the two types of surge are incentive compatible in opposing regimes. When $\frac{\ltwoone}{\lonetwo}$ is large, surge is comparatively rare and short, and so short trips are naturally under-valued -- accepting them decreases the time spent in the surge state -- and additive surge is incentive compatible. With long-lasting surge (small $\frac{\ltwoone}{\lonetwo}$), on the other hand, the world almost seems unchanging during surge, and so multiplicative surge becomes incentive compatible.

The short, in-frequent surge setting  -- in which additive surge is preferable -- is pre-dominant in the RideAustin data used in Section~\ref{sec:empiricalevidence}.	Nevertheless, our analysis suggests that when surge is expected to last throughout the day, such as with a predictable demand shock, multiplicative surge may be preferable. (However, switching between different payment functions may be undesirable for transparency and communication reasons).

\section{Empirical Comparison of Surge Mechanisms}
\label{sec:empiricalevidence}

We now study how the various surge mechanisms affect driver earnings in practice using publicly available trips data from RideAustin, a nonprofit ride-hailing company based and operating in Austin, Texas. We show that additive surge effectively balances the relative value of short and long surged trips, in contrast to the multiplicative surge pricing scheme used in practice by the platform, which comparatively undervalues short surged trips.

After reverse-engineering the functional form of the actual driver payments, we calculate both status quo (with multiplicative surge) and simulated (with additive surge) driver earnings. For each payment scheme, we estimate the driver's value in receiving and accepting a given trip request, as a function of the trip---where ``value'' is the increase (or decrease) in the driver's earnings over the next 90 minutes as a result of accepting the given request.

We note that this data is not the result of an experiment with additive surge, and thus our analysis describes what changes would occur in driver earnings with the new pricing function {\em if driver behavior does not change}.\footnote{We are not concerned with rider behavior changing, as with decoupled pricing the rider pricing can remain the same even as the driver payments change.} Thus, the additive surge exercise is a calibrated simulation for such pricing functions in a realistic setting: such as when surge has more than two levels and may not evolve in a Markovian manner, the driver is not paid for the time it takes to drive to the rider, and where location plays a role. Furthermore, as the data observed is at the {\em completed} trip level (i.e., requests which the driver accepted), results showing that the driver would be better off accepting the same trip in the counter-factual world should directionally hold even as driver behavior changes.

This section is organized as follows: Section~\ref{sec:datadescription} describes the data,  context, and analysis, and Section~\ref{sec:dataanalaysis} contains results. %
Appendix Section~\ref{appsec:empiricalevidence} contains supporting details, and both the data~\citep{rideaustin} and our replication code is available online.\footnote{\url{https://github.com/nikhgarg/driver_surge_rideaustin}
}

\subsection{Data setting and analysis description}
\label{sec:datadescription}

This analysis is enabled by the rich dataset, spanning from June 2016 to April 2017, during which RideAustin experienced tremendous growth and was one of the largest ride-hailing marketplaces serving the area. The data is at the completed trip level.~\citet{komanduri2018assessing} study the same dataset and provide useful statistics about driver earnings, platform growth, and the service's relationship to public transportation.

We consider the period from February 16, 2017, to April 10, 2017, as (1) we can reliably reverse engineer the platform's payment function during this period, and (2), the underlying marketplace was fairly stable during this period, except for one week of high, atypical demand and surge, corresponding to the SXSW Music Festival held in Austin. (Figure~\ref{fig:tripsperday} in the Appendix shows the trips per day during this period). We discard trips longer than 1 hour or shorter than 30 seconds and other trips with data errors; \num{6440} such trips were discarded. We analyze \num{503383} completed trips by \num{3811} drivers. (For analyses aggregating multiple trips, such as driver earnings in a given time period, we discard aggregations that include a discarded trip). The full pre-processing sequence is described in the appendix.

Several dataset features make it attractive for our analysis when compared to other publicly available ride-hailing datasets. Most importantly, there are consistent {\em driver ID}s attached to each trip. Second, for each trip, there is a value for the {\em total fare} paid by the rider, along with terms that contribute to this calculated fare: {\em trip duration (in time and distance)}, {\em payment rate (in time and distance)}, {\em surge factor}, {\em standard additive fare (Pickup)}, and {\em trip class} (Regular vs Luxury vs SUV).\footnote{Our results include trips from all trip classes, as a given driver may be cross-dispatched across trip classes.} These features allow us to track a driver's trajectory and earnings over a day and the entire year, reverse engineer how RideAustin calculates payments, and simulate additive surge payments.

\subsubsection{Constructing payment functions}
\label{sec:empiricalpricingfunc}
To simulate driver earnings with additive surge, we must first reverse engineer how the platform's actual {\em total fare} was calculated, a non-trivial task as the calculation changes over time in the dataset and is not documented. We find that this \textbf{status quo fare} is approximately:\footnote{The payment includes a multiplier of 1.01 and an additive value of 2.02. From publicly available information, we assume that the platform takes a fixed commission independent of trip length, and so the driver receies everything but the $\$2.02$~\citep{rideaustinrates}. On average, this reversed engineered fare differs from {\em total fare} by less than $1$ cent.}
\[
\max(B + \text{\em Pickup},\ \text{\em MinFareForClass})\times \text{\em SurgeFactor}.
\]
$B \triangleq (DistanceRate\times Distance) + (TimeRate \times Time)$ is the trip time and distance fare, only counting when the rider is in the car (recall that current practice deviates from the theory in that driving to the rider is typically unpaid). {\em MinFareForClass} is $\$4$ for Regular trips and $\$10$ otherwise. {\em SurgeFactor} of $1$ indicates no surge, comprising $70\%$ of trips. It increments in multiples of $0.25$, and  $97\%$ of surged trips have a factor of at most $3$. Each of the above payment components are given as columns in the dataset.

Then, we construct the following payment for each trip, to simulate how the driver would be paid with additive surge, i.e., \textbf{Additive surge with base fare}:
\begin{align*}
\max(B + \text{\em Pickup},\ \text{\em MinFareForClass}) + \left[(\text{\em SurgeFactor}-1)\times A_{SurgeFactor}.\right]
\end{align*}
$A_{SurgeFactor}$ are (calculated) surge factor dependent constants that are set such that this alternative payment function spends the same amount of money overall for each surge factor as does the status quo fare. In other words, the alternative payment does not change the mean trip payment conditional on the surge factor, but does change how money is allocated to various trips within that surge. This choice reflects our theory in assuming an exogenous $R_i$ and removes any degrees of freedom in setting $A_{SurgeFactor}$. If instead we used a single constant across surge factors, Additive surge with base fare may pay different amounts on average for the same surge factor than does the status quo fare.

\subsubsection{Matching open drivers}
\label{sec:matchingdescription}

We are interested in the value of a trip request to a driver $A$; to calculate this value, we need a measure of the counter-factual: what would have happened if the driver does not accept (or does not receive) a trip of length $\tau$. We \textit{match} the focal driver $A$ of each given completed trip to a nearby driver $B$ who is also open to receive a trip request at the time of the request. Driver $B$'s earnings then serve as a counter-factual for focal driver $A$'s earnings had driver $A$ rejected the request.

We estimate matches for each focal driver $A$ as follows. We observe trip start and end times and locations but not driver locations when they are not on a trip or even whether they still have their app open. We also observe the time at which a driver received a given trip request but not their location at this time, due to what seems like a data export bug.

This data does not allow us to simply query for other open drivers nearby who could have (but did not) receive a given trip request, as we do not directly observe drivers' movements while they are not on a trip.
Instead, we leverage recent, nearby completed trips to identify drivers who must still be nearby, as follows.

First, we define a ``matching distance'' between pairs of ({\em date-time}, {\em location}) tuples. Events with small matching distances occur nearby and at similar times. The exact function with how time and geographic distance are weighted is specified in the appendix. For driver's $A$'s time and location, we use the trip's {\em start} location (where the rider was) and the {\em dispatch} time (when the rider's request was accepted).
Then, we find a driver $B$ who recently {completed} a trip nearby and has yet to start another trip.  We do so by calculating the matching distance between driver $A$ and each recent completed trips' {\em destination} time and location. We choose the closest match, filtering out drivers who are the same as the given trip's driver, who have started another trip before the given trip's start time, or who ended their session (did not start any trip in the next hour).

 In the appendix, we provide results from a different but complementary matching method, as well as additional information about the matches and their quality. %

\subsubsection{Calculating the value of a trip to a driver}

\begin{figure*}[t!]
	\centering
		\includegraphics[width=\linewidth]{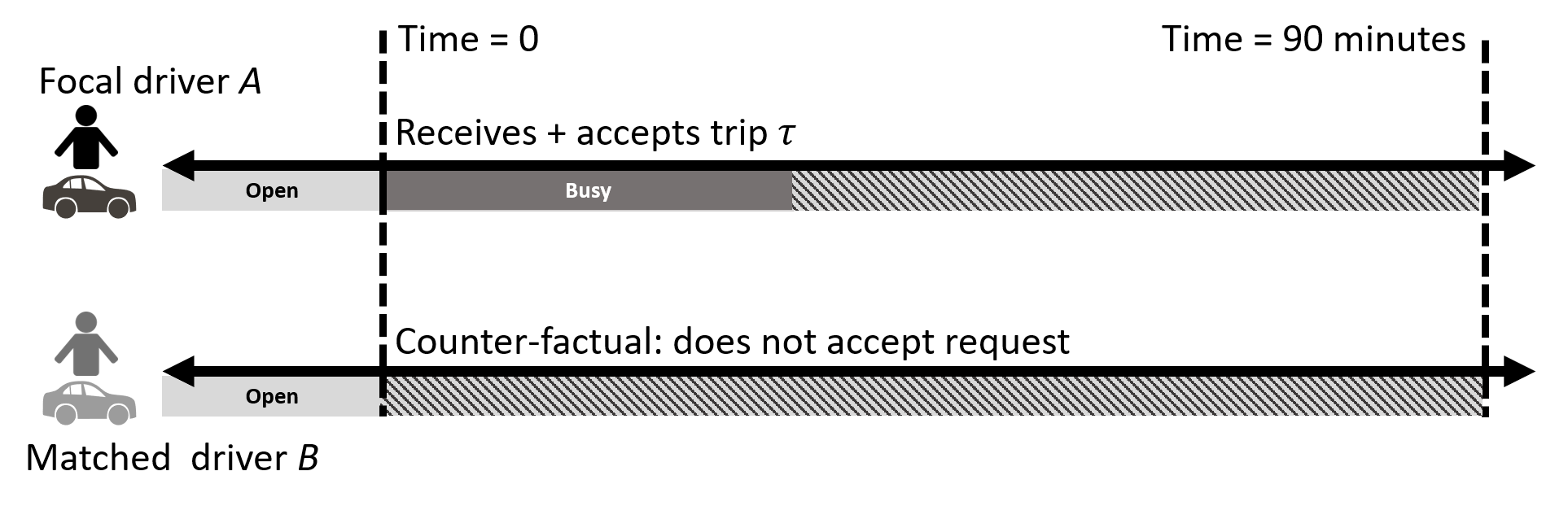}
		\caption{Earnings of focal vs matched driver. The value of a trip $\tau$ to the focal driver is the difference (conditional on $\tau$) between the 90 minute earnings of the focal and matched drivers. Diagonal stripes represent earnings unknown at the time the focal driver $A$ starts trip $\tau$ but are later observed in the data.}
		\label{fig:matching}
\end{figure*}

We now measure how valuable a trip is to a driver, through a notion we call {\em trip indifference}: given a specific trip request length $\tau$, in expectation the driver is at least as well off accepting the request as rejecting it, assuming some future behavior. Given focal driver $A$ with trip request $\tau$ and a matched driver $B$, we estimate this measure as illustrated in Figure~\ref{fig:matching}: we compare the two drivers' future earnings over the 90 minutes after the accepted trip begins---the higher driver $A$'s earnings over that of matched driver $B$, the more valuable the given trip request $\tau$. If there is no difference, i.e., the matched driver in expectation earns the same amount, then the given driver should be ``indifferent'' between accepting or rejecting the request.\footnote{{Trip indifference} is related to our theoretical notion of incentive compatibility as follows. Suppose the given driver accepts all future requests over the next 90 minutes. Then, if a payment scheme is incentive compatible, the earnings difference between the given driver who accepts trip $\tau$ and the matched driver will be at least $0$ for all $\tau$.}

Suppose trips are mis-priced and do not fully incorporate the drivers' temporal externalities. Then, trips of different lengths $\tau$ would vary in the value delivered to drivers. We would expect to see the average earnings differential, conditional on trip length, to vary as a function of the trip length; i.e., receiving a long trip during surge may be more valuable to a driver than is receiving a short trip.\footnote{Bias in the matching process may shift the expected earnings difference, but should not differentially affect the measurement for each payment function: the same matches are used for each. As robustness checks, in the appendix we vary both the matching function and the length of time over which we calculate the two drivers' earnings.}

\subsection{Results: value of short versus long trips}
\label{sec:dataanalaysis}

\begin{figure*}[t!]
	\centering
		\includegraphics[width=.8\linewidth]{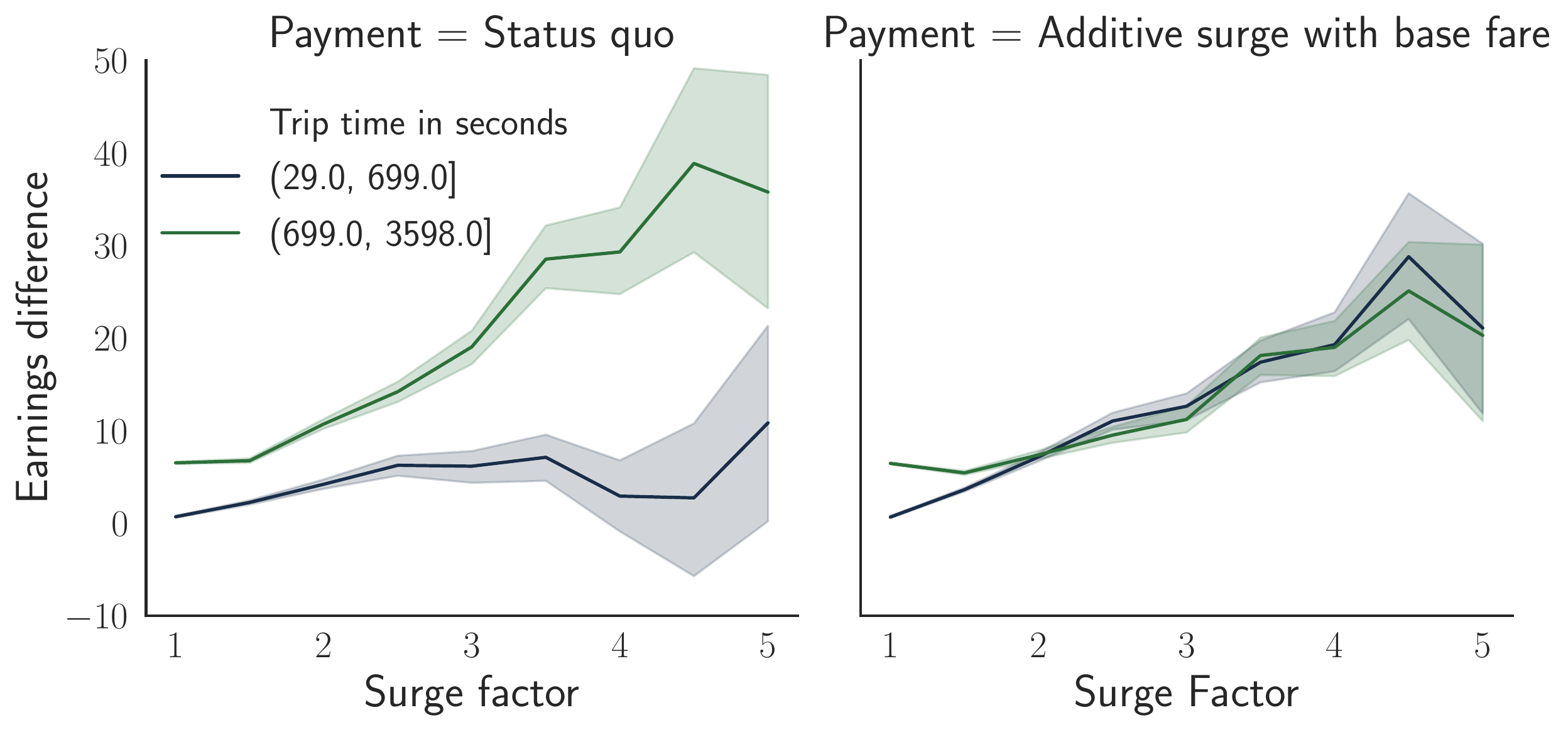}

	\caption{Difference in earnings over the next 90 minutes for the driver of a given accepted trip request, and a matched driver who also was open nearby at the time of the request, conditional on surge factor (rounded to nearest $0.5$) and length of trip. Error bars are $95\%$ bootstrapped confidence intervals.
		 }
			\label{fig:earningsdifferencebytriptype}
\end{figure*}

Figure~\ref{fig:earningsdifferencebytriptype} shows the difference in value between short (below the median trip length) and long (above the median) trips, as it changes with surge. As expected, it is more beneficial for drivers to receive trips with higher surge factors. However, with the platform's existing multiplicative surge payment function, only long trips become more valuable as the surge factor increases; even at high surge factors, drivers would have often had higher earnings had they rejected short trip requests. With additive surge, in contrast, trips of all lengths become more beneficial on average as surge increases. During high surge times, additive surge increases the value of short trips by about $\$15$ per hour.

In the appendix, we further simulate a world with the RideAustin data, but with surge being common and extremely valuable (we ``flip'' the surge factor). This analysis illustrates that our other insights also extend to practice, with there being settings where non-surge periods cannot be made incentive compatible, and where neither multiplicative nor additive surge correctly balance the value of short and long trips.
We also show how hourly driver earnings during a single ``shift'' change with additive and multiplicative surge, and how the former leads to more stable earnings. Overall, this analysis suggests the substantial difference that changing the structure of payments can make, and the comparative benefits of additive surge in practice under common regimes in ride-hailing.

\section{Conclusion}
In this work, we studied the problem of designing incentive compatible mechanisms for ride-hailing marketplaces. We presented a dynamic model to capture essential features of these environments. Even-though our model is simple and stylized, it highlights how driver incentives and subsequently dynamic pricing strategies would change in the presence of stochasticity. Our numeric and empirical analysis suggests the importance of such components in practice. We hope our work inspires other researchers in this area to incorporate such uncertainty in their models, as it is one of the biggest challenges faced in practice.

An important direction for extending our work is studying matching and pricing polices jointly, i.e., how to best match open drivers to riders in the presence of such effects, cf. ~\citep{ozkan_dynamic_2016,banerjee2017pricing,banerjee_segmenting_2017,feng2017we,zhang2017taxi,banerjee_state_2018,hu2018dynamic,korolko_dynamic_2018,ozkan2018joint,ashlagi2018maximizing,kanoria_near_2019}. In this work, we look at incentive compatible pricing. The platform, in addition to pricing, can use matching policies to align incentives.

\FloatBarrier
\bibliographystyle{plainnat} %
\bibliography{bib} %

\pagebreak\newpage

\appendix
 \textbf{APPENDIX TABLE OF CONTENTS}

\iftoggle{tableofcontentsapp}{
	\etocsettocstyle{}{}

	\textbf{Appendix \ref{appsec:additionaldiscussion} Additional discussion and information}
	\tableofcontents \ref{toc:additionaldiscussion}

	\textbf{Appendix \ref{appsec:empiricalevidence} More empirics}
	\tableofcontents \ref{toc:empirics}

	\textbf{Appendix \ref{appsec:proofssinglestate} Proofs of single state model results}
	\tableofcontents \ref{toc:proofsinglestate}

	\textbf{Appendix \ref{appsec:proofsdynamic} Proofs of dynamic model results}
	\tableofcontents \ref{toc:proofsdynamic}
}{}

\newpage
\section{Additional discussion and information}
\label{appsec:additionaldiscussion}
\iftoggle{tableofcontentsapp}{
\invisiblelocaltableofcontents \label{toc:additionaldiscussion}
}{}
\FloatBarrier

\subsection{{Platform objective}}
\label{sec:platformobjdetail}
{Our focus in this work is on designing incentive compatible payment functions for drivers. Here, we establish that this task is a sub-problem of the comprehensive platform pricing problem---one that can be studied separately given the components we considered exogenous in our model description. We work with the dynamic model, and suppose that the platform's primary objective is profit (our argument also trivially holds for revenue, trips served, welfare, or other objectives).} With our assumption of a single, earnings-maximizing driver, the platform's overall challenge is as follows.

On the rider side, we suppose that the two world state periods, $i\in \{1, 2\}$, are induced by latent demand shocks. The platform's design lever is the pricing policy $p = \{p_1, p_2\}$, where $p_i(\tau)$ indicates the rider price for trip length $\tau$ in world state $i$. Rider demand depends on the prices, inducing request rates and distributions $\lambda^p_i, F^p_i$ through a standard demand model for each trip: a rider with latent demand for trip $\tau$ requests a ride if the price is no more than their valuation for the trip (without substituting for trips of different lengths).

On the driver side, as detailed in our model formulation, the driver chooses a strategy $\sigma$ to maximize earnings rate $R(w, \sigma, \lambda^p_i, F^p_i)$, where the additional arguments emphasize that earnings depend on rider prices through induced demand. Further, the driver has an outside option earnings rate of $R$, and will participate in the system only if it is possible to achieve earnings rate $R(w, \sigma,\lambda^p_i, F^p_i) \geq R$ with some strategy $\sigma$.

The set $S$ of rides \textit{served} by the platform are those that are both \textit{requested} by riders (as induced by pricing $p$ and denoted by $\text{Support}(F^p_i)$) and \textit{accepted} by the driver (denoted by driver strategy $\sigma$):
\[S = \text{Support}(F^p_i) \cap \sigma. \]
	Let $\text{Rev}(p,\lambda^p_i, F^p_i, \sigma)=\lim\inf_{t\to\infty} \frac{\text{Rev}(p,\lambda^p_i, F^p_i, \sigma, t)}{t}$ denote the resulting \textit{revenue} rate for the platform, i.e., the rate paid by riders for served trips.

{Putting things together, the platform's profit maximization problem is as follows.}
\begin{equation}
\begin{aligned}
& \underset{p, w}{\text{maximize}}
& & \text{Rev}(p, \lambda^p_i, F^p_i, \sigma^*) - R(w, \sigma^*, \lambda^p_i, F^p_i)\\
& \text{subject to}
& & R(w, \sigma^*, \lambda^p_i, F^p_i) \geq R\\
& & & \sigma^* \in \argmax_\sigma \ \ R(w, \sigma, \lambda^p_i, F^p_i)\\
\end{aligned}
\label{eqn:platformopt}
\end{equation}

Where the first constraint is for driver participation, and the second for incentive compatibility (where the $\arg\max$ is not unique, assume that the driver chooses the policy $\sigma$ with largest measure.). With this formulation, the platform must jointly optimize prices $p$ and payments $w$, as both together determine the set of trips served and the profit for each such trip. Such a tightly connected optimization would preclude the approach taken in this work, where we focus on designing the payment functions for drivers, holding prices $p$ fixed. However, the optimization can be rewritten.

\begin{restatable}{proposition}{propplatformobj}
	\label{prop:platformobj}
Program~\eqref{eqn:platformopt2} below yields the same optimal value as Program~\eqref{eqn:platformopt}. For each solution, the same set of rides $S$ are served at the same prices $p$ as in a matching solution of Program~\eqref{eqn:platformopt}.
\begin{equation}
\begin{aligned}
& \underset{p, w}{\text{maximize}}
& & \text{Rev}(p,\lambda^p_i, F^p_i, \sigma^*)- R\\
& \text{subject to}
&& \sigma^* = \text{Support}(F^p_i)\\
&& & R(w, \sigma^*, \lambda^p_i, F^p_i) = R\\
& & & \sigma^* \in \argmax_\sigma \ \ R(w, \sigma, \lambda^p_i, F^p_i)
\end{aligned}
\label{eqn:platformopt2}
\end{equation}

\end{restatable}

{The reformulation in Proposition~\ref{prop:platformobj} follows from a simple insight: in our model with no driver private information, a driver rejecting a request is equivalent to the rider not requesting the trip in terms of how it affects the set $S$ of trips served -- and the platform can predict such rejections perfectly. Then, for any optimal solution of Program~\eqref{eqn:platformopt} in which a rider requests a trip $\tau$ but the driver rejects it, the platform can equivalently raise rider prices until no rider requests such a trip, $F^p_i(\tau) = 0$, and so the driver accepts all requested trips lengths. Further, the driver earnings constraint $R(w, \sigma^*, \lambda^p_i, F^p_i) \geq R$ is of course tight: driver payments can otherwise be proportionally scaled down, as scaling $w$ does not affect incentive compatibility.
}

With Program~\eqref{eqn:platformopt2}, the driver payment function $w$ and induced driver strategies $\sigma$ just appear in the constraints. Given each potential choice of rider pricing function $p$ and induced demand $\lambda^p_i, F^p_i$ (i.e., which trips to service at what prices), the platform must determine how to pay drivers such that they accept every request, i.e., the platform must choose payments $w_i$ such that the participation and IC constraints are met.
{In this work, we focus on this challenge, holding rider prices $p$ and thus demand $\lambda_i \triangleq \lambda^p_i$, $F_i \triangleq F^p_i$, fixed. Note that in the main text we denote the challenge as finding payments such that $\sigma^* = \{(0, \infty), (0, \infty)\}$, instead of $\sigma^* = \text{Support}(F^p_i)$. The two notations are equivalent: we can trivially add to a driver's policy trips lengths where the measure under $F^p_i$ is zero, as such trips do not affect driver earnings. We use the former notation for convenience.

}

\subsection{Driver earnings in each state}
\label{sec:eachstateearningsdetail}
Recall that in Lemma~\ref{lem:statingreward}, we decompose the driver reward into reward rates for each world state, $R_i(w_i, \sigma_i)$, denoting the earnings rate while the driver is either open in $i$ or on a trip that started in $i$. In our theoretical pricing results in Section~\ref{sec:IC}, we show how to construct incentive compatible pricing given choices of average earnings in each state, i.e., setting $R_i(w_i, \sigma_i) = R_i$ for some $R_1,R_2$. These rates, subject to the participation constraint that overall earnings $R(w, \sigma) \geq R$, is a design choice for the platform. Here, we provide some intuition for how to make this choice.\footnote{The rider-side pricing problem of setting average prices and thus revenue $\text{Rev}_i$, given the latent demand, is potentially easier as the primary goal is a short-term allocation of the supply (drivers) to the riders who most value the service. The driver side problem, as discussed, is trickier as there are both short- and long-term effects.}

{\bf Business constraint from revenues}. The platform's revenue rate can be decomposed just like the driver earnings rate, with
state $i$ revenue rate, $\text{Rev}_i(p_i, \sigma_i,\lambda^p_i, F^p_i) = \frac{\frac{1}{F_i(\sigma_i)}{\int_{\tau \in \sigma_i} p_i(\tau) dF_i(\tau)}}{T_i(\sigma_i)}$.
Latent demand and the choice of prices $p_i$ together induce platform revenue rates for each world state. Then, the per-state driver earnings rates $R_i$ could be approximately set as a fixed fraction of revenue
\[
R_i = \alpha \ \text{Rev}_i(p_i, \sigma_i,\lambda^p_i, F^p_i)
\]
for some $\alpha$. While in our model we have a constraint on per-state driver earnings assuming the driver accepts every trip, in practice a platform may desire to constrain realized payments with revenue. However, neither the revenue nor driver decisions (and hence actual payments) can be predicted perfectly ahead of time, and so the platform must either dynamically adjust $\alpha$ or otherwise work with approximations that are correct in expectation. How to do so well is in practice an interesting machine learning prediction problem.

 The above choice passes on the revenue earned in each state to drivers, and so represents a \textit{partially} decoupled setting: at the trip level, the amount paid to drivers may deviate from that paid by the rider, but prices are coupled on average at the level of a surge state. In practice this simple rule helps ensure that individual prices for a rider and driver do not differ by too much, which may be desirable for transparency and driver satisfaction reasons.

{\bf Driver positioning}. However, the question of at what level to best decouple prices, and e.g., how to potentially transfer money between different surge states across time and space, is an interesting one for future work. Here, we describe one potential rationale for optimizing $R_i$.

	Empirically,~\citet{lu_surge_2018} find that drivers respond to real-time surge prices (displayed through a heat-map) by re-positioning themselves to surge areas, an effect that is in addition to drivers choosing to drive (activating) in times and places where they expect to see surge. Thus, a higher surge earnings rate $R_2$ translates to more drivers during surge, as a result of both (a) short term, real-time movement toward surge due to seeing the heat-maps as in Figure~\ref{fig:heatmaps}, and (b) drivers logging on when and where there tends to be surge.
	A platform could thus choose the relative values of $R_i$ as a lever for this type of re-positioning. For example,~\citet{freund_escrow_lyft2020} describe how Lyft manages an incentive budget over time and space to incentivize driver re-positioning.
	We further refer the reader to \citet{besbes_surge_2018} for theoretical insight on short-term driver positioning, in a setting with coupled rider prices and driver payments.

\vspace{5pt}
Our model does not directly capture the above ways a platform could set and optimize $R_i$, as it has a single driver and geographic location, and we do not optimize rider prices and thus revenue. However, note that both effects above are mediated through the average earnings (i.e., $R_i$), either predicted by the driver or communicated through a heat-map, and do not depend directly on trip specific earnings, i.e., $w_i$. Thus, these effects can be incorporated by adding the constraints $R_i(w_i, \sigma_i) = R_i$ in Program~\eqref{eqn:platformopt2}, with target earnings rate $R_i$ optimized elsewhere.  An interesting avenue of future work is indeed to optimize $R_i$ over both space and time, given these effects.

We take this approach in this work, analyzing for what values of $R_i$ the constraints $R_i(w_i, \sigma_i) = R_i$ are compatible with incentive compatible pricing. In our main result, Theorem~\ref{thm:ICpolicy}, we cannot construct IC prices that induce all relative values of $R_1(w_1, \sigma_1)=R_1$ and $R_2(w_2, \sigma_2) = R_2$: if the platform tries to make the surge state $i=2$ is too valuable compared to regular times $i=1$, $R_2 \gg R_1$, then drivers will reject long trips in the non-surge state.

\subsection{Model's relationship to practice}
\label{sec:limitations}

Several of our theoretical model choices emerge from common ride-hailing practice; other choices -- such as not considering spatial heterogeneity -- differ from practice, and so we consider the generalizability of our insights to practice in Section~\ref{sec:empiricalevidence}, using real ride-hailing data. See also Section~\ref{sec:moreempiricalfacts} where we justify our choices in the numerical section with RideAustin data and provide more information on, e.g., surge evolution.

{\bf Heat-map constraint and affine pricing.} When drivers are not on a trip, they see a heat-map of the current surge values, indicated as a multiplier or additive value, cf. Figure~\ref{fig:heatmaps}; this has important implications for practice, and for the pricing functions we consider in this work.

First, in our numerical and empirical sections we focus on multiplicative and additive surge, as opposed to other general surge payment schemes. Two rationales for this choice are that these are the schemes considered by platforms in practice, and that they naturally serve as approximations of our IC scheme. More fundamentally, however, such schemes can be directly displayed on the heat-map. With such single-parameter schemes, the driver can connect their surge payment to knowledge available to them before the trip starts. This is an important feature in practice, where platforms must be as transparent as possible regarding how they pay drivers. Consider for example, if the platform instead displayed on the heat-map some expected payment over all trips taken in that spatio-temporal spot (e.g., the equivalent of $R_i$); the driver would not be able to verify that the platform in fact did pay out that amount on average, without data from other drivers.

Second, in this work we consider only pricing functions that depend on the world state when the trip starts, but do not incorporate information from what happens during the trip. Again, this is an important practical constraint: incorporating on-trip information would require the platform to perform a path-integral over surge values in the driver's spatio-temporal path from the origin to the destination, which would be difficult to implement and for the driver to verify. More fundamentally, however, the surge payment is partially an incentive for drivers to re-locate to a surge area, cf.~{\citet{lu_surge_2018}}, and modeled by $R_i$ in our work. Updating surge payments based on what happens when a driver is on-trip would change such incentives.

{\bf Surge evolution.} Surge is clearly non-Markovian and non-binary in practice, with strong intra-day patterns -- for example, rush hours have predictably higher surge values: see Appendix Figure~\ref{fig:surgetripbyhour}.

However, evolution of surge on finer time scales, on the level of individual trips, is more volatile, and believably Markovian: see Appendix Figure~\ref{fig:surgeover3days}, which shows the (spatially-averaged) surge factor in a small region around the Texas Capitol building every ten minutes over 3 days. Thus, from the perspective of a single driver who has decided to drive at a certain time block (for example, 5-8pm), surge is believably Markovian on the time order that they are making decisions for whether to accept certain trips.

The main theoretical difficulty with analyzing non-Markovian updates is that, then, the driver optimal policy is dependent on the time index as well as the state index; then, results will very strongly depend on the specific trip length distribution chosen, and in particular the interaction between the trip length distribution and the surge pattern structure. This interaction prevents any generalizable insights from emerging. However, as detailed above, our empirical analysis suggests that our results hold up even under more realistic surge.

{\bf Driver activation.} We do not endeavor to explain \textit{why} surge pricing might be useful in this paper: in our view, riders respond to rider prices, and drivers activate based on expected mean earnings (i.e., $R_i$ and $R$, as discussed above), which we take as exogenous. These aspects are well studied in the ride-hailing literature. Rather, our paper studies the orthogonal question of how to pay a driver for trips \textit{once they are online}, not how to induce drivers to drive when and where there is high demand.

{\bf Single driver and equilibrium effects.} Our model considers a \textit{single} driver, when in reality there are of course many drivers on the road. We do not believe that doing so affects the results, as the number of other drivers on the road affects average surge dynamics and activation, but presumably not individual trip decisions, except as mediated through future expectations of surge.

The main theoretical difficulty with analyzing multiple drivers is it would add historical state to the system not captured by just the current surge state, pertaining to the number of currently open drivers and the distribution of when currently busy drivers will next become open. This difficulty is similar to that of modeling non-Markovian surge evolution. It would also lead to an implausible driver behavior model -- each earnings maximizing driver would have to keep track of the number of other open drivers (and the distribution of when currently busy drivers will next become open).

\subsection{Supplementary figures}

\begin{figure*}[t!]
	\centering
	\centering
	\includegraphics[width=.5\linewidth]{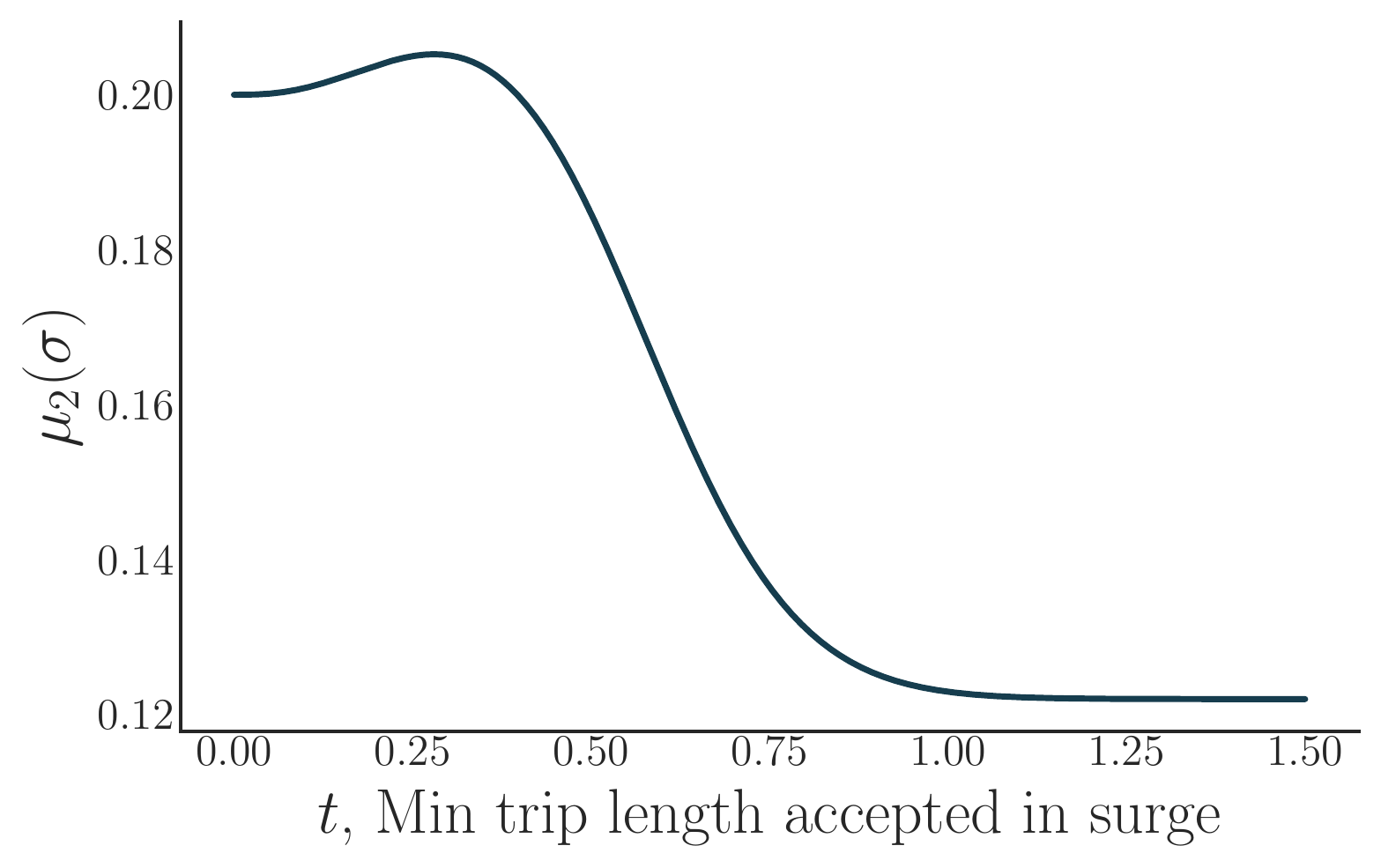}
	\caption{Fraction of time spent in surge state, $\mu_2(\sigma)$, with driver policy $\sigma = \{\sigma_1= (0, \infty), \sigma_2\}$, where $\sigma_2 = (t,\infty)$, i.e., $t$ is the minimum trip length accepted in the surge state. The primitives are as follows: $\lambda_1 = \lambda_2 = 12, \lonetwo = 1,\ltwoone = 4$; in both states, trip lengths are distributed according to a Weibull distribution with shape $2$ and mean $\frac{1}{3}$. These parameters reflect realistic average trip to wait time values, and that surge tends to be short-lived compared to non-surge times. Note that the driver can increase the time spent in the surge state by rejecting short surge trips.}
	\label{fig:percenttimespentinstate2}
\end{figure*}

\begin{figure*}[t!]
	\centering
	\begin{subfigure}[t]{0.48\textwidth}
		\centering
		\includegraphics[width=\linewidth]{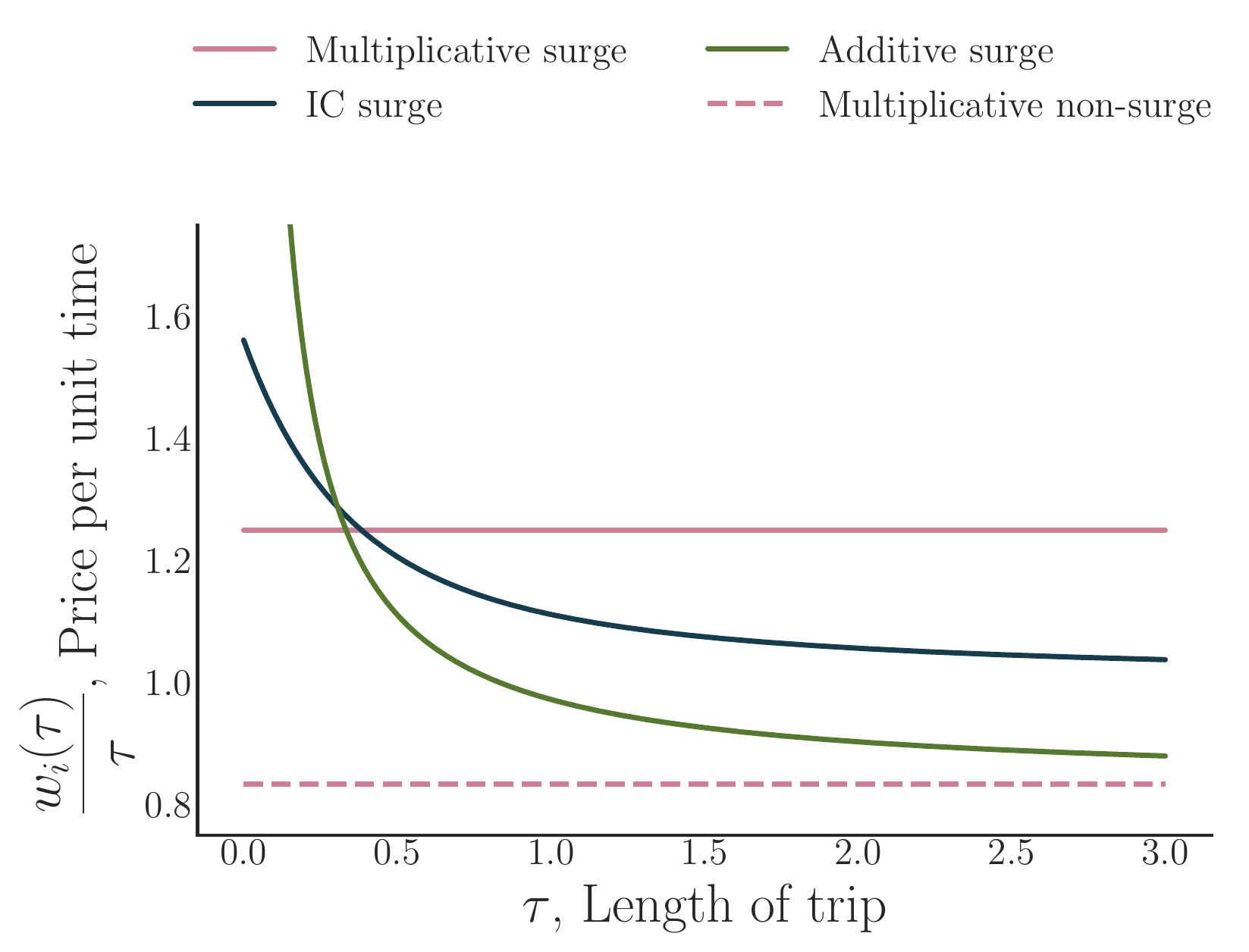}
		\caption{Price per unit time $\frac{w_i(\tau)}{\tau}$}
		\label{fig:surgeplotpricingcomparative_overtauaffine}
	\end{subfigure}
	~ \hfill
	\begin{subfigure}[t]{0.48\textwidth}
		\centering
		\includegraphics[width=\linewidth]{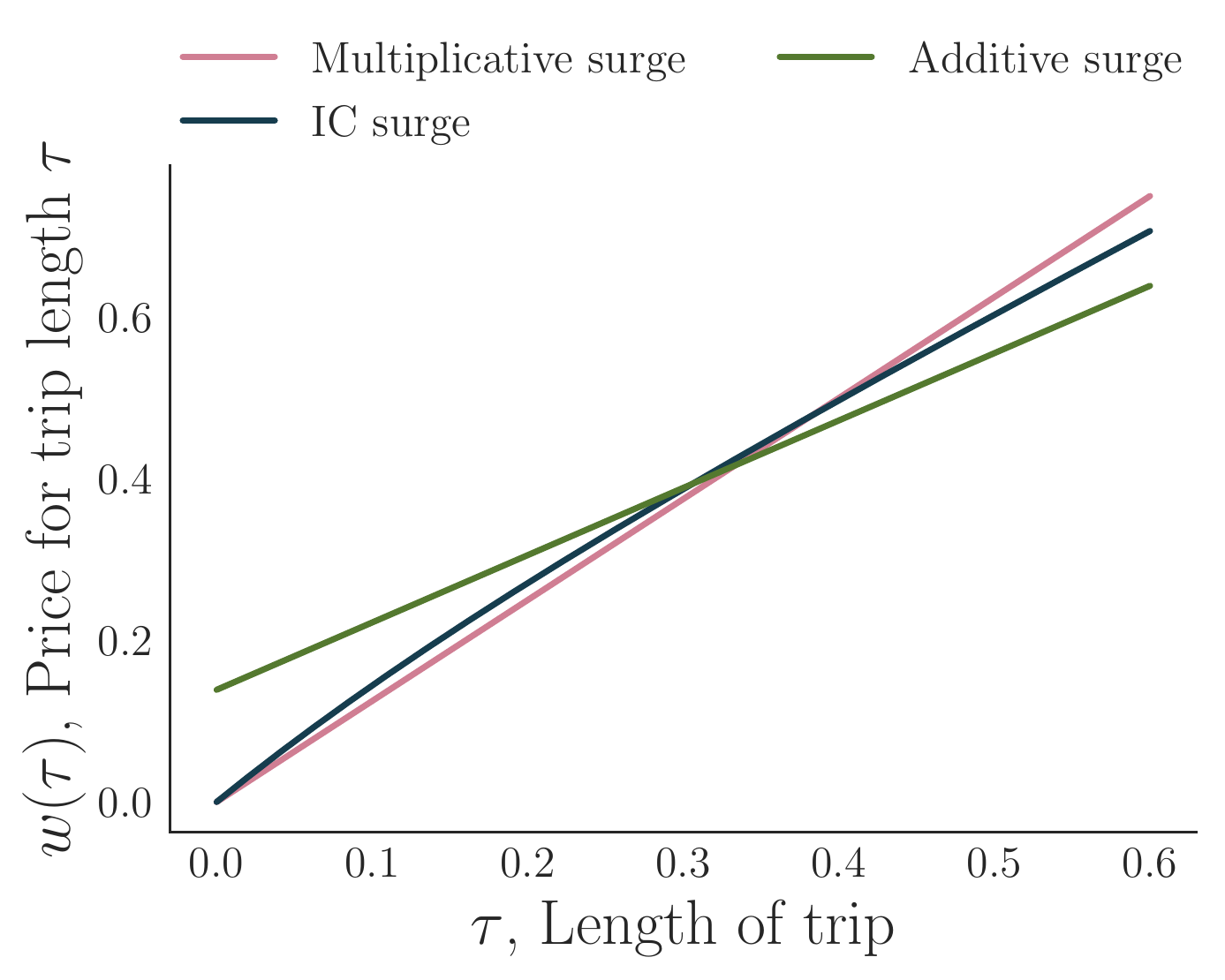}
		\caption{Price ${w_2(\tau)}$ }
		\label{fig:surgeplotpricingcomparative_tauaffine}
	\end{subfigure}
	\caption{Using the same model primitives as in Figure~\ref{fig:percenttimespentinstate2}: the payment function $w_i(\tau)$ for various surge mechanisms plotted two ways, when $R_2 = 1$ and $R_1 = \frac{2}{3}$ for drivers who accept every trip.}
	\label{fig:surgepolicyexample_affine}
\end{figure*}

Figure~\ref{fig:percenttimespentinstate2} shows in an example $\mu_2(\sigma)$ as it changes with the surge driver policy $\sigma_2 = (t, \infty)$, for some $t$. Figure~\ref{fig:surgepolicyexample_affine} compares IC surge pricing to multiplicative and additive surge.

\FloatBarrier
\newpage
\section{Supplementary empirical information}
\label{appsec:empiricalevidence}
\iftoggle{tableofcontentsapp}{
\invisiblelocaltableofcontents \label{toc:empirics}
\FloatBarrier
}{}

This section contains supplementary empirical information. Section~\ref{sec:moreempiricalfacts} contains new findings related to the model validity and the variance of driver earnings with the various payment functions. Section~\ref{sec:empiricalrobustness} contains more detail and robustness checks for the primary empirical analysis presented in the main text.

\subsection{Additional results and facts}
\label{sec:moreempiricalfacts}

We now detail new results and empirical findings discussed briefly in the Appendix. Section~\ref{sec:appmodelvalidity} validates our model choices and claims in the numerical analysis in Section~\ref{sec:additivesurgesims}. Section~\ref{sec:frequentvaluablesurge} presents an analysis of a simulated scenario in which surge is frequent and highly valuable, as opposed to rare and moderately valuable. Finally, Section~\ref{sec:earningsvarianceempirical} shows that additive surge pricing has the additional benefit of reducing drivers' earning variance in practice.

\subsubsection{Model validity}
\label{sec:appmodelvalidity}

\begin{figure*}[t!]
	\begin{subfigure}[t]{0.48\textwidth}
		\centering
		\includegraphics[width=\linewidth]{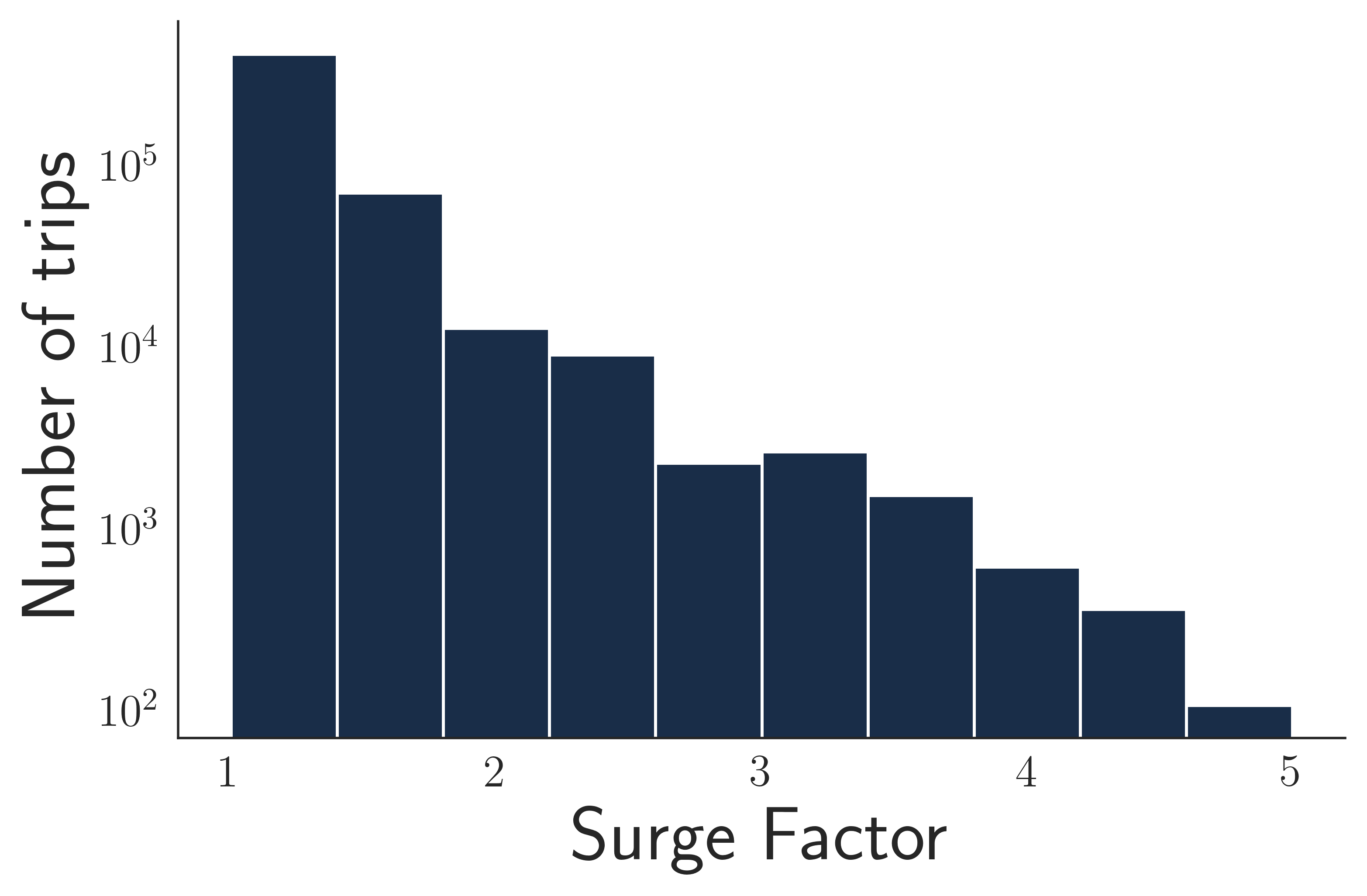}
		\caption{Histogram of surge, in log scale.}
		\label{fig:surgehistogram}
	\end{subfigure}
	~ \hfill
	\begin{subfigure}[t]{0.48\textwidth}
		\centering
		\includegraphics[width=\linewidth]{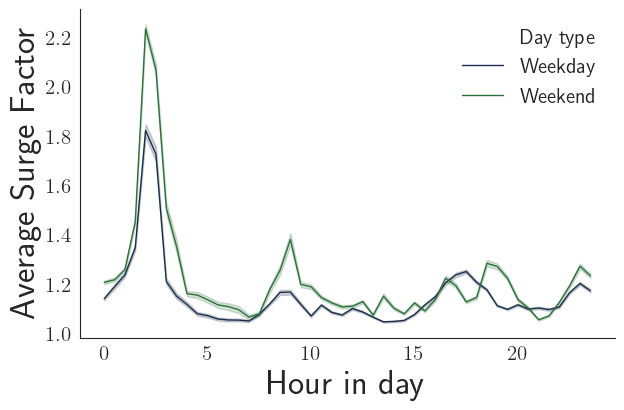}
		\caption{Average surge factor in each 30 minute period of the day}
		\label{fig:surgetripbyhour}
	\end{subfigure}

	\begin{subfigure}[t]{0.48\textwidth}
		\centering
		\includegraphics[width=\linewidth]{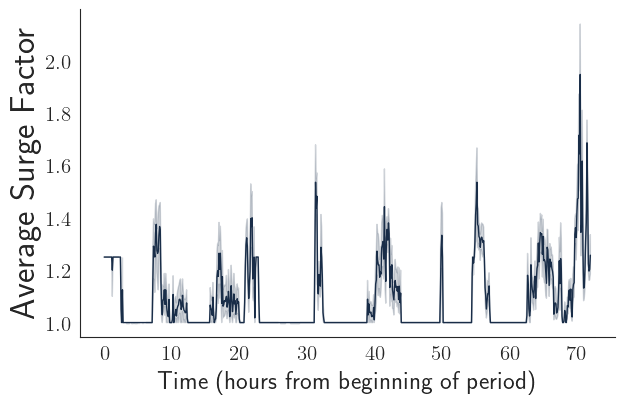}
		\caption{Average surge in each 10 minute period over three days on trips that start within 5 miles from Texas Capitol building.}
		\label{fig:surgeover3days}
	\end{subfigure}~\hfill
	\begin{subfigure}[t]{0.48\textwidth}
		\centering
		\includegraphics[width=\linewidth]{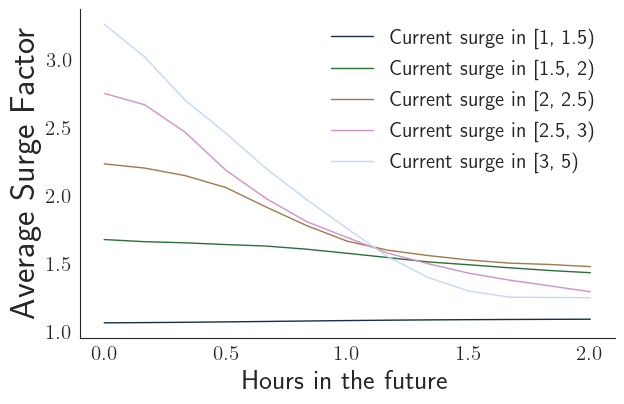}
		\caption{Divide the 2 months into periods of 10 minutes each. Then, this plot shows the mean surge factor $x$ hours in the future, split by bucket of the current surge factor.}
		\label{fig:surgeautocorr}
	\end{subfigure}

	\begin{subfigure}[t]{\textwidth}
		\centering
		\includegraphics[width=.5\linewidth]{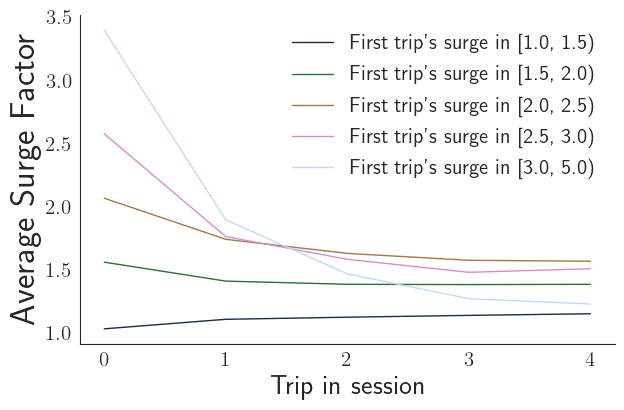}
		\caption{For each driver session that has at least 5 trips, the average surge factor of each trip in the session, split by the surge factor of the first trip.}
		\label{fig:surgeautocorrbytrip}
	\end{subfigure}~\hfill

	\caption{Surge facts from RideAustin marketplace}
\end{figure*}

\begin{figure*}[t!]
	\centering

		\begin{subfigure}[t]{0.48\textwidth}
		\centering
		\includegraphics[width=\linewidth]{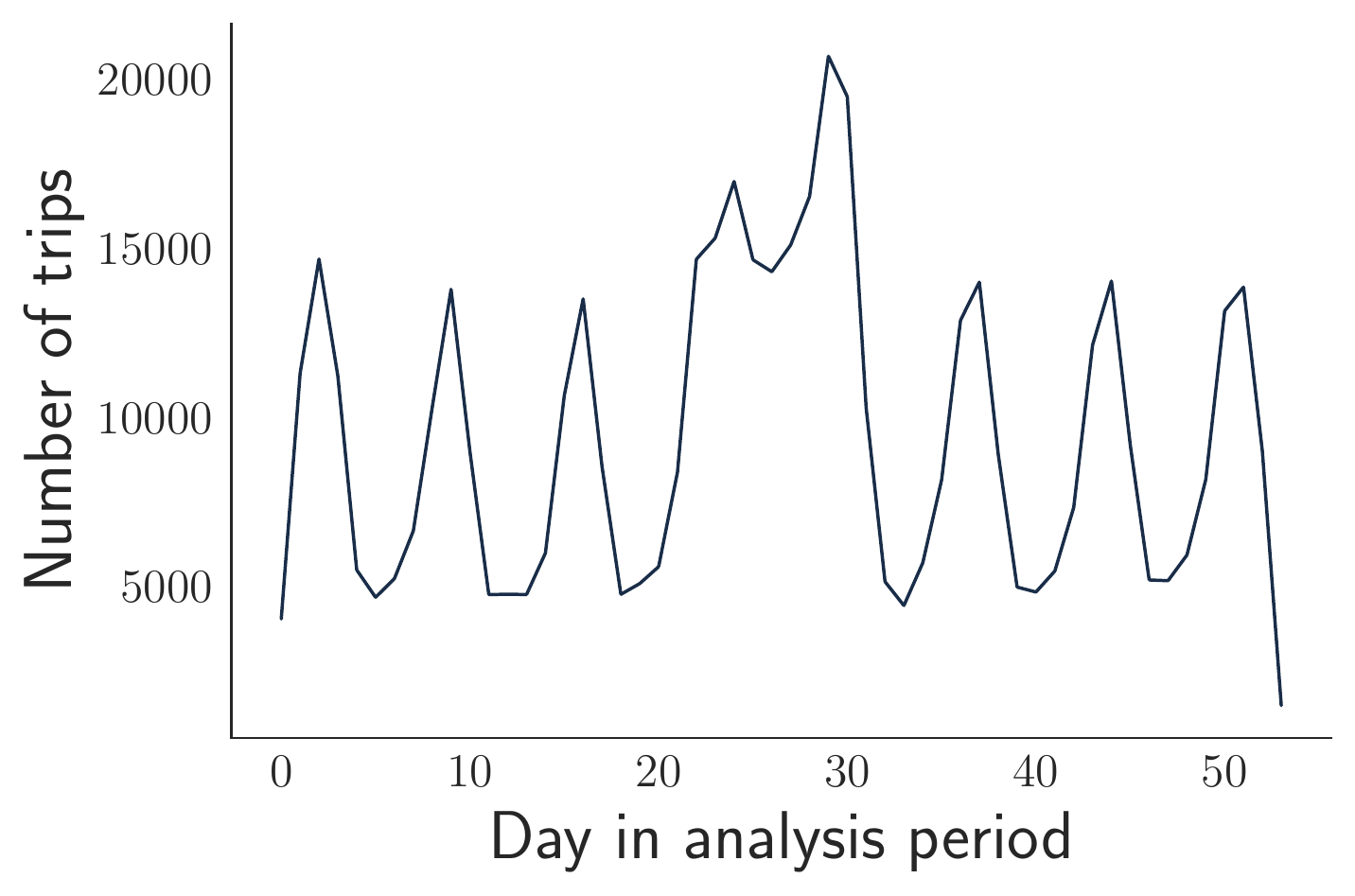}
		\caption{Trips per day}
		\label{fig:tripsperday}
	\end{subfigure} 	~ \hfill
	\begin{subfigure}[t]{0.48\textwidth}
		\centering
		\includegraphics[width=\linewidth]{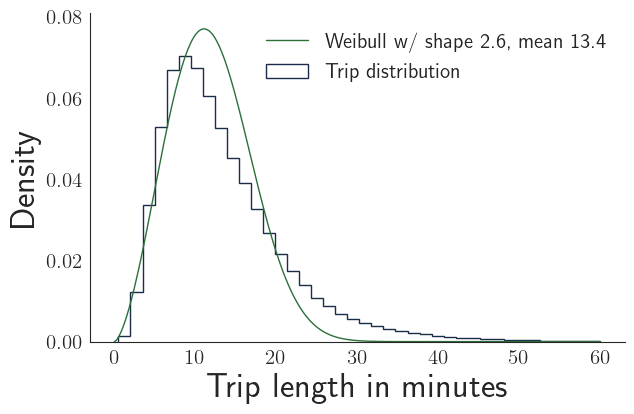}
		\caption{Trip length distribution (non-surge)}
		\label{fig:triplengthdist}
	\end{subfigure}

	\begin{subfigure}[t]{0.48\textwidth}
		\centering
		\includegraphics[width=\linewidth]{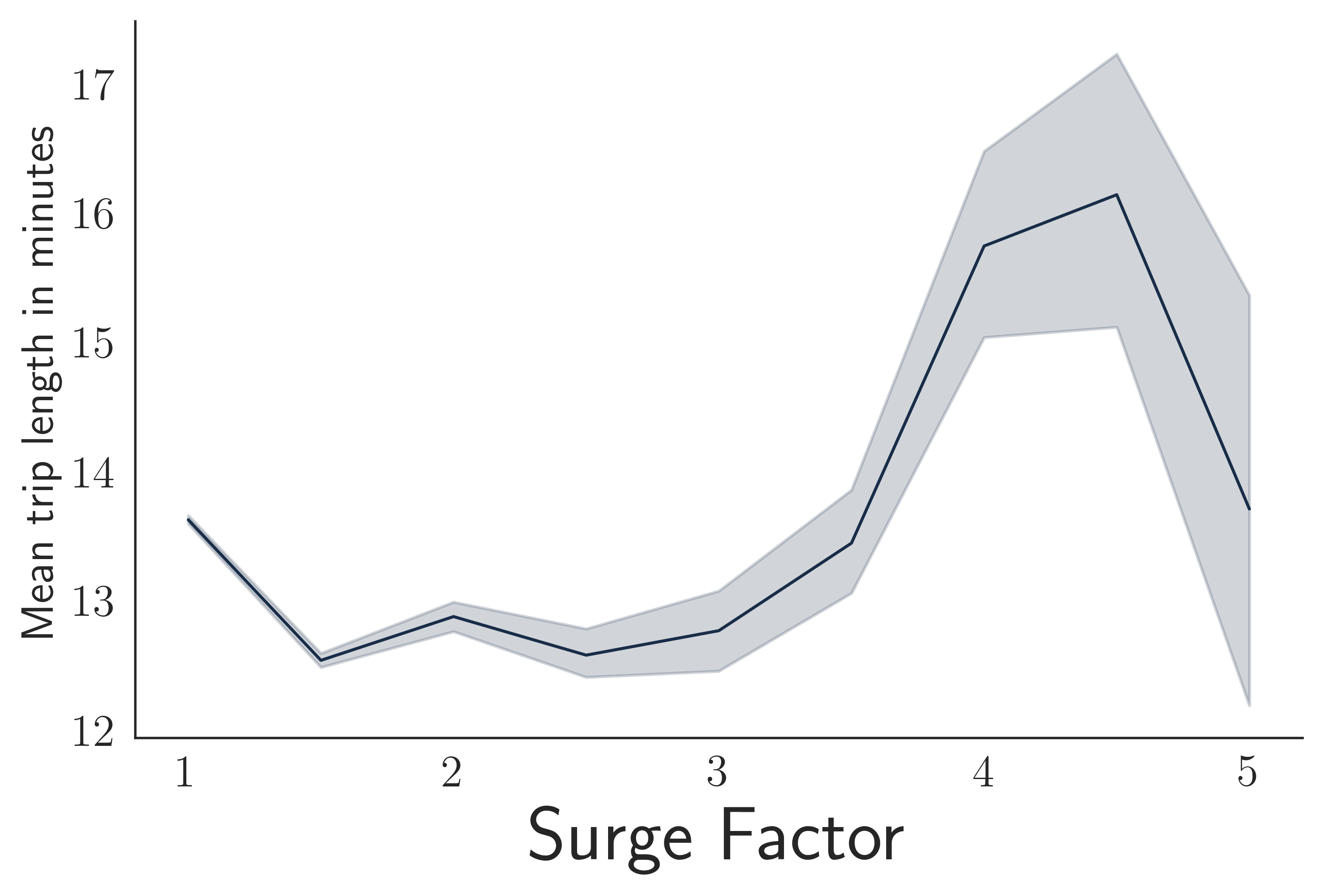}
		\caption{Mean trip length by surge factor}
		\label{fig:tripbysurgefactor}
	\end{subfigure}

	\begin{subfigure}[t]{0.48\textwidth}
		\centering
		\includegraphics[width=\linewidth]{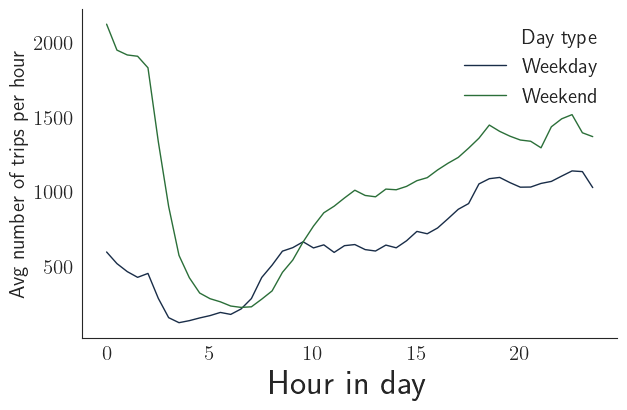}
		\caption{Average number of trips per hour period in the day. Note the discontinuities at midnight are due to weekdays becoming weekends and vice versa (Friday night becomes Saturday morning).}
		\label{fig:avgnumtrips}
	\end{subfigure}~ \hfill
	\begin{subfigure}[t]{0.48\textwidth}
	\centering
	\includegraphics[width=\linewidth]{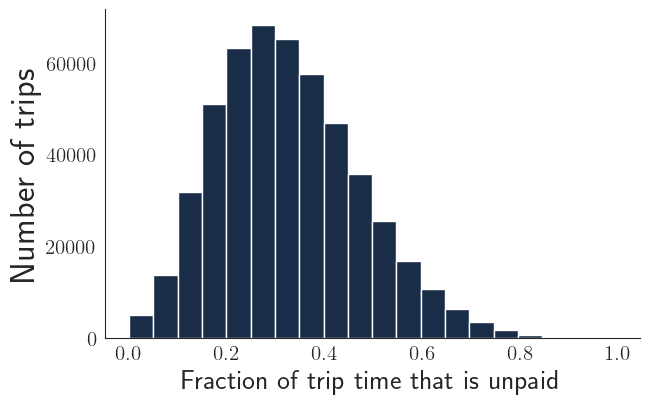}
	\caption{Fraction of a job's total time that is the time to drive to the rider, i.e., unpaid.}
	\label{fig:dispatchratio}
\end{subfigure}

	\caption{Basic trip facts from RideAustin marketplace}
\end{figure*}

Here, we discuss how various components of the model relate to ride-hailing marketplaces in practice, using the RideAustin data from the rest of the empirics. We also justify the three claims we make in the numerics regarding the common parameter regimes for ride-hailing platforms.

\begin{description}
\item [Surge is non-binary, and between $1.1$ and $3$ times more valuable than non-surge]

Figure \ref{fig:surgehistogram} contains a histogram of the surge factor. Surge in the RideAustin marketplace during the time period analyzed takes values divisible by $0.25$, between $1$ and $5$. The mean surge factor is 1.19, only $30\%$ of trips are surged, and more than $97\%$ of surged trips have a surge factor in $(1, 3]$.

\item [In the model, surge evolves according to a continuous time markov chain.]

Figure~\ref{fig:surgetripbyhour} breaks down the average surge factor in each 30 minute period in a day, split up by weekdays and weekends. Surge is clearly not Markovian -- there are clear, expected patterns in surge that correlate with rush hours and early morning times when there may be few drivers on the road.

However, there is substantial additional volatility in addition to the non-Markovian daily patterns. Figure~\ref{fig:surgeover3days} shows average surge in each 10 minute period, over 3 days for trips starting near the Texas Capitol building. The lengths, peak, and start/end times of each surge period differ -- on a ten minute time scale, i.e., on the order of trip lengths, surge is not very predictable, and so a Markovian assumption may be reasonable on a small time scale.

\item [Surge is short-lived compared to non-surge periods ($\ltwoone \gg \lonetwo$)]

High-surge periods are indeed short-lasting compared to low surge periods, and peak surge tends to be short lasting. Figure~\ref{fig:surgeautocorr} shows the mean surge factor in the future, based on the current surge factor. Without surge, the average surge even an hour in the future remains close to $1$. With high surge, however, the average surge in the future decays -- and the higher the surge, the faster the decay. Reality deviates from the model with low surge, with factor in $[1.5, 2)$ -- average surge even an hour later tends to stay in this region, suggesting that such levels of surge are durable on this platform and surge trips may be more common than non-surge during such times.

\item [In a typical surge a driver may only be able to complete one or two such trips.] ($\frac{1}{\ltwoone} \approx$ mean trip length). By jointly analyzing Figures~\ref{fig:surgeautocorr} and~\ref{fig:tripbysurgefactor}, we can see that drivers are indeed only be able to complete a few trips during surge before it dissipates. On-trip times (with rider in the car) are on the order of 10-15 minutes, and the driver must also wait for a new request and then drive to the rider. Surge has typically decreased substantially after an hour.

More directly, Figure~\ref{fig:surgeautocorrbytrip} shows for each driver session that has at least 5 trips, the average surge factor of each trip in the session, split by the surge factor of the first trip. Indeed, a driver is only able to complete a few trips with peak surge. We note, however, that this plot is susceptible to selection effects -- a driver may choose to drive a different amount of time based on surge conditions.

\item [In the model, on-trip time and time driving to the rider are combined.] In practice, a job is typically split up into two components: the time it takes to drive to the rider, and the time that the ride is in the car -- and only the second part is paid. Figure~\ref{fig:dispatchratio} shows a histogram of the resulting fraction of the total job time that is unpaid. Note that this time is substantial in the RideAustin data, on average about $30\%$.

\item [In the numerics, trip lengths are distributed as a Weibull distribution with shape 2.]  Figure~\ref{fig:triplengthdist} shows the distribution of trip lengths for trips without surge. The shape approximation is reasonable, as a Weibull distribution with shape $2.6$ best fits the data (with mean set to the empirical mean). Figure~\ref{fig:tripbysurgefactor} shows the mean length distribution by surge factor. Perhaps interestingly, this mean length is non-monotonic in the surge factor, first decreasing and then increasing with the surge factor.

\end{description}

We cannot directly test the claim in the numerics that in a typical surge the driver will be able to receive and reject several trip requests ($\frac{\lambda_2}{\ltwoone}> 1$, but small) -- we do not observe drivers being open to receive a request, unless they actually received a trip request. Unlike in the matching technique for trip indifference, we cannot use drivers who completed a trip as a proxy -- the measurement would be sensitive to drivers logging off, and the end-locations of trips not being representative of all trips.

Despite the ways reality deviates from the model, the insights regarding additive vs multiplicative surge extend to the empirics.

\subsubsection{Regime with frequent, valuable surge}
\label{sec:frequentvaluablesurge}
\begin{figure*}[t!]
	\centering
	\includegraphics[width=.8\linewidth]{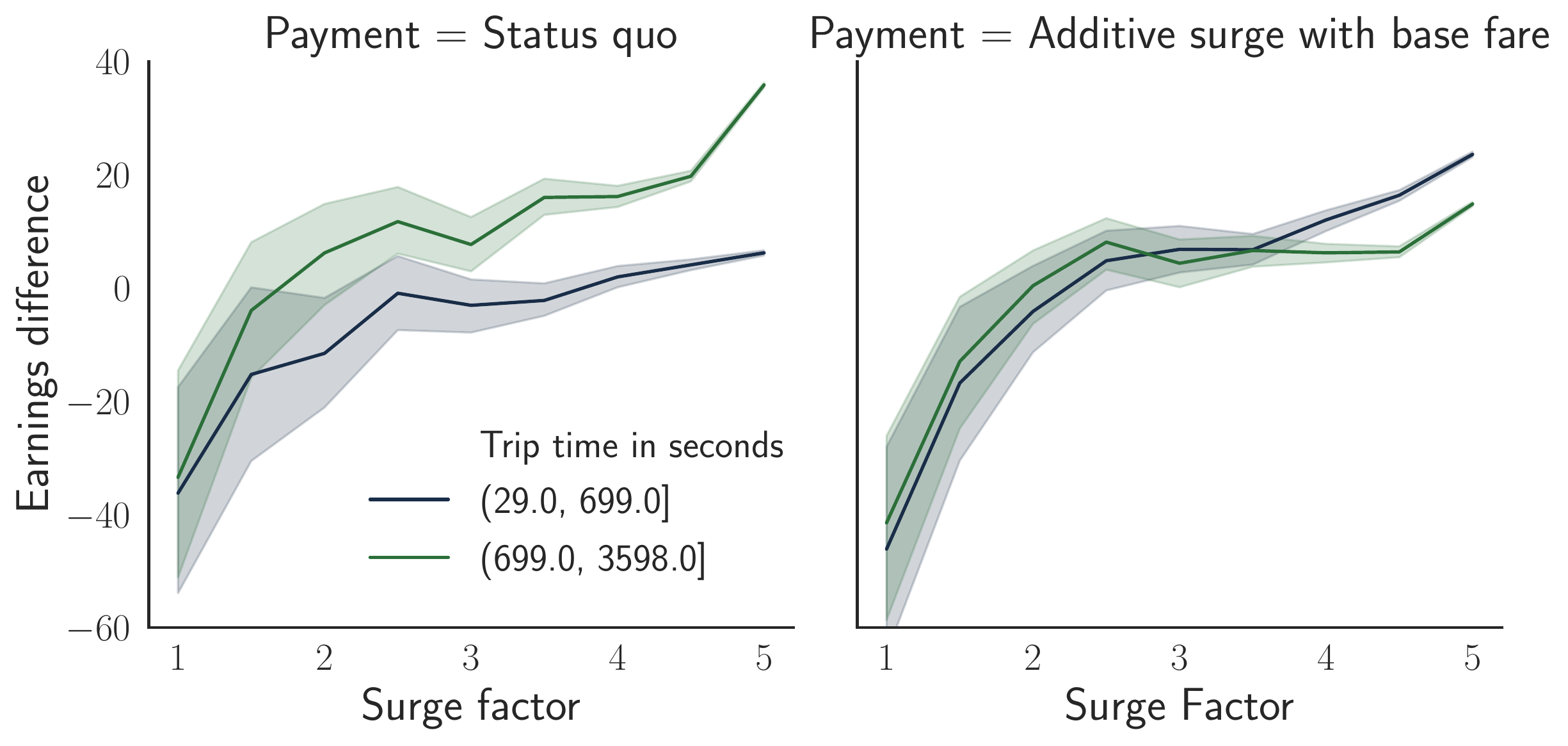}
	\caption{Same as Figure~\ref{fig:earningsdifferencebytriptype}, except with the surge factor flipped to simulate a world with frequent, valuable surge.
	}
	\label{fig:earningsdifferencebytriptype_flipsurge}
\end{figure*}

Recall that one of the theoretical insights from Theorem~\ref{thm:ICpolicy} is that our incentive compatible pricing scheme only works in a certain regime, if surge is not too valuable compared to regular periods on average, that $\frac{R_1}{R_2}\in (C,1)$. This general insight extends to arbitrary pricing functions (i.e., as $\frac{R_1}{R_2} \to 0$, then no pricing function $w_1$ during regular periods will induce drivers to accept non-surged trips).

Here, we show that this insight also extends to practice, with non-binary surge. We simulate the following world: we ``flip'' the surge factor
\[ \text{Simulated surge}  = 6 - \text{Actual surge}.\]

With this flipped surge, 97\% of surged trips have a surge factor in $[3, 5]$, and 30\% of the trips have a surge factor of $5$: surge is now the default, and extremely valuable compared to non-surge periods.

Then, we calculate the driver's payment according to each such pricing function. Figure~\ref{fig:earningsdifferencebytriptype_flipsurge} shows the resulting plots for earnings difference by trip length, using the status quo payment function (but with the simulated surge factor) and with an equivalent additive surge. Two insights emerge:

\begin{itemize}
	\item With low surge (factor in $[1, 3]$), drivers are better off on average rejecting most trip requests, regardless of whether payments are additive or multiplicative.
	\item A more complex pricing function may be needed: multiplicative surge over-values long trips with high surge, and additive surge over-values short trips.
\end{itemize}

\subsubsection{Driver earnings variance}
\label{sec:earningsvarianceempirical}
\begin{figure*}[t!]
	\centering
		\includegraphics[width=.5\linewidth]{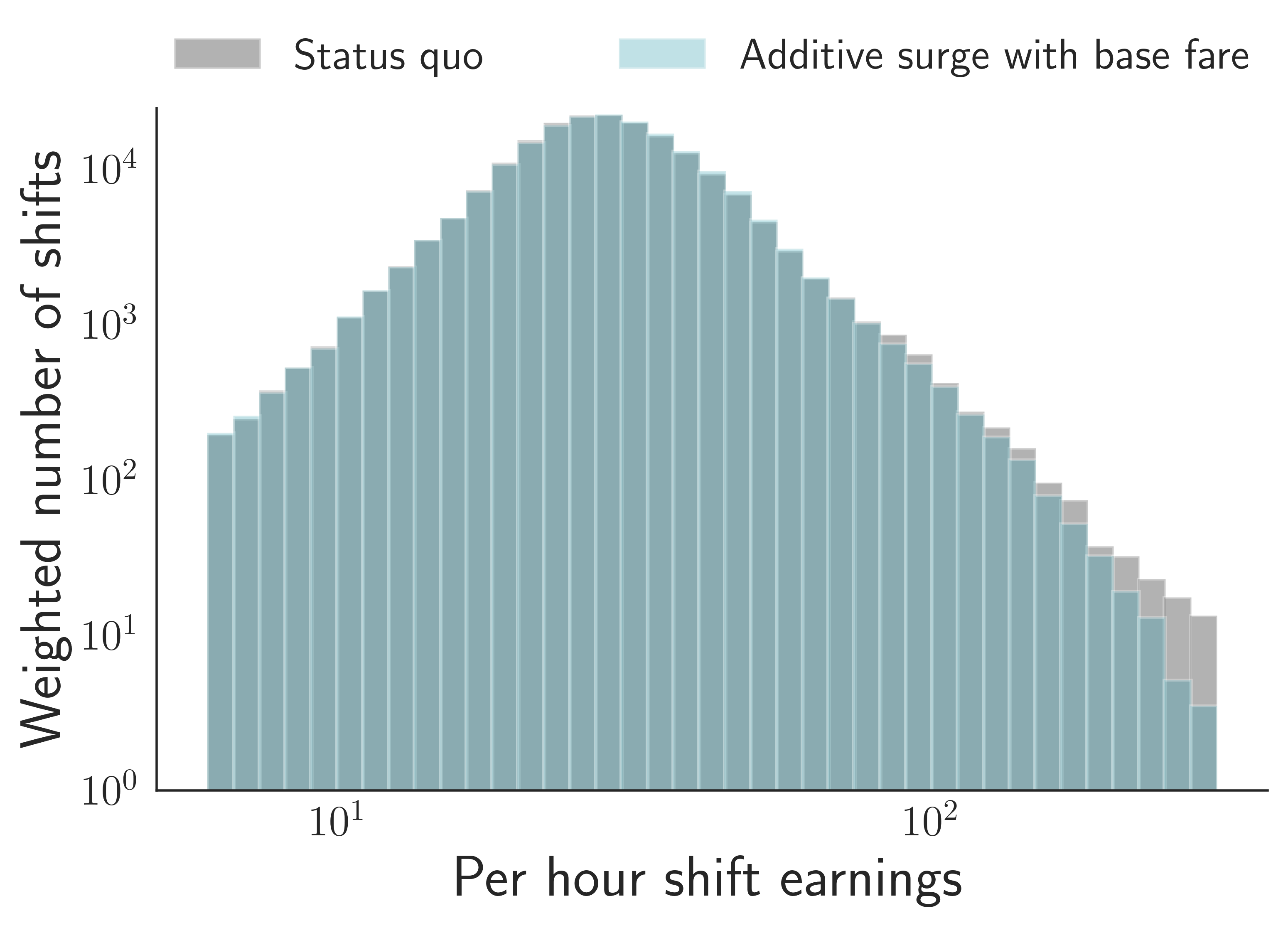}
	\caption{Histogram of per-shift driver earnings per hour. Note that the y-axis is in log scale.}
	\label{fig:rideaustinvariance}
\end{figure*}
We now calculate statistics regarding the average amount drivers earn during a single driving ``shift,'' ideally defined as the time between which drivers turn on their app and when they turn it off. To group trips together into a single driver shift, we use a data column called \textit{active driver ID}, which is a refinement of \textit{driver ID} and seems to correspond to a shift as defined internally by RideAustin.

The ``length'' of a shift is defined as the time between the first time the driver was dispatched for a trip during the shift, and the end time of the last completed trip during the shift. Note that this value is an underestimate of the true shift length, as it does not contain the time it took to receive the first trip request or the time it takes for the driver to go home after their last trip. Thus, our estimated shift per hour earnings are biased upwards.

The driver's total earning during the shift is simply the sum of the payments from each trip, under the payment function being analyzed. Then, the earnings per hour in a single shift is the total earnings divided by the shift length.

Figure~\ref{fig:rideaustinvariance} shows a weighted histogram of the per hour shift earnings, where the weights are the shift lengths in hours. Additive surge leads to a lower variance of per hour shift earnings (but the same mean, as constructed). The standard deviation of per-hour earnings are, respectively: $\$16.97$ (Status quo fare), and $\$15.83$ (Additive surge with base fare) with mean hourly earnings of about $\$32.22$. If we instead remove the minimum fare and pickup fare components and simulate pure additive or multiplicative surge, the standard deviations are:
$\$16.59$ (Additive surge), $\$18.35$ (Multiplicative surge).

\FloatBarrier
\subsection{Empirical analysis additional information}
\label{sec:empiricalrobustness}

We now provide additional detail for each step of the primary analysis presented in Section~\ref{sec:empiricalevidence}.

\subsubsection{Pre-processing}

There are $509,823$ rows (trips) in the time period analyzed.

\begin{itemize}
	\item $4626$ trips were longer than 1 hour or shorter than 30 seconds and were discarded.
	\item $3780$ were longer than 100 miles or shorter than 0.25 miles and were discarded (some overlap with those discarded for time).
	\item $26$ trips had clearly erroneous \textit{total fare} (null, or too high for mileage/distance by multiple orders of magnitude) and were not used to calibrate the reverse engineered fare.
\end{itemize}

We end up with $503,383$ trips in our analysis.

\subsubsection{Payment functions}

\begin{figure*}
		\begin{subfigure}[t]{0.48\textwidth}
	\centering
	\includegraphics[width=\linewidth]{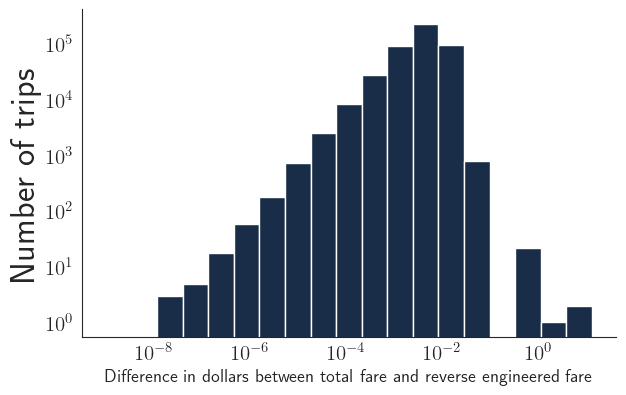}
	\caption{Histogram of difference between \textit{total fare} and the reverse engineered fare.}
	\label{fig:paymentdiffreverseengineer}
		\end{subfigure}
		~ \hfill
	\begin{subfigure}[t]{0.48\textwidth}
	\centering
\includegraphics[width=\linewidth]{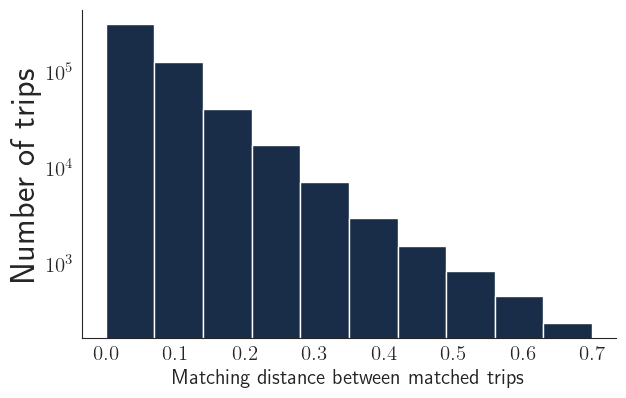}
\caption{``Matching distance'' between matched trips used for the counter-factual earnings. } %
\label{fig:matchingdistanceopendriver}
	\end{subfigure}
\end{figure*}

Figure~\ref{fig:paymentdiffreverseengineer} shows a histogram of the difference between the \textit{total fare} available as a column, and the reverse engineered fare derived from the functional form in the main text. The fit is good, with a mean difference of $\$0.005$.

\begin{figure*}[t!]
	\centering
		\includegraphics[width=.7\linewidth]{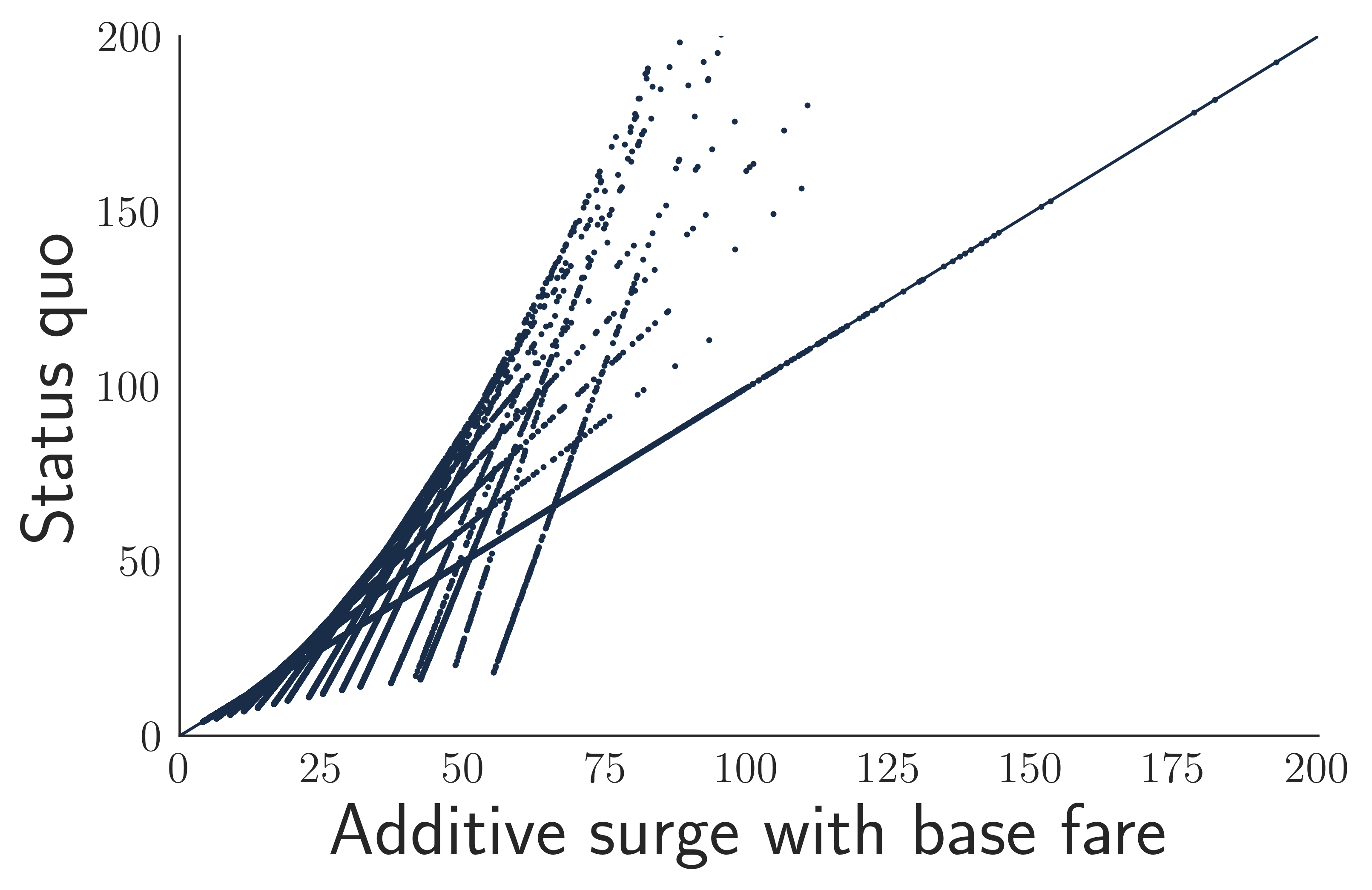}
	\caption{Constructed payment function (Additive surge with base fare) vs the reverse engineered Status quo fare payments at the trip level. As expected, additive surge tends to pay higher for shorter trips and lower for longer trips.}
	\label{fig:surgevsreverse}
\end{figure*}

Figure~\ref{fig:surgevsreverse} plots the constructed Additive surge fare versus the status quo payments, at the trip level. As expected, additive surge pays more for short surged trips, and less for long surged trips.

\subsubsection{Matching trips}
\label{sec:matchingappendix}
The ``matching distance'' as described in the main text between pairs of (\textit{date-time}, \textit{location}) tuples is:
\begin{align*}
\text{distance}((\text{time}_1, \text{location}_1), (\text{time}_2, \text{location}_2)) ={} & \text{difference in hours}(\text{time}_1, \text{time}_2) \\ &+\frac{1}{20} \text{difference in miles}(\text{location}_1, \text{location}_2)
\end{align*}

Figure~\ref{fig:matchingdistanceopendriver} shows the distribution of these distances between a given trip and the matched trip used for counter-factual earnings, for the matching technique described in the main text.

For robustness, we also use an alternate way to find a match for a given trip: using the \textit{next} driver who accepted a trip nearby. We calculate the matching distance between the given trip's \textit{start} time and location, and each future trips' \textit{start} time and location, and choose the driver of the closest match. As with the previous method, we filter out recent trips with drivers who are the same as the given trip's driver. Note that with this method, the expected earnings difference should be close to zero, as both drivers match at about the same time and place. However, the variances may vary with the payment function.

\subsubsection{Trip indifference}

\begin{figure*}[t!]
	\centering
		\includegraphics[width=.8\linewidth]{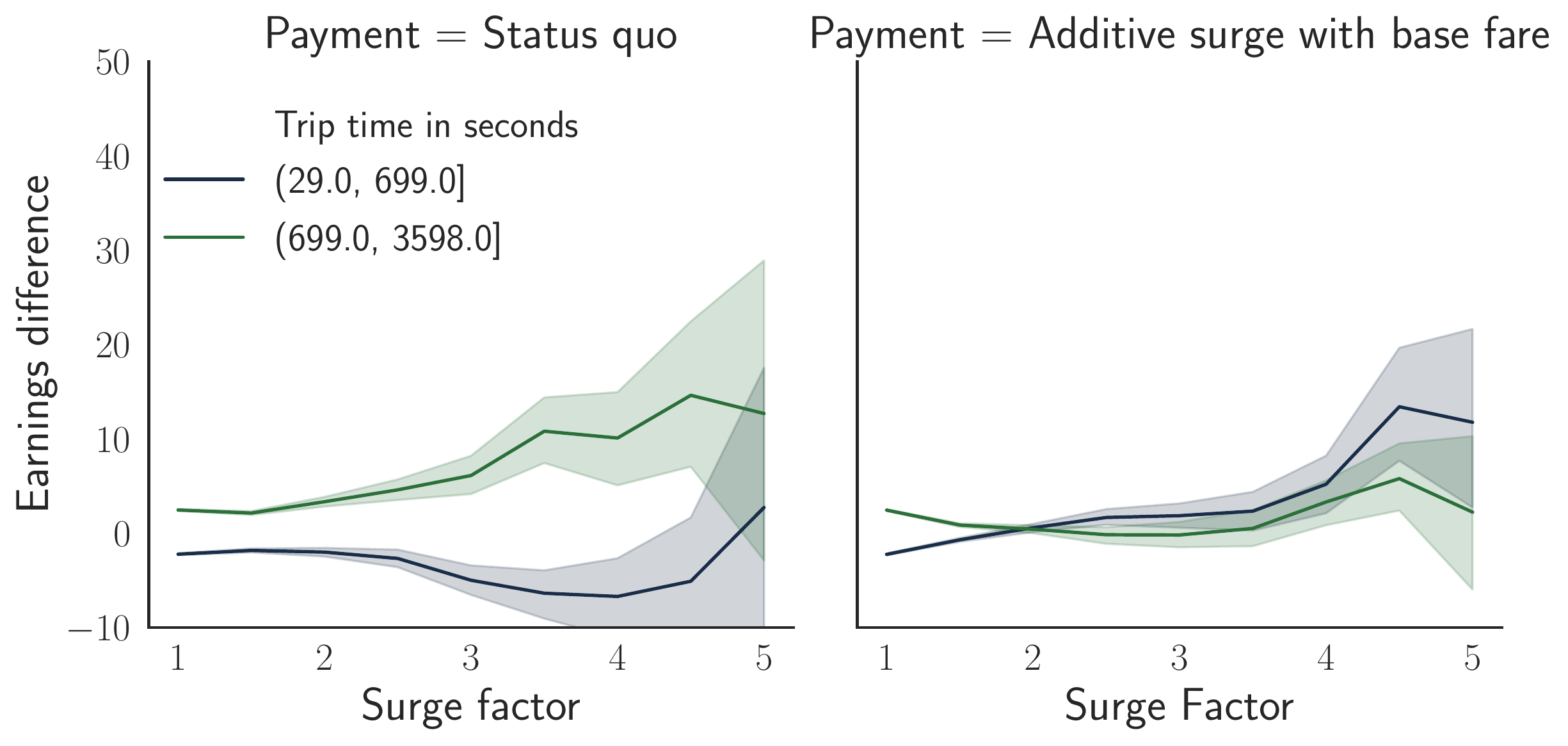}
		\caption{Using next nearby driver with an accepted trip as the counter-factual match.}
				\label{fig:tripindifferencenextnearbydriver}
\end{figure*}

\begin{figure*}[t!]
	\centering
			\includegraphics[width=.8\linewidth]{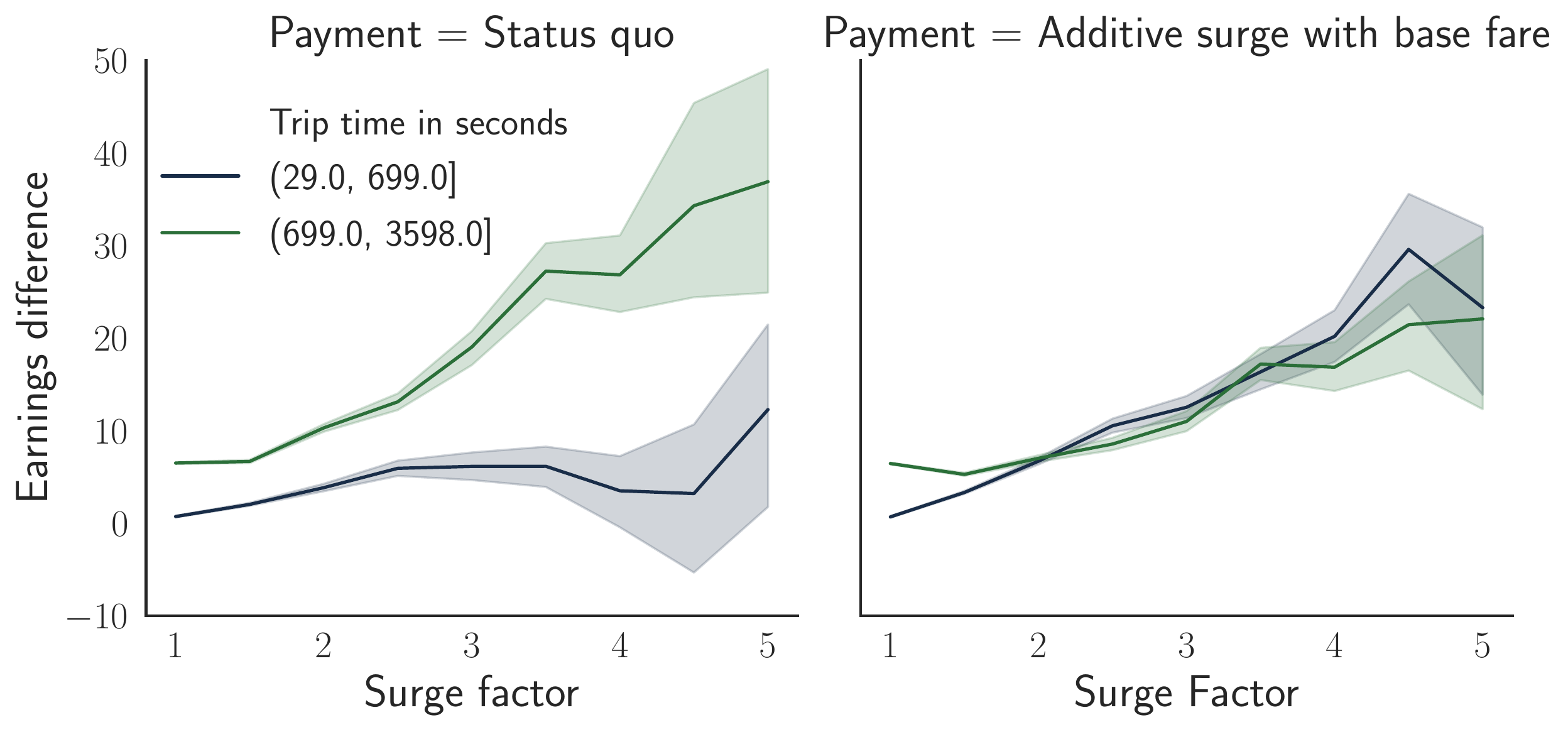}
	\caption{Using period length of next 1 hour (instead of 1.5 hours).}
	\label{fig:tripindifference15hours}
\end{figure*}

\begin{figure*}[t!]
	\centering
	\includegraphics[width=.8\linewidth]{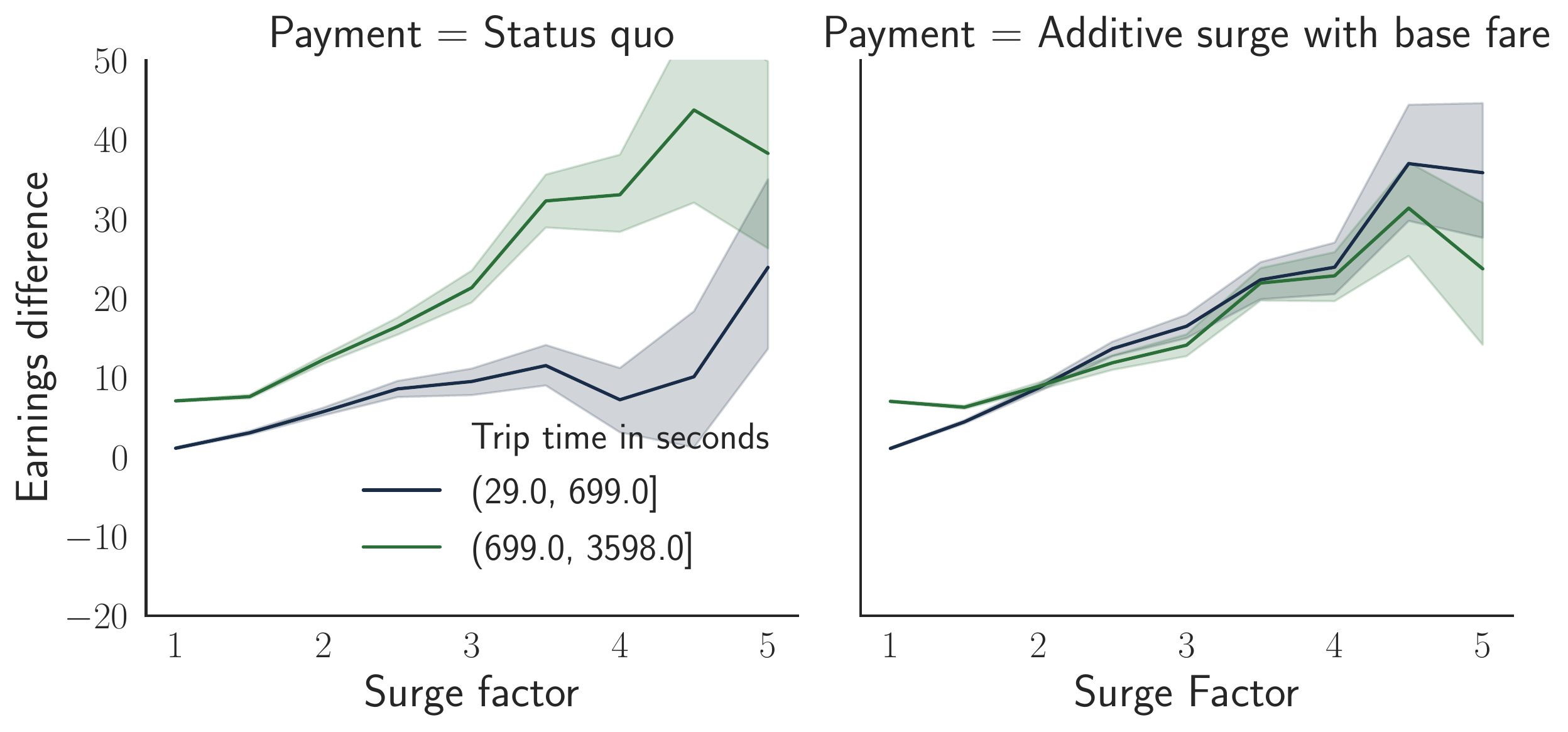}
	\caption{Starting measurement from dispatch time instead of trip start time, i.e., taking into account the first part of the trip that is unpaid for the driver.}
	\label{fig:tripindifferencedispatchtime}
\end{figure*}

\begin{figure*}[t!]
	\centering
	\includegraphics[width=.8\linewidth]{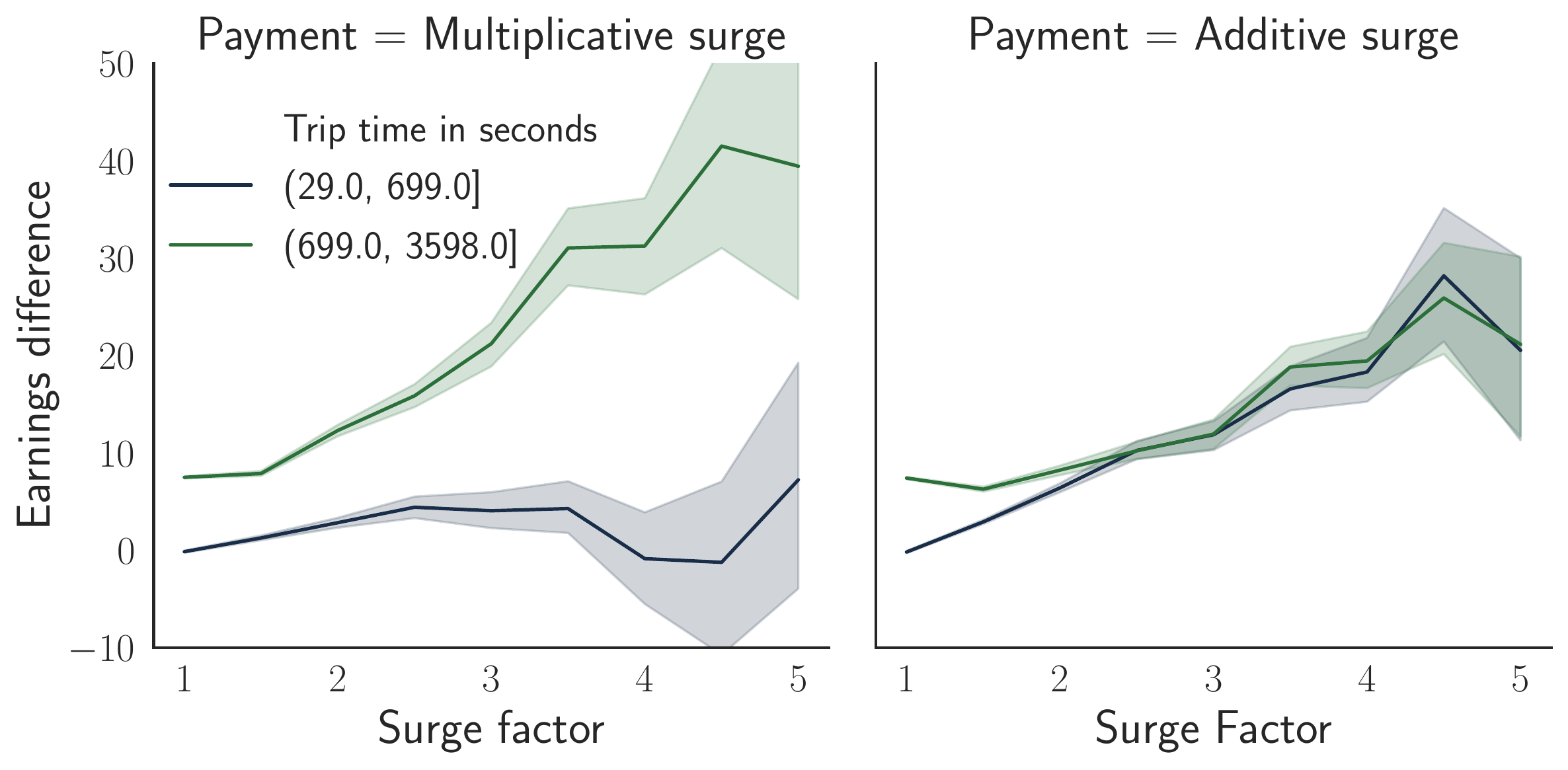}
	\caption{With pure multiplicative and additive surge, respectively (no min fare). }
	\label{fig:tripindifferencepurefuncs}
\end{figure*}

We now carry out some robustness checks for the trip indifference results, and present supplementary results.

 Figure~\ref{fig:tripindifferencenextnearbydriver} shows the same figure as in the main text, but instead using the next driver with an accepted trip matching function described in Section~\ref{sec:matchingappendix}. The means of the trip indifference (unconditional on trip length) are close to zero, as expected, but additive surge better balances the relative value of short and long trips, as before.

 Figure~\ref{fig:tripindifference15hours} shows the same figure as in the main text with the same matching function, but instead calculating the driver's earnings over the next 1 hour. Results are identical.

 Figure~\ref{fig:tripindifferencedispatchtime} starts counting the earnings of drivers starting at the given driver's \textit{dispatch} time instead of trip \textit{start} time; results are qualitatively identical, demonstrating that the fact that in practice there are two components to a trip  -- time from dispatch to the rider (unpaid typically), and time with the rider to the destination (paid) -- do not substantively affect the results.

Finally, Figure~\ref{fig:tripindifferencepurefuncs} shows the same figure but with how the driver would be paid under the \textit{pure} multiplicative and additive surge functions studied in the rest of this work, defined as follows:
\begin{align*}
\text{\textbf{Multiplicative surge}}: \,\,\,\,\,\,\, & \left[B \times M_{SurgeFactor}\right] \times SurgeFactor \\
\text{\textbf{Additive surge}}: \,\,\,\,\,\,\, & \left[B\times M_{SurgeFactor}\right] + \left[(SurgeFactor-1)\times A_{SurgeFactor}\right]
\end{align*}

$M_{SurgeFactor}$ and $A_{SurgeFactor}$ are surge factor dependent constants that are set such that these alternative payment functions spend the same amount of money overall for each surge factor as does the status quo fare. As with the additive surge with a minimum fare, these alternative payments do not change the mean trip payment conditional on the surge factor, but do change how money is allocated to various trips within that surge. If instead we used a single constant across surge factors, this feature would not hold, and the payment functions may pay different amounts on average for the same surge factor. %

\FloatBarrier

\newpage
\section{Proofs of single state model results}
\label{appsec:proofssinglestate}
\iftoggle{tableofcontentsapp}{
\invisiblelocaltableofcontents \label{toc:proofsinglestate}
}{}
In this section, we provide proofs of the theorems and lemmas in the main text regarding the single state model. Section~\ref{sec:appsingledriverreward} formally states the driver reward. Section~\ref{sec:appproofbasicmodelstrategy} contains the proof of Theorem~\ref{thm:basicmodeloptimal}. Section~\ref{sec:appproofsinglestateaffine} contains the proof of Proposition~\ref{lem:affineICstationary}. Finally, Section~\ref{sec:appuniquenesssinglestate} contains a partial uniqueness result regarding optimal driver policies.

\subsection{Driver reward}
\label{sec:appsingledriverreward}
Recall that $R(w,\sigma, t)$ is the total earnings from jobs finished from time $0$ to time $t$, i.e., $R(w,\sigma, t) = \bbE\left[\sum_{k=1}^{N(t)} w(\tau_i) \right]$, where $\tau_i$ is the length of the $i$th job the driver accepts, $e_i$ is time at which that job is accepted, and $N(t) = \left| \{i : 0 \leq e_i + \tau_i \leq t\}\right|$ is the number of accepted jobs up to time $t$.

Let a renewal cycle be the time the driver is open after completing a job to the next time the driver is open after completing a job. As mentioned using the renewal reward theorem in the main text,
\begin{align*}
R(w,\sigma) &\triangleq \lim\inf_{t\to\infty}\frac{R(w,\sigma, t)}{t} = \frac{\text{Expected cycle payment given } \sigma}{\text{Expected cycle length given } \sigma} = \frac{\frac{1}{F(\sigma)}\int_{\tau \in \sigma} w(\tau) dF(\tau)}{\frac{1}{F(\sigma)\lambda} + \frac{1}{F(\sigma)}\int_{\tau \in \sigma} \tau dF(\tau)}
\end{align*}
The $\frac{1}{\lambda F(\sigma)}$ term is the expected value of a exponential random variable with rate $\lambda F(\sigma)$, which is the rate at which a driver accepts ride requests when open.

\subsection{Proof of Theorem~\ref{thm:basicmodeloptimal}}
\label{sec:appproofbasicmodelstrategy}
We now prove Theorem~\ref{thm:basicmodeloptimal}, regarding the form of the optimal policy in the single-state model -- where the \textit{length} of a trip does not matter, only the earnings rate. The optimal policy trades off the earnings rate while on a trip with the driver's utilization rate. At a high level, the proof proceeds as follows: starting from any policy that is not of the appropriate form, we replace trips in the policy with those with a higher earnings rate, while keeping the utilization rate exactly the same. Such replacements result in a policy that is \textit{almost} of the correct form, except there may be an earnings rate $c$ such that only a \textit{subset} of $\{\tau : \frac{w(\tau)}{\tau} = c \}$ is in the policy. The remainder of the proof is showing that such a policy can transformed to a policy of the appropriate form without reducing the reward.

For ease of reading, we re-state each main text result in the appropriate location in the Appendix.

\thmbasicmodeloptimal*
\proof{Proof.}
Let $\gamma(\tau) \triangleq \frac{w(\tau)}{\tau}$ be the per time earning rate for a trip of length $\tau$. 	Assume that $w(\tau)$ is not zero everywhere, i.e., $F(\{\tau: w(\tau) > 0\})>0$. Otherwise any policy is optimal and so the result is trivial.

For each threshold $c$, let $\sigma^>_c$ denote the set  $\{\tau: \gamma(\tau) > c \}$, i.e., a \textit{strict} threshold policy where the threshold inequality is strict. Let $\sigma^\geq_c$ denote the set $\{\tau: \gamma(\tau) \geq c \}$, i.e., a \textit{complete} threshold policy where all trips at the threshold are included. Let $\tilde\sigma_{c} = \sigma^>_{c} \cup C$ for some $C \subseteq \{\tau: \gamma(\tau) = c_0\}$ be a \textit{partial} threshold policy where \textit{some} trips at the threshold are included.

The proof proceeds in three steps. Starting at any set $\sigma \subset (0, \infty)$, each step sequentially replaces $\sigma$ with a set $\sigma'$ closer to the desired form such that $R(\sigma') \geq R(\sigma)$.

\begin{description}
	\item[Step 1] If $\sigma$ is not already at least \textit{partial} threshold policy, then it can be replaced by a partial threshold policy while improving the reward: If there does not exist $c \in \bbR^+, C \subseteq \{\tau: \gamma(\tau) = c\}$ such that $\sigma = \sigma^>_c \cup C$, then there exists $\tilde\sigma_{c_0}$ such that $R(w, \tilde\sigma_{c_0}) > R(w, \sigma)$, where $\tilde\sigma_{c_0} = \sigma^>_{c_0} \cup C_0$ for some $c_0 \in \bbR^+, C_0 \subseteq \{\tau: \gamma(\tau) = c_0\}$.

	The rest of the proof is devoted to showing that a partial threshold policy can be replaced by a threshold policy where all trips at the threshold are included.

	\item [Step 2] A partial threshold policy is weakly dominated by either a strict or complete threshold policy: for any $\tilde\sigma_{c_0}$ of the form $\tilde\sigma_{c_0} = \sigma^>_{c_0} \cup C_0$ for some $c_0 \in \bbR^+, C_0 \subseteq \{\tau: \gamma(\tau) = c_0\}$, at least one of the following is true: $R(w, \sigma^>_{c_0}) \geq R(w, \tilde\sigma_{c_0})$ or $R(w, \sigma^\geq_{c_0}) \geq R(w, \tilde\sigma_{c_0})$.

	\item [Step 3]  There exists an optimal complete threshold policy: $\exists c^*$ such that for all $c$: \[R(w, \sigma^\geq_{c^*}) \geq \max( R(w, \sigma^\geq_{c}), R(w, \sigma^>_{c}))\]
\end{description}

Thus there exists $c^*$, such that for all $\sigma$, we have $R(w, \sigma^\geq_{c^*}) \geq R(w, \sigma)$.

Note that if $\sigma = (0, \infty)$, $F(\{\tau: \gamma(\tau) > 0\}\cap\sigma) = 0$, or $F(\{\tau: \gamma(\tau) > 0\}\cap\sigma) = 1$, then we can skip the first two steps, with the set $\sigma_c = (0, \infty)$.%

\vspace{.5em}

\parbold{Step 1} If there does not exist $c \in \bbR^+, C \subseteq \{\tau: \gamma(\tau) = c\}$ such that $\sigma = \sigma^>_c \cup C$, then there exists $\tilde\sigma_{c_0}$ such that $R(w, \tilde\sigma_{c_0}) > R(w, \sigma)$, where $\tilde\sigma_{c_0} = \sigma^>_{c_0} \cup C_0$ for some $c_0 \in \bbR^+, C_0 \subseteq \{\tau: \gamma(\tau) = c_0\}$.

	For given $\sigma,c$, let
	\begin{align*}
	A_c &= \{\tau: \tau \notin \sigma, \gamma(\tau) \geq c \}\\
	B_c &= \{\tau: \tau \in \sigma, \gamma(\tau) < c \}\\
	L(X) &= \int_{\tau\in X} \tau dF(\tau) & X\subseteq (0, \infty)
	\end{align*}

	$A_c$ is a set of trips that pay at least $c$ per unit time but are not in $\sigma$, and $B_c$ is the set of the trips that pay less than $c$ per unit time and are in $\sigma$. $L(X)$ is the mean extra utilization that trips in $X$ contribute in a renewal cycle. 	The idea is that if we find sets $A,B$ such that the marginal utilizations are equal ($L(A) = L(B) > 0$) and the earnings rate in set $A$ dominate those in set $B$ ($\gamma(a)> \gamma(b), \forall a\in A, b\in B$), then we can replace $B$ in the policy with $A$:  $\sigma' = \sigma \cup A \setminus B \implies R(w, \sigma') > R(w, \sigma)$. The denominator of the reward stays the same, and the numerator increases.

	A few facts that follow from the assumptions and definitions:
	\begin{align*}
		L(A_0) > 0 && \sigma \neq (0, \infty), \,\,\,\, F(\{\tau: \gamma(\tau) > 0\}\cap\sigma) < 1\\
		\exists c: L(B_c) > 0 && F(\{\tau: \gamma(\tau) > 0\}\cap\sigma) > 0 \\
		L(A_c) \text{ is non-increasing in $c$} && A_c \text{ contracts as $c$ increases}\\
		L(B_c)\text{ is non-decreasing in $c$} && B_c \text{ expands as $c$ increases} \\
		\lim_{c\to\infty} L(A_c)= 0 && \gamma(\tau) \text{ asymptotically bounded by defn}\\
		L(B_0)= 0 && \gamma(\tau) \text{ non-negative} \\
		\text{$L(A_c)$, $L(B_c)$ are left-continuous in $c$} &&
    \end{align*}
    To help see the last claim, notice that $L(A_c)$ and $L(B_c)$ are discontinuous only where $F(\{ \tau : \gamma(\tau) = c\})$ is non-zero, and even in such cases are continuous from the left.

The non-increasing/non-decreasing properties imply that $\exists c'$ such that $L(A_c) < L(B_c), \forall c > c'$.
This fact, along with the left-continuity and the same points of discontinuity for $L(A_c),L(B_c)$, implies that \[\exists c_0 \,\,\,\,\, \text{such that }\,\,\, c_0 = \max \{c' : L(A_{c'}) \geq L(B_{c'}) \}\]

If $L(A_{c_0}) = L(B_{c_0})$, then we are done with this part: let $\tilde\sigma_{c_0} = \sigma \cup A_{c_0} \setminus B_{c_0} = \{\tau: \gamma(\tau)\geq c_0\}$, and we have that $R(w, \tilde\sigma_{c_0}) > R(w, \sigma)$.

Otherwise if $ L(A_{c_0}) > L(B_{c_0})$ (which can happen if $F(\{ \tau : \gamma(\tau) = c_0\})$ is non-zero.), we need to select a subset of $\{ \tau : \gamma(\tau) = c_0\}$ such that the overall utilization of the constructed set remains the same. We can do so as follows:%
	\begin{itemize}
		\item 		 By the definition of $c_0$, for all $c > c_0$ we have $L(A_{c}) < L(B_{c})$. Then %
		\begin{align*}
		L(B_{c_0}) < L(A_{c_0}) < L(B_{c_0} \cup \{\tau: \tau \in \sigma, \gamma(\tau) = c_0\})
		\end{align*}
		\item Let $C \subseteq \{\tau: \tau \in \sigma, \gamma(\tau) = c_0\}$ such that $L(B_{c_0} \cup C) = L(A_{c_0})$. Such $C$ exists because $F$ is continuous.
		\item Let $\tilde\sigma_{c_0} = \sigma \cup A_{c_0 } \setminus \left(B_{c_0} \cup C\right)$, which is equal to $\tilde\sigma_{c_0} = \sigma^>_{c_0 } \cup C_0$ for some $C_0 \subseteq \{\tau: \gamma(\tau) = c_0\}$.
	\end{itemize}
	We now have $R(w, \tilde\sigma_{c_0}) > R(w, \sigma)$, as the utilization rates of both sets are the same, and each trip in $\tilde\sigma_{c_0}$ is at least as valuable per unit time as a corresponding trip in $\sigma$.
\vspace{3em}
\parbold{Step 2} For any $\tilde\sigma_{c_0}$ of the form $\tilde\sigma_{c_0} = \sigma^>_{c_0} \cup C_0$ for some $c_0 \in \bbR^+, C_0 \subseteq \{\tau: \gamma(\tau) = c_0\}$, at least one of the following is true: $R(w, \sigma^>_{c_0}) \geq R(w, \tilde\sigma_{c_0})$ or $R(w, \sigma^\geq_{c_0}) \geq R(w, \tilde\sigma_{c_0})$.

Let $C_0' = \{\tau \in  \{\tau: \tau \in \sigma, \gamma(\tau) = c_0\} \setminus C_0\}$, i.e., the set of trips such that $\gamma(\tau) = c_0$ but that are not in $C_0$.

We prove this step by reasoning about the value of $R(w, \tilde\sigma_{c_0})$ in comparison to the marginal value threshold $c_0$.

	\begin{itemize}
		\item Suppose $c_0 \geq R(w, \tilde\sigma_{c_0})$. Then, we can add trips to the set:
		\begin{align}
		R(w, \{\tau: \gamma(\tau) \geq c_0 \}) &= \frac{\lambda\int_{\tau\in{\tilde\sigma_{c_0}}} w(\tau)dF(\tau) + \lambda\int_{\tau \in C_0'} w(\tau)dF(\tau)}{1 + \lambda\int_{\tau\in{\tilde\sigma_{c_0}}} \tau dF(\tau) + \lambda\int_{\tau \in C_0'} \tau dF(\tau)}\nonumber\\
		&\geq R(w, \tilde\sigma_{c_0}) \label{eqnpart:addmassimprove}
		\end{align}
		Where the inequality follows from $
		R(w, \tilde\sigma_{c_0}) = \frac{\lambda\int_{\tau\in{\tilde\sigma_{c_0}}} w(\tau)dF(\tau) }{1 + \lambda\int_{\tau\in{\tilde\sigma_{c_0}}} \tau dF(\tau) }$,

		$\frac{\lambda\int_{\tau\in{C_0'}} w(\tau) dF(\tau)}{\lambda\int_{\tau\in C_0'} \tau dF(\tau)} = \frac{\lambda  \int_{\tau\in{C_0'}} \frac{w(\tau)}{\tau}\tau dF(\tau)}{\lambda\int_{\tau\in C_0'} \tau dF(\tau)} = c_0$, and
		$ \frac{x}{z}  \geq \frac{w}{y}\implies \frac{w + x}{y + z} \geq \frac{w}{y}$.

		\item Alternatively, suppose $c_0 < R(w, \tilde\sigma_{c_0})$. Then, we can remove trips from the set:
		\begin{align}
		R(w, \{\tau: \gamma(\tau) > c_0 \}) &= \frac{\lambda\int_{\tau\in{\tilde\sigma_{c_0}}} w(\tau)dF(\tau) - \lambda\int_{\tau \in C_0} w(\tau)dF(\tau)}{1 + \lambda\int_{\tau\in{\tilde\sigma_{c_0}}} \tau dF(\tau) - \lambda\int_{\tau \in  C_0} \tau dF(\tau)}\nonumber\\
		&> R(w, \tilde\sigma_{c_0}) \label{eqnpart:removemassimprove}
		\end{align}
		Where the inequality follows from
		$\frac{w}{y} > \frac{x}{z}  \implies \frac{w - x}{y - z} > \frac{w}{y}  $ when $w-x\geq0, y-z\geq 0$.

	\end{itemize}

\vspace{2em}
\parbold{Step 3} $\exists c^*$ such that for all $c$: \[R(w, \sigma^\geq_{c^*}) \geq \max( R(w, \sigma^\geq_{c}), R(w, \sigma^>_{c}))\]

In the first subpart, we simply need to prove that there exists a maximizer $c^*$ for the function $\max( R(w, \sigma^\geq_{c}), R(w, \sigma^>_{c}))$: this fact is not immediate because $\sigma$ are infinite sets.
The following are true
\begin{itemize}
	\item By assumption that $w(\tau)/\tau$ is asymptotically bounded, we have that the reward is bounded: there exists $\bar R$ such that for all $\sigma$, we have $R(w, \sigma) \in [0, \bar R]$.

	\item $F$ is a continuous distribution, and so $ \lim_{c\to\infty} F(\{\tau: \gamma(\tau) \geq c\}) = 0$. There exists $C$ such that $\forall c > C$: $R(w, \sigma^>_{c})< R((0, \infty)), R(w, \sigma^\geq_{c}) < R((0, \infty))$

	\item $R(w, \sigma^>_c)$ is continuous from the right in $c$, and $R(w, \sigma^\geq_c)$ is continuous from the left in $c$, and the two functions have the same points of discontinuities: $c$ such that $F(\{\tau: \gamma(\tau) = c\}) > 0$ (and these are also their only points of disagreement). To see these facts, observe that $F(\sigma^>_c)$ and $F(\sigma^\geq_c)$ have the same properties, respectively.

\end{itemize}

Thus, the function $\max (R(w, \sigma^\geq_c), R(w, \sigma^>_c))$ of $c$ attains its maximum at some $c^* \in [0,C]$. In other words, there exists $c^{*}$ such that $\forall c, \max (R(w, \sigma^\geq_{c^*}), R(w, \sigma^>_{c^*})) \geq \max (R(w, \sigma^\geq_c), R(w, \sigma^>_c))$.

In the second subpart, we finish by proving that $R(w, \sigma^\geq_{c^{*}}) \geq R(w, \sigma^>_{c^{*}})$, i.e., that we can include trips at the margin of per-time value to the policy.

\begin{itemize}
	\item Suppose ${c^{*}} \geq R(w, \sigma^>_{{c^{*}}})$. Then, by the same argument as Line~\eqref{eqnpart:addmassimprove}, $ R(w, \sigma^\geq_{{c^{*}}}) \geq R(w, \sigma^>_{{c^{*}}})$, including the marginal trips increases the reward.

	\item Suppose ${c^{*}} < R(w, \sigma^>_{{c^{*}}})$. %

	\begin{itemize}
		\item If $\exists B: {c^{*}} < B$ such that the mass $F(\{ \tau: \gamma(\tau) \in ({c^{*}}, B]\})= 0$, then note that $\sigma^>_{c^{*}}$ is equal to $\sigma^\geq_{B}$ up to a set of measure $0$, and so $R(w, \sigma^>_{c^{*}}) = R(w, \sigma^\geq_{B})$.%
		\item Otherwise, let $B:  {c^{*}} < B < R(w, \sigma^>_{{c^{*}}})$, and note that $F(\{ \tau: \gamma(\tau) \in ({c^{*}}, B]\}) > 0$. Then, by the same argument as in Line~\eqref{eqnpart:removemassimprove}, $R(w, \sigma^>_{c^{*}}) < R(w, \sigma^>_B) \leq \max (R(w, \sigma^\geq_{c^*}), R(w, \sigma^>_{c^*})) = R(w, \sigma^\geq_{c^*})$: we can remove the subset $(c^*, B)$ from the policy $\sigma^>_{c^*}$ and improve reward, and so $\sigma^\geq_{c^*}$ must be optimal. %
	\end{itemize}
\end{itemize}

Thus there exists $c^*$, such that for all $\sigma$, we have $R(w, \sigma^\geq_{c^*}) \geq R(w, \sigma)$. \qed

\subsection{Proof of Proposition~\ref{lem:affineICstationary}}
\label{sec:appproofsinglestateaffine}

\lemaffineICstationary*
\proof{Proof.}
Let $T = \int_{\tau\in (0, \infty)} \tau dF(\tau)$. Let $\sigma' = (0, \infty)\setminus \sigma$, for some $\sigma$.
\begin{align*}
R((0, \infty))&= \frac{\lambda\int_{\tau\in(0, \infty)} w(\tau) dF(\tau)}{1 + \lambda T}\\
R(\sigma') &= \frac{\lambda\int_{\tau\in(0, \infty)} w(\tau) dF(\tau) - \lambda\int_{\tau\in \sigma} w(\tau) dF(\tau)}{1 + \lambda T - \lambda\int_{\tau\in\sigma} \tau dF(\tau)}\\
\implies R((0, \infty)) \geq R(\sigma')   &\iff \frac{\lambda\int_{\tau\in(0, \infty)} w(\tau) dF(\tau)}{1 + \lambda T} \leq \frac{\int_{\tau\in \sigma} w(\tau) dF(\tau)}{\int_{\tau\in\sigma} \tau dF(\tau)}
\end{align*}
The last line follows from $\frac{w}{y} \geq\frac{w-x}{y-z} \iff \frac{w}{y}\leq\frac{x}{z}$.

Thus, a necessary and sufficient condition for incentive compatibility is that
\begin{align*}	\frac{\lambda\int_{\tau\in(0, \infty)} w(\tau) dF(\tau)}{1 + \lambda T} &\leq \frac{\int_{\tau\in \sigma} w(\tau) dF(\tau)}{\int_{\tau\in\sigma} \tau dF(\tau)} &  \forall \sigma. \end{align*}

Suppose $w(\tau) = m\tau + a$. Then, for $0 \leq a \leq \frac{m}{\lambda}$:
\begin{align}	\frac{\lambda\int_{\tau\in(0, \infty)} w(\tau) dF(\tau)}{1 + \lambda T}  &= \frac{\lambda(mT + a)}{1 + \lambda T}\nonumber\\ &\leq \frac{m(\lambda T + 1)}{1 + \lambda T} = m & a \leq \frac{m}{\lambda}\nonumber\\
&\leq m + a\left[\frac{F(\sigma)}{\int_{\tau\in\sigma} \tau dF(\tau)}\right] \,\,\,\,\,\,\,\, \forall \sigma & a \geq 0 \label{eqpart:loosenec}\\
&= \frac{\int_{\tau\in \sigma} w(\tau) dF(\tau)}{\int_{\tau\in\sigma} \tau dF(\tau)}\nonumber
\end{align}

Note that the condition is sufficient but not necessary. Proving necessary conditions requires tightening Line~\eqref{eqpart:loosenec} under assumptions on the distribution $F$.

\qed

\subsection{Uniqueness of optimal policy for single-state model}
\label{sec:appuniquenesssinglestate}
\begin{lemma}
	Consider the single-state model. There exists an optimal policy $\sigma^*$ of the form $\sigma^* = \{\tau : \frac{w(\tau)}{\tau} \geq c^*\}$ such that $R(\sigma^*) = c^*$. Furthermore, this policy is the unique optimal policy, up to sets of measure 0 and up to modifications (subtractions) of sets $\{\tau : \frac{w(\tau)}{\tau} = c^*\}$.

	\label{lem:basicoptimalformunique}
\end{lemma}
\proof{Proof.}
	By Theorem~\ref{thm:basicmodeloptimal}, there exists an optimal policy of the form $\sigma^* = \{\tau : \frac{w(\tau)}{\tau} \geq c^*\}$, for some $c^*$. Here, we show (1) that there exists an optimal policy $\sigma^*$ of that form such that $R(\sigma^*) = c^*$, and (2) this is the unique optimal policy up to sets of measure 0 and up to modifications (subtractions) of sets $\{\tau : \frac{w(\tau)}{\tau} = c^*\}$. 
	
	\begin{enumerate}
		\item Start with any optimal policy $\sigma^*$ of the form $\sigma^* = \{\tau : \frac{w(\tau)}{\tau} \geq c\}$, for some $c$, and let $c^* = R(\sigma^*)$ be the optimal reward. Then, $\sigma_{c^*}^\geq = \sigma^*$ up to sets of measure 0, where $\sigma_{c^*}^\geq = \{\tau : \frac{w(\tau)}{\tau} \geq c^*\}$. If $c^* = c$, this is trivial. Otherwise,

		Suppose $R(\sigma^*) = c^* > c$. Then, note that $\sigma_{c^*}^\geq \subseteq \sigma^*$. If $F(\sigma^*\setminus \sigma_{c^*}^\geq) > 0$:
		\begin{align*}
		R(\sigma_{c^*}^\geq) &= \frac{\lambda\int_{\tau\in{\sigma^*}} w(\tau)dF(\tau) - \lambda\int_{\tau \in \sigma^*\setminus \sigma_{c^*}^\geq } w(\tau)dF(\tau)}{1 + \lambda\int_{\tau\in{\sigma^*}} \tau dF(\tau) - \lambda\int_{\tau \in \sigma^*\setminus \sigma_{c^*}^\geq} \tau dF(\tau)}\nonumber\\
		&> R(\sigma^*) %
		\end{align*}
		Which follows from $\frac{\lambda\int_{\tau \in \sigma^*\setminus \sigma_{c^*}^\geq } w(\tau)dF(\tau)}{\lambda\int_{\tau \in \sigma^*\setminus \sigma_{c^*}^\geq} \tau dF(\tau)} < c^* = R(\sigma^*) = \frac{\lambda\int_{\tau\in{\sigma^*}} w(\tau)dF(\tau)}{1 + \lambda\int_{\tau\in{\sigma^*}} \tau dF(\tau)}$, and $\frac{x}{z} < \frac{w}{y}  \implies \frac{w-x}{y-z} > \frac{w}{y}$ when $w-x\geq0, y-z\geq 0$.
		 This contradicts that $\sigma^*$ is optimal.
		
		Similarly, suppose $R(\sigma^*) = c^* < c$. Then, note that $\sigma^*\subseteq\sigma_{c^*}^\geq $. If $F(\sigma_{c^*}^\geq\setminus\sigma^*) > 0$:
		\begin{align*}
		R(\sigma_{c^*}^\geq) &= \frac{\lambda\int_{\tau\in{\sigma^*}} w(\tau)dF(\tau) + \lambda\int_{\tau \in \sigma_{c^*}^\geq\setminus \sigma^* } w(\tau)dF(\tau)}{1 + \lambda\int_{\tau\in{\sigma^*}} \tau dF(\tau) + \lambda\int_{\tau \in \sigma_{c^*}^\geq\setminus \sigma^*} \tau dF(\tau)}\nonumber\\
		&> R(\sigma^*) %
		\end{align*}
		Which follows from $\frac{\lambda\int_{\tau \in \sigma_{c^*}^\geq\setminus \sigma^* } w(\tau)dF(\tau)}{\lambda\int_{\tau \in \sigma_{c^*}^\geq\setminus \sigma^*} \tau dF(\tau)} > c^* = R(\sigma^*) = \frac{\lambda\int_{\tau\in{\sigma^*}} w(\tau)dF(\tau)}{1 + \lambda\int_{\tau\in{\sigma^*}} \tau dF(\tau)}$, and $\frac{x}{z} > \frac{w}{y} \implies \frac{w+x}{y+z} > \frac{w}{y}$. This contradicts that $\sigma^*$ is optimal.

	\item The first part above proves uniqueness among policies of the form $\sigma_c^\geq = \{\tau : \frac{w(\tau)}{\tau} \geq c\}$, for some $c$. Step 1 of the proof of Theorem~\ref{thm:basicmodeloptimal} further shows that only policies of the form $\tilde\sigma_{c_0} = \sigma^\geq_{c_0} \setminus C_0$ for some $c_0 \in \bbR^+, C_0 \subseteq \{\tau: \gamma(\tau) = c_0\}$ can be optimal. Arguments near identical to that above and to Step 2 of the proof of Theorem~\ref{thm:basicmodeloptimal} will finish the proof.

	\end{enumerate}

\qed

\newpage
\section{Proofs of dynamic model results}
\label{appsec:proofsdynamic}
\iftoggle{tableofcontentsapp}{
\invisiblelocaltableofcontents \label{toc:proofsdynamic}
}{}

In this section, we provide proofs of the theorems and lemmas in the main text regarding the dynamic model. Section~\ref{appsec:breakingdownmu} contains proofs for the dynamic model lemmas regarding driver reward and time spent in each state, Lemmas~\ref{lem:statingreward},~\ref{lem:findingq}, and~\ref{lem:breakingdownmui}.

Section~\ref{sec:dynamicproofstrategyandoverview} contains an overview of the proof strategy for both Theorems~\ref{prop:multnotoptimal} and~\ref{thm:ICpolicy}, and in particular contains the main technical lemma used to prove both theorems.

Section~\ref{sec:ICapplemmas} contains the statements of several auxiliary lemmas that are used to prove the main results. Proofs for these lemmas are deferred to Section~\ref{sec:appdynamiclemmasproofstedious}, as they are algebraically tedious.

Finally, Section~\ref{sec:proofsmaindynamicresults} contains the proofs for our main results, Theorems~\ref{prop:multnotoptimal} and~\ref{thm:ICpolicy}.

\subsection{Driver reward}
\label{appsec:breakingdownmu}

\lemfindingq*
\proof{Proof.} Given the state dynamics in the model, $q_{i\to j}(s)$ is determined by the evolution of a CTMC in time $s$, given that the current state is $i$. We can use standard CTMC results here. Let $Q$ denote the $Q$-matrix for the world state CTMC. From the model definition,
$$Q = \begin{bmatrix}
	-\lambda_{1\to 2}       & \lambda_{1\to 2}  \\
	\lambda_{2\to 1} & -\lambda_{2\to 1}\end{bmatrix} $$
Recall that the state transition matrix after time $t$ is then given by the matrix exponential $e^{Qt}$, which is equal to the inverse of the Laplace transform of the inverse of the resolvent of $Q$:
\begin{align*}
	q_{i\to j}(\tau) &= (e^{Q\tau})_{ij}\\
	&= \mathcal{L}^{-1}((wI - Q)^{-1}_{ij})(\tau) & \text{$w$ is a Laplace transform parameter}\\
	&=  \frac{\lambda_{i\to j}}{\lambda_{i\to j}+\lambda_{j\to i}}\left[1 - e^{-(\lambda_{i\to j} + \lambda_{j\to i})\tau}\right]
\end{align*}
where the closed form in the last line emerges due to the 2 state model assumption. \qed

Then, Lemma~\ref{lem:statingreward} and Lemma~\ref{lem:breakingdownmui} are proven together next.

\lemstatingreward*
\lembreakingdownmui*

\proof{Proof.} Consider the renewal process (with cycles and sub-cycles) defined in the main text. A single reward renewal cycle is: the time between the driver is open in state 1 to the next time the driver is open in state 1 \textit{after} being open in state 2 at least once. In other words, \textit{each} renewal cycle is composed of some number (potentially zero) of sub-cycles in which the driver is open in state 1 and then is open in state 1 again after a completed trip; one sub-cycle starting with the driver open in state 1 and ending with being open in state 2 (either after a completed trip or a state transition while open); some number (potentially zero) of sub-cycles in which the driver is open in state 2 and then is open in state 2 again after a completed trip; and finally one sub-cycle starting in state 2 and ending with the driver open in state $1$.

We use the following notation
\begin{itemize}
	\item $M(t)$ is the total number of cycles that have been completed up to time $t$
	\item $N_j(M)$ is the number of sub-cycles in state $j$ in the $M$th cycle -- i.e., in the $M$th cycle of the single renewal process described above, the number of times that the driver is open in state $j$ (after transitioning from the other state, or finishing a trip that started in the same state $j$)
	\item $S_j(k,M)$ is the length of the $k$th such sub-cycle in the $M$th cycle, with expected length $S_j(\sigma_j)$. Let $\tilde{S}(\sigma)$ be the expected length of one of the overall cycles.
	\item $W_j(k,M)$ is the earnings of the driver in the $k$th such sub-cycle in the $M$th cycle, with expected value $\hat W_j(\sigma_j)$
	\item $p_{ji}(\sigma_j)$ is the probability that the current sub-cycle is the last in state $j$ for the current cycle -- as the next sub-cycle starts in the other state.
	\item $R_j(w_j, \sigma_j, M)$ is the total amount earned in state $j$ after $M$ such cycles.
	\item Define the earnings rate in state $j$ as $R_j(w_j, \sigma_j) = \frac{\hat W_j(\sigma_j)}{S_j(\sigma_j)}$, the expected earnings in a sub-cycle over the expected length.
\end{itemize}

Then:
\begin{align*}
R_j(w_j, \sigma_j, M(t)) &= \sum_{M=1}^{M(t)} \sum_{k=1}^{N_{j}(M)} W_j(k,M)\\
\lim_{t\to\infty} \frac{R_j(w_j, \sigma_j, M(t))}{M(t)}
&= \lim_{t\to\infty}\frac{1}{M(t)}\left[ \sum_{M=1}^{M(t)} \sum_{k=1}^{N_{j}(M)} W_j(k,M)\right] \\
&= \frac{\hat W_j(\sigma_j)}{p_{ji} (\sigma_j) }  & \text{almost surely}
\end{align*}
by the mean of a geometric random variable ($\bbE[N_j(M)] = \frac{1}{p_{ji}(\sigma_j)}$ is the expected number of sub-cycles in $j$ in a given cycle)  and the basic law of large numbers for renewal processes.

Similarly, we know that $\frac{M(t)}{t}$ converges to its mean almost surely as $t\to\infty$, where the mean is based on the length of time in each state in each cycle. Then:
\begin{align*}
\lim_{t\to\infty} \frac{M(t)}{t} &= \frac{1}{\tilde{S}(\sigma)}\\
\tilde{S}(\sigma) &= \bbE\left[\sum_{k=1}^{N_{1}(1)} S_1(k,1)\right] + \bbE\left[\sum_{k=1}^{N_{2}(1)} S_2(k,1)\right] \\
&= \bbE[N_{1}(1)] \bbE[S_1(k,1)] + \bbE[N_{2}(1)] \bbE[S_2(k,1)]& \text{Wald's identity}\\
&= \frac{1}{p_{12}(\sigma_1)} S_1(\sigma_1) + \frac{1}{p_{21}(\sigma_2)} S_2(\sigma_2)\\
\implies \lim_{t\to\infty} \frac{M(t)}{t} &= \frac{p_{21}(\sigma_2)p_{12}(\sigma_1)}{p_{21}(\sigma_2)S_1(\sigma_1) + p_{12}(\sigma_1)S_2(\sigma_2)}
\end{align*}

Then, by standard algebra on multiplication with almost sure convergence
\begin{align*}
\lim_{t\to\infty} \frac{R_j(w_j, \sigma_j, M(t))}{t}%
&= \lim_{t\to\infty} \frac{R_j(w_j, \sigma_j, M(t))}{M(t)} \frac{M(t)}{t}\\
&= \frac{1}{p_{ji} (\sigma_j) } \hat W_j(\sigma_j) \left[\frac{p_{21}(\sigma_2)p_{12}(\sigma_1)}{p_{21}(\sigma_2)S_1(\sigma_1) + p_{12}(\sigma_1)S_2(\sigma_2)}\right]\\
&=  \left[\frac{p_{ij}(\sigma_i)S_j(\sigma_j)}{p_{21}(\sigma_2)S_1(\sigma_1) + p_{12}(\sigma_1)S_2(\sigma_2)}\right] R_j(w_j, \sigma_j)
\end{align*}

Let $\mu_j(\sigma) \triangleq \frac{p_{ij}(\sigma_i)S_j(\sigma_j)}{p_{21}(\sigma_2)S_1(\sigma_1) + p_{12}(\sigma_1)S_2(\sigma_2)} $.
Putting the above together:
\begin{align*}
\lim \inf_{t\to\infty} \frac{R(w, \sigma, t)}{t} &= \lim \inf_{t\to\infty} \frac{R_1(w_1, \sigma_1, M(t))}{t} + \lim \inf_{t\to\infty} \frac{R_2(w_2, \sigma_2, M(t))}{t}\\
&= \mu_1(\sigma) R_1(w_1, \sigma_1) + \mu_2(\sigma) R_2(w_2, \sigma_2)
\end{align*}

To finish the proofs, we will further derive the form of the appropriate quantities. Recall that $S_i(\sigma_i)$ is the expected length of the time between being open in a state $i$ to being open again, either after a state transition or after finishing a job; and $p_{ij}(\sigma_i)$ is the probability that the driver is next open in state $j$ given they are  currently open in state $i$. These are:
\begin{align*}
S_i(\sigma_i) &= \frac{1}{\lambda_i F_i(\sigma_i) + \lambda_{i\to j}} + \frac{\lambda_iF_i(\sigma_i)}{{\lambda_i F_i(\sigma_i) + \lambda_{i\to j}}}\int_{\tau\in\sigma_i} \tau \frac{f_i(\tau)}{F_i(\sigma_i)} d\tau
\\&= \frac{1}{\lambda_i F_i(\sigma_i) + \lambda_{i\to j}}\left[1 + \lambda_i \int_{\tau\in\sigma_i} \tau dF_i(\tau)\right]\\
&= \left[\frac{\lambda_i F_i(\sigma_i)}{\lambda_i F_i(\sigma_i) + \lambda_{i\to j}}\right]T_i(\sigma_i) & T_i(\sigma_i) \triangleq \frac{1}{\lambda_iF_i(\sigma_i)} + \frac{1}{F_i(\sigma_i)} \int_{\tau\in\sigma_i} \tau dF_i(\tau)
\end{align*}
The first part of the sum $\frac{1}{\lambda_i F_i(\sigma_i) + \lambda_{i\to j}}$ is the expected time until either the driver receives and accepts a request, or the world state transitions to the other state. This form emerges because there are two competing independent exponential clocks -- that for a request and that for the world state changing. The second part of the sum is the probability of receiving an accepted trip request before a state transition, times the expected length of an accepted trip.

Similarly, the expected earning in a sub-cycle in state $j$ is:
\begin{align*}
	 \hat W_j(\sigma_j) &= \frac{\lambda_iF_i(\sigma_i)}{{\lambda_i F_i(\sigma_i) + \lambda_{i\to j}}}\int_{\tau\in\sigma_i} w_i(\tau) \frac{f_i(\tau)}{F_i(\sigma_i)} d\tau \\
	 &= \left[\frac{\lambda_i F_i(\sigma_i)}{\lambda_i F_i(\sigma_i) + \lambda_{i\to j}}\right]W_i(\sigma_i)
\end{align*}
and so $R_j(w_j, \sigma_j) = \frac{\hat W_j(\sigma_j)}{S_j(\sigma_j)} = \frac{ W_j(\sigma_j)}{T(\sigma_j)}$.

The next step is to find an expression for $p_{ij}(\sigma_i)$, the probability that the next renewal cycle is at state $j$, given the current one is at state $i$. We find it for $j\neq i$, and then $p_{ii}= 1-p_{ij}$.
\begin{align*}
p_{ij}(\sigma) &= \frac{\lambda_{i\to j}}{\lambda_iF_i(\sigma_i) + \lambda_{i\to j}} + \frac{\lambda_iF_i(\sigma_i)}{\lambda_iF_i(\sigma_i) + \lambda_{i\to j}} \frac{1}{{F_i(\sigma_i)}}\int_{\sigma_i} q_{i\to j}(\tau) dF_i(\tau)\\
&= \left[\frac{1}{\lambda_i F_i(\sigma_i) + \lambda_{i\to j}}\right]Q_i(\sigma_i)
\end{align*}

The first part of the summation is the probability that the world state transitions to state $j$ before the driver accepts a trip request. The second part is the probability that the driver accepts a trip request before the state transitions, times the probability $q_{i\to j}(\sigma_i) =  \frac{1}{{F_i(\sigma_i)}}\int_{\sigma_i} q_{i\to j}(\tau) dF_i(\tau)$ that the world will be in state $j$ when the driver's trip ends. The result follows.
\qed

\subsection{Proof strategy for incentive compatible pricing and structural results}
\label{sec:dynamicproofstrategyandoverview}

We now give an overview of the proof strategy for both Theorems~\ref{prop:multnotoptimal} and~\ref{thm:ICpolicy}. The key step to both is Lemma~\ref{thm:derivativepropertycombined} below, which shows how to use properties of the derivative of a reward function with respect to an element of a driver policy, to establish the structural properties of optimal driver policies -- this Lemma is the primary theoretical result. Next, Section~\ref{sec:ICapplemmas} provides lemmas that help us establish the properties of this derivative as they depend on the pricing function. We put things together in Section~\ref{sec:proofsmaindynamicresults} to prove Theorems~\ref{prop:multnotoptimal} and~\ref{thm:ICpolicy}. Proofs of the lemmas in Section~\ref{sec:ICapplemmas} are in Section~\ref{sec:appdynamiclemmasproofstedious}.

Lemma~\ref{thm:derivativepropertycombined} shows how the structure of an optimizer $\sigma^* = \cup_k (\ell_k, u_k)$ of a set function $\hat R(\sigma)$ depends on the derivative of the set function with respect to the endpoints of the sets that make up the policy, $\paru {\hat R(\sigma)}$. The main idea is that as long as the derivative can be shown to be positive for some $u$ that is an endpoint of $\sigma_i$, that policy can be locally modified to accept more trips and increase the overall reward function. We work with a function $r(u, \sigma)$ that has the same sign as the derivative $\paru {\hat R(\sigma)}$. Given $r(u, \sigma)$, we analyze the sign of $\paru {\hat R(\sigma)}$ as it depends on the structure of $\sigma$, and in turn can characterize the structure of the optimal $\sigma^*$.

In particular, we will show how properties of $r(u, \sigma)$  -- whether for a fixed $\sigma$ it is always positive, strictly increasing, strictly decreasing, strictly quasi-convex, or strictly quasi-concave in $u$ -- lead to different optimal $\sigma^*$. The rest of the appendix section applies Lemma~\ref{thm:derivativepropertycombined} to our context, by showing how different pricing functions induce different properties of $r(u, \sigma)$ and thus different optimal policies for each state $\sigma_i^*$.

\begin{restatable}{lemma}{thmderivativepropertycombined}
	\label{thm:derivativepropertycombined}

	Consider a function ${\hat R}(\sigma)$ that maps open, measurable subsets $\sigma = \cup_k^\infty (\ell_k, u_k) \subseteq (0, \infty)$ to the non-negative reals, and probability measure $F$ such that $F$ is continuous, i.e. $f$ is bounded.

	Let ${\paru {\hat R({\sigma})}}$ denote the partial derivative of $\hat R$ with respect to an upper end-point $u$ of the intervals that make up $\sigma = \cup_k^\infty (\ell_k, u_k)$, i.e., it is the infinitesimal gain in the value of $\hat R(\sigma)$ by adding $u$ to the set $\sigma$.
 Suppose,
	\begin{enumerate}
		\item $\hat R(\sigma)$ is continuous in $\sigma$, and ${\paru {\hat R({\sigma})}}$ exists, for all $\sigma$ and $u$.

		\item ${\paru {\hat R({\sigma})}}$ is continuous in $u$, for each fixed $\sigma$.
		\item ${\paru {\hat R({\sigma})}}$ is continuous in $\sigma$, for each fixed $u$.
	\end{enumerate}

Suppose that there exists a function $r(u, \sigma)$ that has the same sign as ${\paru {\hat R({\sigma})}}$, for all $\sigma, u$ where $f(u) > 0$. Consider any open measurable subset $\sigma' \subseteq (0, \infty)$, where $F(\sigma') > 0$, and $\hat R(\sigma') > \hat R(\emptyset)$.

Then the following statement holds for each of the below specific cases: ``Suppose $\exists \epsilon >0$ s.t. $r(u, \sigma)$ is \textbf{[Property]}  in $u$ (for a fixed $\sigma$), for all $\sigma$ such that $\hat R(\sigma) \geq \hat R(\sigma') - \epsilon$. Then, there exists a policy $\sigma^*$ of the form \textbf{[Form]} such that $R(\sigma^*) \geq R(\sigma')$, and the inequality is strict unless $\sigma'$ is also of the same form.''

	\begin{table}[H]
		\begin{tabular}{l|l}
			\textbf{Property} & \textbf{Form} \\ \hline
			$r(u, \sigma)>0$ (is \textbf{positive})   &      $\sigma^* = (0, \infty)$      \\

		$r(u, \sigma)$ is \textbf{strictly increasing}	&  $\sigma^* = (\ell^*, \infty)$, for $\ell^*\in\bbrp$ \\

		$r(u, \sigma)$ is \textbf{strictly decreasing}	&  $\sigma^* = (0, u^*)$, for $u^*\in\bbrp\cup\{\infty\}$             \\

		$r(u, \sigma)$ is \textbf{strictly quasi-convex}	&  $\sigma^* = (0,\ell^*) \cup (u^*, \infty)$, for $\ell^*, u^* \in\bbrp \cup \{\infty\}$             \\

		$r(u, \sigma)$ is \textbf{strictly quasi-concave}	&  $\sigma^* = (\ell^*,u^*)$, for $\ell^*, u^* \in\bbrp \cup \{\infty\}$
		\end{tabular}
	\end{table}

\end{restatable}

\proof{Proof.} The facts that policies not of the appropriate form cannot be optimal follow directly from the respective structures of the derivatives, as we will show below. However, since the domain of the set function $\hat R $ is any subset of $\bbR^+$, we need to also prove {\em existence} of a maximizer of the appropriate form. The proof is structured around proving existence of a maximizer in each case, but we will point out where the given facts imply necessity of having the appropriate form.

The general approach is as follows: Start at subset $\sigma' \subseteq (0, \infty) = \cup_k^\infty (\ell_k, u_k) = \cup_k^\infty \zeta_k$, where the intervals are disjoint and $\zeta_k = (\ell_k, u_k)$ denotes the $k$th interval. (recall that any open subset of $\bbR$ can be uniquely written as the countable union of such disjoint intervals).

Then, do the following:

\begin{enumerate}

	\item Create a sequence $\sigma'_\delta \to \sigma'$ (as $\delta \to 0$), where, for each $\delta$, the set $\sigma'_\delta$ is $\delta$-close to $\sigma'$: $F((\sigma'\setminus\sigma'_\delta )\cup (\sigma'_\delta\setminus \sigma')) < \delta$.
	\item Show that there exists a $\sigma^*$ of the appropriate form (according to the property that holds above), such that $\hat R(\sigma'_\delta) \leq R(\sigma^*), \forall \delta$.

\end{enumerate}

By continuity of the set function ${\hat R}$, this implies that $\hat R(\sigma') \leq R(\sigma^*)$.

The second step is the main proof step and the only one that depends substantially on the given property of the function $r(u, \sigma)$.

\paragraph{Step one: a sequence $\sigma'_\delta \to \sigma'$.}
Each $\sigma_\delta'$ will be of the form $\sigma_\delta' =(0, L) \cup \left(\cup_{k=1}^K (\ell_k, u_k)\right) \cup (B, \infty)$, for some $K,B,L$ that depend on $\delta$. We construct a $\sigma'_\delta$ such that $F(\sigma'\setminus\sigma'_\delta \cup \sigma'_\delta\setminus \sigma') < \delta$ as follows:

\begin{itemize}
	\item $F$ is a finite (probability) measure, and so there exists $K$ such that $F(\cup_{k=K+1}^\infty (\ell_k, u_k)) < \delta/2$. (Since $F(\sigma')\leq 1$, it follows by the Cauchy condition).
	\item Let $B\in\bbR$ s.t. $F((B, \infty)) < \delta/4$. Let $L\in\bbR$ s.t. $F((0, L)) < \delta/4$. Such $B,L$ exist by condition on $F$.
	\item Set $\sigma_\delta' = (0, L) \cup \left(\cup_{k=1}^K (\ell_k, u_k)\right) \cup (B, \infty)$.
	\item For convenience, we re-index the disjoint intervals $\{\zeta_k\}_{k=1}^{K+2}$ such that they are in increasing order, i.e. $u_k>\ell_k\geq u_{k-1}, \forall k>1$, starting at $(0, L)$, with the last interval $(B, \infty)$. If there exist any intervals such that $\ell_k = u_{k-1}$, replace them with the combined interval $(\ell_{k-1}, u_k)$. If $\{B, \infty\}$ overlaps with the last interval, combine them.
\end{itemize}

Note that, by the suppositions, in each case $\exists \delta_0$ small enough such that $r(u, \sigma)$ maintains the appropriate property for all $\sigma$ such that $\hat R({\sigma}) \geq \hat R(\sigma'_\delta)$, $\forall \delta < \delta_0$.%

\paragraph{Step 2: showing that ${\hat R}({\sigma}^*) \geq \hat R(\sigma')$, where ${\sigma}^*$ is of the appropriate form.}

Now, starting at $\sigma = \sigma'_\delta =(0, L) \cup \left(\cup_{k=1}^K (\ell_k, u_k)\right) \cup (B, \infty)$, we describe a sequence of modifications to $\sigma$, such that each modification does not reduce the reward $\hat R(\sigma)$. The limit of this sequence of modifications is a policy $\sigma^*$ of the appropriate form, regardless of the starting $\sigma'_\delta$.

We now carry out this step separately for each case. The general argument is that the properties force $r(u, \sigma)$ to be positive at certain points, which allows expanding the policy until a policy of the appropriate form is reached.

Let $r_L(\ell, \sigma) = -r(u, \sigma)$, i.e., it is a function that has the same sign as the derivative of a lower endpoint of $\sigma_i$ (the same sign as the infinitesimal loss as removing the point $\ell$ from the set $\sigma$).

~\hrule

\parbold{Setting where $r(u, \sigma) > 0$ (is positive)} By supposition that ${\paru {\hat R({\sigma})}}$ is {positive} in $u$, we can increase $u_1$ (merging with other intervals) while increasing $\hat R(\sigma)$. Thus, we can keep increasing $u_1$, and $u_1 \to B$, and so $R((0, \infty)) \geq R(\sigma'_\delta)$. For any set $\sigma' \neq (0, \infty)$, we can increase the reward by expanding an interval, and so it cannot be optimal. Thus, $(0, \infty)$ is the unique optimal set.

~\hrule

\parbold{Setting where $r(u, \sigma)$ is strictly increasing}

Now, starting at $\sigma = \sigma'_\delta$, the limit of the sequence of modifications is a policy $\sigma^* = (\ell^*, \infty)$.

By the supposition that $r(u, \sigma)$ strictly increasing in $u$, we have:
\begin{align*}
	r_L(\ell, \sigma)	&{} \text { strictly decreasing}\\
	r_L(\ell_1, \sigma)	\leq 0 &\implies r(u_1, \sigma)	> 0 & \ell_1 < u_1\\
	\equiv \partildeR{\ell_1}	\leq 0 &\implies \partildeR{u_1}	> 0 \\
	&\\
	r_L(\ell_1, \sigma)	> 0 &\impliedby r(u_1, \sigma)	\leq 0 \\
	\equiv \partildeR{\ell_1}	> 0 &\impliedby \partildeR{u_1}	\leq 0
\end{align*}
~\\
\textbf{Case 1: $\exists \zeta_1,\zeta_2 \subset \sigma$ such that $\ell_2>u_1, |\zeta_1|,|\zeta_2|$, i.e. there is more than one interval that makes up $\sigma$}, and $\zeta_1,\zeta_2$ are the first and second such intervals, respectively, with positive mass. \\

Then we make the following sequence of changes (forming new $\sigma$), depending on $\partildeR{\ell_1}, \partildeR{u_1}$:
\begin{description}
	\item [Subcase 1A, $\partildeR{u_1} > 0$:] Increase $u_1$ until $u_1 = \ell_2$ (exit Case 1), or $\partildeR{u_1} \leq 0$ (go to Case 1B).
	\begin{description}
		\item [Sub-subcase 1AA, $\partildeR{\ell_1} < 0$, $\ell_1 > 0$: ] Simultaneously, decrease $\ell_1$.
		\item [Sub-subcase 1AB, $\partildeR{\ell_1} \geq 0$ or $\ell_1 = 0$: ] Hold $\ell_1$ fixed.
	\end{description}
	\item [Subcase 1B, $\partildeR{u_1} \leq 0 \implies \partildeR{\ell_1}	> 0$:] Increase $\ell_1$ until  $\ell_1 = u_1$ (exit Case 1), or $\partildeR{\ell_1}	\leq 0$ (which implies $\partildeR{u_1}	> 0$, i.e. go to Case 1A).
\end{description}
Each of these changes increases ${\hat R({\sigma})}$, due to the direction of the changes in $u_1, \ell_1$ and the signs of the appropriate derivatives. Note that these subcases are mutually-exclusive, and one is true as long as there is more than one disjoint interval, $\exists \zeta_1, \zeta_2 \subset \sigma, \ell_2>u_1$. Further, note that $u_1$ is increasing in Subcase 1A and constant in Subcase 1B. Thus, with $\ell_2$ fixed and bounded, eventually:
\begin{itemize}
	\item $\ell_1 \to u_1$, in Subcase 1B (i.e. the first interval collapses to mass 0). OR
	\item $u_1 \to \ell_2$, in Subcase 1A (i.e. the first interval merges with the second).
\end{itemize}
Thus, this sequence of changes increases the reward, and results in there being one fewer interval than before (after combining the bottom 2 intervals by adding the point $u_1 = \ell_2$ of 0 measure). Case 1 can be iteratively applied until there is just a single interval $\sigma = (\ell', \infty)$.

That the changes can increase the reward for any other $\sigma$ implies that such $\sigma$ cannot be optimal.

~\\\noindent \textbf{Case 2: $\sigma = (\ell', \infty)$, i.e. there is a single interval that makes up $\sigma$}

By supposition, $\hat R(\sigma') > \hat R(\emptyset)$ and so $\hat R((\ell', \infty)) > \hat R(\emptyset)$. Further $\hat R((\ell, \infty))$ is a continuous function in $\ell$. Thus, there exists $L$ such that $\forall \ell > L$, $  \hat R((\ell', \infty)) > \hat R((\ell, \infty))$.

Thus, there exists $\ell^*\in [0,L]$ such that $\hat R((\ell^*, \infty)) \geq \hat R((\ell, \infty)), \forall \ell\in\bbrp \cup \{\infty\}$ (continuous functions in a compact domain have a maximum).
\vspace{.8em}
~\hrule

\parbold{Setting where $r(u, \sigma)$ is strictly decreasing}

The proof is extremely similar to the strictly increasing case, with two differences.

First we now need to modify the starting $\sigma'$ so it \textit{does not} contain an interval $(B, \infty)$: set $\sigma_\delta' = (0, L) \cup \left(\cup_{k=1}^K (\ell_k, u_k)\right) \setminus (B, \infty)$.

Second, each case from above is duplicated but move the policy in the opposite different to increase the reward. We omit the details of this case for brevity.

~\hrule
\parbold{Setting where $r(u, \sigma)$ is strictly quasi-convex}

We show that there exists a $\sigma^* = (0,\ell^*) \cup (u^*,\infty)$, for some $u^*,l^* \in \bbrp$, such that $\hat R(\sigma'_\delta) \leq R(\sigma^*), \forall \delta$.

The key is noting that quasi-convexity of $r(u, \sigma)$ in $u$ implies that any $\sigma$ with three intervals $\zeta_1, \zeta_2, \zeta_3$ can be improved by eliminating the middle interval (or joining it with one of the others).

\textbf{Case 1: $\exists$ disjoint $\zeta_1 = (0, u_1), \zeta_2 = (\ell_2, u_2), \zeta_3 = (\ell_3, u_3)$, s.t. $|\zeta_1|,|\zeta_2|,|\zeta_3|>0$}, i.e. $\sigma$ is composed of at least three intervals, and $\zeta_1,\zeta_2,\zeta_3$ are the first three such intervals with positive mass. ($u_3$ may be $\infty$).

By supposition, $r(u_k, \sigma)$, is strictly quasi-convex in $u$, and so $r_L(\ell_k, \sigma)$, is strictly quasi-concave in $\ell$.%

Then, we have:
\begin{align*}
	\partildeR{u_1}	\leq 0 \text{ and } \partildeR{\ell_3}
	\geq 0  &\implies \partildeR{\ell_2}	> 0 \text{ and } \partildeR{u_2}	< 0 \\
	\partildeR{\ell_2}	\leq 0 \text{ or } \partildeR{u_2}
	\geq 0
	&\implies \partildeR{u_1}	> 0 \text{ or } \partildeR{\ell_3}
	< 0
\end{align*}
~\\
Then we make the following sequence of changes (forming new $\sigma$):
\begin{description}
	\item [Subcase 1A, $\partildeR{u_1}	\leq 0 \text{ and } \partildeR{\ell_3}	\geq 0 \implies \partildeR{\ell_2}	> 0 \text{ and } \partildeR{u_2}	< 0$:] Increase $\ell_2$ and decrease $u_2$ simultaneously until $\ell_2 = u_2$ (exit Case 1), $\partildeR{u_2} \geq 0, \text{ or } \partildeR{\ell_2} \leq 0 $ (go to \textbf{1B} or \textbf{1C}).
	\item [Subcase 1B, $\partildeR{u_1} > 0$:] Increase $u_1$ until  $u_1 = \ell_2$ (exit Case 1), or $\partildeR{u_1} \leq 0$ (go to \textbf{1A} or \textbf{1C})
	\item [Subcase 1C, $\partildeR{\ell_3} < 0$:] Decrease $\ell_3$ until  $u_2 = \ell_3$ (exit Case 1), or $\partildeR{\ell_3} \geq 0$ (go to \textbf{1B} or \textbf{1A}).

\end{description}
Each of these changes strictly increase $\hat{R}({\sigma})$. \textbf{1B} and \textbf{1C} may both be true, in which case we arbitrarily decide between them. At least one of the three subcases is true as long as the Case 1 condition holds.
Thus, eventually:
\begin{itemize}
	\item $\ell_2 = u_2$, in Subcase 1A (i.e. the middle interval collapses to mass 0). OR
	\item $u_1 = \ell_2$, in Subcase 1B (i.e. the first interval merges with the second). OR
	\item $u_2 = \ell_3$, in Subcase 1C (i.e. the third interval merges with the second).
\end{itemize}
Thus, this sequence of changes cannot decrease the reward, and results in there being one fewer interval than before. Case 1 can be iteratively applied until there are just two intervals $\sigma = (0, t_1) \cup (t_2, \infty)$.

That the changes can increase the reward for any other $\sigma$ implies that such $\sigma$ cannot be optimal.

~\\\noindent \textbf{Case 2: $\sigma = (0, t_1) \cup (t_2, \infty)$.} We need to show that there exists a maxima $(t_1^*, t_2^*)$. Specifically, we need to eliminate the possible cases where $t_1$ or $t_2$ increase to infinity, but the asymptotic values at $\infty$ produce lower rewards, which would have implied that the maximum is not achieved.

By supposition, $\hat R(\sigma') > \hat R(\emptyset)$ and so $\hat R((0, t_1) \cup (t_2, \infty)) > \hat R(\emptyset)$ for $ t_1>0$ or $t_2 < \infty$.

Further $\hat R((0, t_1) \cup (t_2, \infty))$ is a continuous function in $t_1,t_2$. Thus $\hat R((0, t_1) \cup (t_2, \infty)) \to \hat R(\emptyset)$ as $t_1\to 0, t_2\to\infty$ together.

Further, $\hat R((0, t_1) \cup (t_2, \infty)) \to \hat R((0,\infty))$ as $t_1 \to \infty$, regardless of how $t_2$ behaves. Similarly, fixing $t_1$, $\hat R((0, t_1) \cup (t_2, \infty)) \to \hat R((0,t_1))$ as $t_2 \to \infty$.
\begin{itemize}
	\item If $\hat R((0, t_1) \cup (t_2^*(t_1), \infty))$ is increasing for $t_1 > T_1$, for however $t_2^*(t_1)$ behaves as a function of $t_1$ then $\hat R((0,\infty)) \geq \hat R((0, t_1) \cup (t_2, \infty)), \forall t_1>T_1,t_2$.
	\item For any fixed $t_1$, if $\hat R((0, t_1) \cup (t_2, \infty))$ is increasing for $t_2 >T_2$, then $\hat R((0,t_1)) \geq \hat R((0, t_1) \cup (t_2, \infty)), \forall t_2$.
\end{itemize}
 Thus, the maximum is achieved: either
\begin{enumerate}
	\item $\exists t_1^* \in (0, \infty): \hat R((0,t_1^*)) \geq \hat R((0, t_1) \cup (t_2, \infty)), \forall t_1,t_2$
	\item $\exists t_1^*, t_2^* \in [0, \infty):  \hat R((0, t_1^*) \cup (t_2^*, \infty)) \geq \hat R((0, t_1) \cup (t_2, \infty)), \forall t_1,t_2$
\end{enumerate}
~
~\hrule
\parbold{Setting where $r(u, \sigma)$ is strictly quasi-concave}

 The proof is extremely similar to the strictly quasi-convex case. However, we now need to modify the starting $\sigma'$ so it \textit{does not} contain an intervals $(0, L)$ or $(B, \infty)$, and the subsequent modifications also differ directionally.

 Let $\sigma_\delta'  = \left(\cup_{k=1}^K (\ell_k, u_k)\right) \setminus (0, L) \setminus (B, \infty)$. %
 Now, the key step is noting that strict quasi-concavity of $r(u, \sigma)$ implies that any $\sigma$ with two intervals $\zeta_1, \zeta_2$ can be improved by eliminating one (or joining the two).

 \textbf{Case 1: $\exists$ disjoint $\zeta_1 = (\ell_1, u_1), \zeta_2 = (\ell_2, u_2)$, s.t. $|\zeta_1|,|\zeta_2|>0$}, i.e. $\sigma$ is composed of at least two intervals with positive mass, and $\zeta_1,\zeta_2$ are the first two such intervals.

 By supposition, $r(u_k, \sigma)$, is strictly quasi-concave in $u$. Then, $r_L(\ell_k, \sigma)$, is strictly quasi-convex in $u$.%

 Then, we have:
 \begin{align*}
 	\partildeR{\ell_1}	\leq 0 \text{ and } \partildeR{\ell_2}
 	\leq 0  &\implies \partildeR{u_1}	> 0 \\
 	\partildeR{u_1}	\leq 0
 	&\implies \partildeR{\ell_1}	> 0 \text{ or } \partildeR{\ell_2}
 	> 0
 \end{align*}
 Then we make the following sequence of changes (forming new $\sigma$):
 \begin{description}
 	\item [Subcase 1A, $\partildeR{\ell_1}	\leq 0 \text{ and } \partildeR{\ell_2}
 	\leq 0  \implies \partildeR{u_1}	> 0$:] Increase $u_1$ until $u_1 = \ell_2$ (exit Case 1) or $\partildeR{u_1} \leq 0 $ (go to \textbf{1B} or \textbf{1C}).
 	\item [Subcase 1B, $\partildeR{\ell_1} > 0$:] Increase $\ell_1$ until  $u_1 = \ell_1$ (exit Case 1), or $\partildeR{\ell_1} \leq 0$ (go to \textbf{1A} or \textbf{1C})
 	\item [Subcase 1C, $\partildeR{\ell_2} > 0$:] Increase $\ell_2$ until  $u_2 = \ell_2$ (exit Case 1), or $\partildeR{\ell_2} \leq 0$ (go to \textbf{1B} or \textbf{1A}).

 \end{description}
 Each of these changes strictly increase ${R({\sigma})}$. \textbf{1B} and \textbf{1C} may both be true, in which case arbitrarily decide between them. At least one of the three subcases is true as long as the Case 1 condition holds.
 Thus, eventually:
 \begin{itemize}
 	\item $\ell_2 = u_1$, in Subcase 1A (i.e. the intervals combine). OR
 	\item $u_1 = \ell_1$, in Subcase 1B (i.e. the first interval collapses to mass 0). OR
 	\item $u_2 = \ell_2$, in Subcase 1C (i.e. the second interval collapses to mass 0).
 \end{itemize}
 Thus, this sequence of changes cannot decrease the reward, and result in there being one fewer interval than before. Case 1 can be iteratively applied until there is just one interval $\sigma_i = (t_1,t_2)$.

That the changes can increase the reward for any other $\sigma$ implies that such $\sigma$ cannot be optimal.

 ~\\\noindent \textbf{Case 2: $\sigma_i = (t_1,t_2)$.} As in the same case in the quasi-convex setting. We need to values eliminate the possible cases where $t_1$ or $t_2$ increasing to infinity, but the asymptotic values at $\infty$ produce lower rewards, which would imply that the maximum is not achieved.

 Note that $\hat R((t_1, t_2))$ is a continuous function in $t_1,t_2$. Thus $\hat R((t_1, t_2)) \to \hat R(\emptyset)$ as $t_1\to t_2 $.

 Further, $\hat R((t_1, t_2)) \to \hat R(\emptyset)$ as $t_1 \to \infty$, regardless of how $t_2\geq t_1$ behaves. Similarly, fixing $t_1$, $\hat R((t_1, t_2)) \to \hat R((t_1, \infty))$ as $t_2 \to \infty$.
 \begin{itemize}
 	\item If $\hat R((t_1, t_2^*(t_1)))$ is increasing for $t_1 >T_1$, for however $t_2^*(t_1)$ behaves as a function of $t_1$ then $\hat R(\emptyset) \geq \hat R((t_1, t_2)), \forall t_1>T_1,t_2$.
 	\item For any fixed $t_1$, if $\hat R((t_1, t_2))$ is increasing for $t_2 > T_2$, then $\hat R((t_1, \infty)) \geq \hat R((t_1, t_2)), \forall t_2>T_2$.
 \end{itemize}
 Thus, either
 \begin{enumerate}
 	\item $\hat R(\emptyset) \geq \hat R((t_1, t_2)), \forall t_1,t_2$
 	\item $\exists t_1^* \in [0, \infty): \hat R((t_1^*, \infty)) \geq \hat R((t_1, t_2)), \forall t_1,t_2$
 	\item $\exists t_1^*, t_2^* \in [0, \infty):  \hat R(t_1^*, t_2^*) \geq \hat R((t_1, t_2)), \forall t_1,t_2$
 \end{enumerate}

\qed

\subsection{Auxiliary lemmas}
\label{sec:ICapplemmas}

Here we present lemmas necessary to prove the main theorems regarding incentive compatibility and optimal driver policies. Proofs are deferred to Section~\ref{sec:appdynamiclemmasproofstedious}, as they are tedious and algebraic.

These lemmas primarily involve properties of derivatives of the reward function $R(w, \sigma)$ and its components in the dynamic model, as a function of the pricing.

\subsubsection{Notation and assumptions}
Recall in the dynamic model that we constrain $\sigma_i$ to be measurable, \textit{open}, subsets of the $\bbrp$. Then, $\sigma_i$ can be written as a countable union of disjoint subsets of $\bbrp$, i.e. $\sigma_i = \cup_{k=0}^\infty (\ell_k, u_k)$. We further assume that $u_k \neq \ell_m$, for any $k,m$; we can do so without loss of generality by making a measure $0$ change to $\sigma_i$, by adding $u_k=\ell_m$ to $\sigma_i$.

Suppose $u$ is an upper-endpoint of $\sigma_i$, ie. $\exists k$ such that $u = u_k$. Then, we use $\paru H(\sigma_i)$ to denote the derivative of the set function $H$ with respect to $u$ at $\sigma_i$. Similarly, $\parl H(\sigma_i)$ is the derivative of $H$ at $\sigma_i$ with respect to a lower-endpoint of $\sigma_i$.

We derive $\paru R(w, \{\sigma_1,\sigma_2\})$, $\parl R(w, \{\sigma_1,\sigma_2\})$. We will make it clear in each instance whether $u$ or $\ell$ is an endpoint of $\sigma_1$ or $\sigma_2$. For all the derivatives in this subsection $\paru$ refers to the derivative with respect to an upper endpoint in $\sigma_i$, and $\parl$ refers to a derivative at a lower endpoint of $\sigma_i$. Note that $$\paru R(w, \{\sigma_1,\sigma_2\}) = -\parl R(w, \{\sigma_1,\sigma_2\}).$$

Furthermore:
\begin{itemize}
	\item We use $\sigma$ in the function argument when the function depends on policies in both states, and $\sigma_i$ when it only depends on the policy in state $i$.
	\item We use $\propto$ to denote that ``two functions of $u$ \textit{have the same sign} except where $f(u) = 0$'', rather than \textit{proportional to}.

	\item All policy equalities are up to measure $0$.
\end{itemize}

Let \[\Delta(\sigma_i, \sigma_{j}) = R_i(w_i, \sigma_i) - {R_j(w_j, \sigma_j)}.\]
be the earnings difference between the two states.
Finally, when $\sigma_i, \sigma_j$ are clear from context, let
\begin{align*}
Q_i &\triangleq Q_i(\sigma_i) = \lambda_{i\to j} + \lambda_i\int_{\sigma_i} q_{i\to j}(\tau) dF_i(\tau)\\
T_i & \triangleq \lambda_i F_i(\sigma_i)T_i(\sigma_i) = 1 + \lambda_i \int_{\tau\in\sigma_i} \tau dF_i(\tau)\\ W_i&\triangleq \lambda_i F_i(\sigma_i)W_i(\sigma_i) = \lambda_i\int_{\tau\in\sigma_i} w_i(\tau) dF_i(\tau)\\
\Delta_{ji} &\triangleq \Delta(\sigma_j, \sigma_{i}) = {R_j(w_j, \sigma_j)} - R_i(w_i, \sigma_i)
\end{align*}

We assume throughout:
\begin{itemize}
	\item Distribution of jobs $F,F_i$ is a continuous probability measure, i.e., $f,f_i$ bounded.
	\item  There exists a policy in state 2 that dominates state 1: $\exists \sigma_2$ such that $\Delta(\sigma_2, \sigma_1) > 0, \forall \sigma_1 \subseteq (0, \infty)$.
	\item $\sigma, \sigma_i$ constrained to be measurable with respect to $F, F_i$, and $\sigma_i$ are open.
\end{itemize}

\subsubsection{Derivative derivation and comments}

\begin{restatable}{lemma}{lemderivatives}
	\label{lem:derivatives}

	Let $R(w, \sigma)$ be as defined in Lemma~\ref{lem:statingreward}. 	Then,
	$\paru R(w, \sigma) \propto r(u, i, w, \sigma)$, where \[r(u, i, w, \sigma) \triangleq \frac{q_{i\to j}(u)}{u}\Delta_{ji} + \frac{w_i(u)}{u}\left(\frac{Q_i}{T_i} + \frac{Q_j}{T_j}\right) - \left(\frac{Q_i}{T_i}R_j + \frac{Q_j}{T_j}R_i\right)\]

\end{restatable}

In other words, $r(u, i, w, \sigma)$ has the \textit{same sign} as the derivative of the overall reward with respect to $u$ (an upper endpoint of $\sigma_i$) at $w$, $\sigma$, but it is not necessarily monotonic with it.

\begin{remark}
	Given assumptions on $F_i$, $w_i$:
	\begin{itemize}
		\item $R_i(\sigma), R(\sigma), \mu_i$ are continuous in $\sigma$
		\item $\paru R(w, \sigma), r(u, i, w, \sigma)$ are both continuous in $u$ (for fixed $\sigma$), and continuous in $\sigma$.
		\item $\frac{q_{i\to j}(u)}{u}$ is strictly decreasing in $u$.
		\item If $\Delta_{ji} < 0$ ($i=2$ the surge state) and $\frac{w_i(u)}{u}$ is non-decreasing in $u$, then $r(u, i, w, \sigma)$ is strictly increasing in $u$ for a fixed $\sigma$. Thus, $\paru R(w, \sigma)$ is negative up to a certain point $U\in (0, \infty) \cup \{\infty\}$ and then positive thereafter.
		\item If $\Delta_{ji} > 0$ ($i=1$ the non-surge state) and $\frac{w_i(u)}{u}$ is non-increasing in $u$, then $r(u, i, w, \sigma)$ is strictly decreasing in $u$ for a fixed $\sigma$. Thus, $\paru R(w, \sigma)$ is positive up to a certain point $U\in (0, \infty) \cup \{\infty\}$ and then negative thereafter.
	\end{itemize}
	\label{rem:basiccontinuityincreasingproperties}
\end{remark}

\subsubsection{Lemmas for driver policy in response to affine pricing}

\begin{restatable}{lemma}{lemaffinesurgequasiconvex}
	\label{lem:affinesurge_quasiconvex}
	Suppose $w_i(\tau) = m\tau + a$, where $m,a > 0$. Then, $r(u, i, w, \sigma)$ is strictly quasi-convex in $u$, for each fixed $\sigma$ where $\Delta_{ji}\leq0$.
\end{restatable}

\begin{restatable}{lemma}{lemaffinesurgequasiconcave}
	\label{lem:affinesurge_quasiconcave}
	Suppose $w_i(\tau) = m\tau + a$, where $m> 0$ and $a<0$. Then, $r(u, i, w, \sigma)$ is strictly quasi-concave in $u$, for each fixed $\sigma$ where $\Delta_{ji}\geq0$.
\end{restatable}

\subsubsection{Lemmas for IC policy}
\label{sec:lemmashelperICpolicy}
\begin{restatable}{remark}{remderivativesimplification}
	\label{rem:derivativesimplification}
	\begin{align*}
\text{Let } \ \ \ \  &w_i(u) = mu + z q_{i\to j}(u)\\
\text{Then } \ \ \ \  & W_i =m (T_i - 1) + z (Q_i - \lambda_{i\to j})\\
\ \ \ \  & \paru R(w, \sigma)\propto q_{i\to j}(u)\left[ (R_j - m) T_jT_i  + mT_j + zQ_jT_i + z T_j\lambda_{i\to j} \right]\\
&\ \ \ \ \ \ \ \ \ \ \ \ \ \ \ \ \ \ \ \ \ \ \ \ \ \ \ \ \ \ \ \ \ \ \ \ \ \ \ \ + u \left[Q_iT_j (m-R_j) + Q_j(m -z Q_i + z\lambda_{i\to j})\right]
\end{align*}
\end{restatable}

\begin{restatable}{remark}{remlimitqij}
	$\lim_{u \to 0} \frac{q_{i\to j}(u)}{u} = \lambda_{i\to j}$. \label{rem:limitqij}
\end{restatable}

\begin{restatable}{remark}{remvaluesmaximizedwhenacceptall}
	$	\lambda_{i\to j}T_i - Q_i \geq 0$ and maximized when $\sigma_i = (0,\infty)$. Similarly, $Q_i \geq 0$ and maximized when $\sigma_i = (0,\infty)$. \label{rem:valuesmaximizedwhenacceptall}
\end{restatable}

In the next lemma, we consider $u$ an upper endpoint of $\sigma_2$, and so $\paru R(w = \{w_1, w_2\}, \sigma = \{\sigma_1, \sigma_2\})$ is a derivative with respect to an upper endpoint of $\sigma_2$.

\begin{restatable}{lemma}{lemrangeformstatetwo}
	Fix arbitrary $\sigma_1$, and thus $Q_1, T_1,R_1$. Let $\bar Q_2, \bar T_2$ be the respective values of $Q_2, T_2$ at $\sigma_2 = (0,\infty)$. Let $w_2(\tau) = m\tau + zq_{2 \to 1}(\tau)$, where $m > R_1$.

If
\begin{align*}
\frac{T_1(\lambda_{2\to 1}\bar T_2 - \bar Q_2) - \left(Q_1 + T_1\lambda_{2\to 1}\right) }{\left(Q_1(\lambda_{2\to 1}\bar T_2 - \bar Q_2) + \lambda_{2\to 1}(Q_1 + T_1\lambda_{2\to 1}) \right)}<  &\frac{z}{m-R_1} < \frac{\bar Q_2 T_1 + Q_1 }{Q_1 (\bar Q_2 - \lambda_{2\to 1})}
\end{align*}

Then $\paru R(w, \sigma) > 0$, for all $u, \sigma_2$. Furthermore, the constraint set is feasible regardless of the primitives.
\label{lem:rangeform2z2}
\end{restatable}

We can now do the same thing for the first state, assuming that $w_1(\tau)$ is of the form $w_1(\tau) = m\tau + zq_{1\to 2}(\tau)$, where now $z\leq0$ and $m = R_2$. In the next lemma, we consider $u$ an upper endpoint of $\sigma_1$, and so $\paru R(w = \{w_1, w_2\}, \sigma = \{\sigma_1, \sigma_2\})$ is a derivative with respect to an upper endpoint of $\sigma_1$. Then,

\begin{restatable}{lemma}{lemrangeformstateone}
	Fix arbitrary $\sigma_2$, and thus $Q_2, T_2,R_2$. Let $\bar Q_1, \bar T_1$ be the respective values of $Q_1, T_1$ at $\sigma_1 = (0,\infty)$. Let $w_1(\tau) = m\tau + zq_{1 \to 2}(\tau)$, where $m = R_2$.

	If
	\begin{align*}
	-\frac{(T_2\lambda_{1\to 2} + Q_2) }{Q_2 (\lambda_{1\to 2}\bar T_1 - \bar Q_1) + \lambda_{1\to 2}(T_2\lambda_{1\to 2} + Q_2)}<\frac{z}{R_2}< \frac{1 }{(\bar Q_1 - \lambda_{1\to 2})}
	\end{align*}

	Then $\paru R(w, \sigma) > 0$, for all $u, \sigma_1$. Furthermore, the constraint set is feasible regardless of the primitives.

	\label{lem:rangeform1z1}
\end{restatable}

\subsection{Proofs of main results, Theorems~\ref{prop:multnotoptimal} and~\ref{thm:ICpolicy}}
\label{sec:proofsmaindynamicresults}
We are now ready to combine the results above to prove our main results. The following theorem subsumes Theorem~\ref{prop:multnotoptimal}, (slightly expanding it to make it useful to prove Theorem~\ref{thm:ICpolicy}).

\begin{restatable}{theorem}{thmsurgemodeloptimal}
	\label{thm:surgemodeloptimal_table}

	Consider pricing function $w = \{w_1, w_2\}$, where $i=2$ is the surge state as defined. Then, there exists an optimal policy $\sigma = \{\sigma_1, \sigma_2\}$ that maximizes $R(w, \sigma)$, with the following properties.

	\begin{itemize}
		\item Non-surge state driver optimal policy $\sigma_1$:
		\begin{itemize}
			\item If $w_1(\tau) = m_1\tau + a_1$, for $a_1 \geq 0$, then $\sigma_1 = (0, t_1)$, for some $t_1 \in [0, \infty) \cup \{\infty\}$.
			\item If $w_1(\tau) = m_1\tau - a_1$, for $a_1 > 0$, then $\sigma_1 = (t_2, t_3)$, for some $t_2,t_3 \in [0, \infty) \cup \{\infty\}$.
			\item If $w_1$ such that $\paru R(w, \sigma' = \{\sigma'_1, \sigma'_2\}) > 0$ for all $\sigma'$, where $u$ is an upper endpoint of an interval that makes up $\sigma'_1$, then $\sigma_1 = (0, \infty)$.
		\end{itemize}

		\item Surge state driver optimal policy $\sigma_2$:
		\begin{itemize}
			\item If $w_2(\tau) = m_2\tau - a_2$, for $a_2 \geq 0$, then $\sigma_1 = (t_4, \infty)$, for some $t_4 \in [0, \infty)$.
			\item If $w_2(\tau) = m_2\tau + a_2$, for $a_2 > 0$, then $\sigma_1 = (0, t_5) \cup (t_6, \infty)$, for some $t_5,t_6 \in [0, \infty) \cup \{\infty\}$.
			\item If $w_2$ such that $\paru R(w, \sigma' = \{\sigma'_1, \sigma'_2\}) > 0$ for all $\sigma'$, where $u$ is an upper endpoint of an interval that makes up $\sigma'_2$, then $\sigma_2 = (0, \infty)$.
		\end{itemize}
	\end{itemize}

Furthermore, only policies of the given forms can be optimal.

\end{restatable}

\proof{Proof.} The proof strategy is as follows:
\begin{itemize}
	\item Start with some arbitrary policy $\sigma = \{\sigma_1, \sigma_2\}$.
	\item With assumption on the surge state providing higher potential earnings, replace $\sigma_2$ with a policy that provides higher earnings in state $2$ than $\sigma_1$ does in state 1, without decreasing total reward.
	\item Using Lemma~\ref{thm:derivativepropertycombined}, replace $\sigma_1$ with policy of the appropriate form, without decreasing total reward.
	\item Using Lemma~\ref{thm:derivativepropertycombined}, replace $\sigma_2$ with policy of the appropriate form, without decreasing total reward.
\end{itemize}

Let $r(u,i,w,\sigma)$ be as defined in Lemma~\ref{lem:derivatives}, a function that has the same sign as $\paru R(w,\sigma)$, where $u$ is an upper endpoint of an interval that is part of $\sigma_i$. Recall that, above, we show

\begin{itemize}
	\item (Remark~\ref{rem:basiccontinuityincreasingproperties}). $\Delta(\sigma_i, \sigma_{j})>0$ and $\frac{w_i(\tau)}{\tau}$ non-decreasing implies $r(u,i,w,\sigma)$ strictly increasing in $u \in \sigma_i$.
	\item (Remark~\ref{rem:basiccontinuityincreasingproperties}). $\Delta(\sigma_i, \sigma_{j})<0$ and $\frac{w_i(\tau)}{\tau}$ non-increasing implies $r(u,i,w,\sigma)$ strictly decreasing in $u \in \sigma_i$.
	\item (Lemma~\ref{lem:affinesurge_quasiconvex}). $w(\tau) = m\tau + a$ for $m,a>0$ and $\Delta(\sigma_i, \sigma_{-i})\geq0$ implies $r(u,i,w,\sigma)$ is strictly quasi-convex in $u \in \sigma_i$
	\item (Lemma~\ref{lem:affinesurge_quasiconcave}). $w(\tau) = m\tau - a$ for $m,a>0$ and $\Delta(\sigma_i, \sigma_{-i})\leq0$ implies $r(u,i,w,\sigma)$ is strictly quasi-concave in $u \in \sigma_i$
\end{itemize}

We need to show that there exists a $\sigma$ of the appropriate form such that $R(w, \sigma) \geq R(w, \sigma')$, for all $\sigma'$.

Start with arbitrary $\sigma' = \{\sigma'_1, \sigma'_2\}$ where $\sigma'_1,\sigma'_2\subseteq \bbrp$ are open, measurable sets, but not of the correct form in the theorem statement. Invoking Lemma~\ref{thm:derivativepropertycombined} as appropriate, we construct a sequence of changes to $\sigma'$ such that the overall reward does not decrease with each change, and the sequence ends with a policy consistent with the theorem statement.

\begin{description}
	\item [Step A] First, if $R_2(\sigma_2') < R_1(\sigma_1')$, then we replace $\sigma_2'$ with a policy $\sigma_2^A$ such that $R_2(\sigma_2^A) > R_1(\sigma_1), \forall \sigma_1$. %

	Let $\sigma_2^A$ be such that $\Delta(\sigma_2^A,\sigma_1'') > 0$, for all $\sigma_1''$ open and measurable. Such $\sigma_2^A$ exists by the definition of the surge state (we in fact define the surge state so that such $\sigma_2^A$ exists).

	Then, let $\sigma^A \triangleq \{\sigma_1^A = \sigma_1', \sigma_2^A\}$.

	Note that $R(w, \sigma^A) \geq R(w, \sigma')$: time spent earning reward at the rate of $R_2(w_2, \sigma_2')$ is replaced by time spent earning at rate $R_1(w_1, \sigma_1')$ or earning at rate $R_2(w_2, \sigma_2^A)$; time spent earning at $R_1(w_1, \sigma_1')$ may be replaced by time earning at rate $R_2(w_2, \sigma_2^A)$.

	\item [Step B] Now, we replace $\sigma_1^A$ with a policy that is of the appropriate form.

	Let $\hat R( \sigma_1) \triangleq R(w, \{\sigma_1, \sigma_2^A\})$.
	By Lemma~\ref{thm:derivativepropertycombined}, there exists $\sigma_1^B$ such that $R(w, \{\sigma_1^B, \sigma_2^A\}) \geq R(w, \{\sigma_1^A, \sigma_2^A\})$, and $\sigma_1^B$ is of the required form. The inequality is strict if $\sigma_1^A$ is not of the required form.

	Note that all the assumptions of Lemma~\ref{thm:derivativepropertycombined} are met for each appropriate case: $\sigma_2^B$ such that $\Delta(\sigma_2^B, \sigma_1') > 0$, $\forall \sigma_1'$, and so $r(u,i,w,\sigma)$ remains decreasing or strictly quasi-concave as necessary.

		Let  $\sigma^B \triangleq \{\sigma_1^B, \sigma_2^B= \sigma_2^A\}$.

	\item [Step C] %
	Now, we replace $\sigma_2^B$ with a policy that is of the appropriate form.

	Let $\hat R( \sigma_2) \triangleq R(w, \{\sigma_1^B, \sigma_2\})$.

	By Lemma~\ref{thm:derivativepropertycombined}, there exists $\sigma_2^C$ such that $R(w, \{\sigma_1^B, \sigma_2^C\}) \geq R(w, \{\sigma_1^B, \sigma_2^B\})$, and $\sigma_2^C$ is of the required form according to the table. The inequality is strict if $\sigma_2^B$ is not of the required form.

	As before, all the assumptions of Lemma~\ref{thm:derivativepropertycombined} are met for each appropriate case.  $\sigma_2^B$ such that $\Delta(\sigma_2^B, \sigma_1') > 0$, $\forall \sigma_1'$, and so $r(u,i,w,\sigma)$ remains strictly increasing / strictly quasi-convex in $u$ for a fixed $\sigma$.

	Let  $\sigma^C \triangleq \{\sigma_1^C = \sigma_1^B, \sigma_2^C\}$.

\end{description}

Thus, we have constructed $\sigma^* = \{\sigma_1^* = \sigma_1^C, \sigma_2^* = \sigma_2^C\}$ such that $\sigma_1^*, \sigma_2^*$ correspond to theorem statement for the appropriate cases, respectively, and $R(w, \sigma^*) \geq R(w, \sigma)$, for all $\sigma = \{\sigma_1, \sigma_2\}$ where $\sigma_1,\sigma_2\subseteq \bbrp$ are open, measurable sets, with the inequality strict if $\sigma$ is not of the required form.

\qed

\thmicpolicy*
\proof{Proof.}	Note that in the theorem statement we defined $Q_i,T_i$ as what we call $\bar Q_i,\bar T_i$ in the helper lemmas in Section~\ref{sec:lemmashelperICpolicy}, i.e., they refer to their respective values when every trip is accepted.

Let $w_2(\tau) =m_2\tau + z_2q_{2\to 1}(\tau)$, and $w_1(\tau) =m_1\tau + z_1q_{1\to 2}(\tau)$.

The following constraints are sufficient such that for these prices, $\paru R(w, \sigma) > 0$, where the derivatives are with respect to upper endpoints $u$ of the intervals that compose either $\sigma_1$ or $\sigma_2$:

From Lemma~\ref{lem:rangeform2z2}, for derivative with respect to $\sigma_2$:
\begin{align*}
\frac{ T_1(\lambda_{2\to 1} T_2 -  Q_2) - \left( Q_1 +  T_1\lambda_{2\to 1}\right) }{\left( Q_1(\lambda_{2\to 1} T_2 -  Q_2) + \lambda_{2\to 1}( Q_1 +  T_1\lambda_{2\to 1}) \right)}<  &\frac{z_2}{m_2-R_1} < \frac{ Q_2  T_1 +  Q_1 }{ Q_1 ( Q_2 - \lambda_{2\to 1})}
\end{align*}

From Lemma \ref{lem:rangeform1z1}, for derivative with respect to $\sigma_1$:
\begin{align*}
-\frac{( T_2\lambda_{1\to 2} +  Q_2) }{ Q_2 (\lambda_{1\to 2} T_1 -  Q_1) + \lambda_{1\to 2}( T_2\lambda_{1\to 2} +  Q_2)}< &\frac{z_1}{R_2} < \frac{1 }{( Q_1 - \lambda_{1\to 2})}
\end{align*}

Now, applying Theorem~\ref{thm:surgemodeloptimal_table}, the policy that accepts everything, $\sigma = \{(0, \infty),(0, \infty)\}$, is the unique optimal policy, given these constraints are satisfied, as the appropriate derivative is always positive.

\parbold{Resulting constraints on $R_1, R_2$} These constraints limit $R_1, R_2$ with respect to each other.

From Remark~\ref{rem:derivativesimplification},
\begin{align*}
W_2 &=m_2 ( T_2 - 1) + z_2 ( Q_2 - \lambda_{2\to 1})\\
W_1 &=m_1 ( T_1 - 1) + z_1 ( Q_1 - \lambda_{1\to 2})
\end{align*}

\textbf{Given $R_2$, what's the range $R_1$ can be in to still satisfy Lemma \ref{lem:rangeform1z1} conditions?}

First, we need
\begin{align*}
\frac{z_1}{R_2} < \frac{1 }{( Q_1 - \lambda_{1\to 2})} & \iff W_1 - m_1(T_1 - 1) < R_2
\end{align*}

Let $m_1 = R_2$. Then, $\frac{R_1}{R_2} < 1$ is satisfies the condition.

Second, we need
\begin{align*}
\frac{z_1}{R_2} &> -\frac{( T_2\lambda_{1\to 2} +  Q_2) }{ Q_2 (\lambda_{1\to 2} T_1 -  Q_1) + \lambda_{1\to 2}( T_2\lambda_{1\to 2} +  Q_2)}\\
\iff W_1 &> R_2 \left[ T_1 - 1 - \left[\frac{( T_2\lambda_{1\to 2} +  Q_2) }{ Q_2 (\lambda_{1\to 2} T_1 -  Q_1) + \lambda_{1\to 2}( T_2\lambda_{1\to 2} +  Q_2)}\right]( Q_1 - \lambda_{1\to 2})\right]\\
\iff \frac{R_1}{R_2}
&>  \frac{1}{ T_1}\left[ T_1 - 1 - \left[\frac{( T_2\lambda_{1\to 2} +  Q_2) }{ Q_2 (\lambda_{1\to 2} T_1 -  Q_1) + \lambda_{1\to 2}( T_2\lambda_{1\to 2} +  Q_2)}\right]( Q_1 - \lambda_{1\to 2})\right]\\
&= 1 - \frac{1}{ T_1}\left[1 + \frac{( T_2\lambda_{1\to 2} +  Q_2)( Q_1 - \lambda_{1\to 2}) }{ Q_2 (\lambda_{1\to 2} T_1 -  Q_1) + \lambda_{1\to 2}( T_2\lambda_{1\to 2} +  Q_2)}\right]\\
&= 1 - \frac{1}{ T_1}\frac{ Q_2 (\lambda_{1\to 2} T_1 -  Q_1) + \lambda_{1\to 2}( T_2\lambda_{1\to 2} +  Q_2) + ( T_2\lambda_{1\to 2} +  Q_2)( Q_1 - \lambda_{1\to 2}) }{ Q_2 (\lambda_{1\to 2} T_1 -  Q_1) + \lambda_{1\to 2}( T_2\lambda_{1\to 2} +  Q_2)}\\
&= 1 - \frac{1}{ T_1}\frac{ Q_2 (\lambda_{1\to 2} T_1 -  Q_1) +  Q_1( T_2\lambda_{1\to 2} +  Q_2) }{ Q_2 (\lambda_{1\to 2} T_1 -  Q_1) + \lambda_{1\to 2}( T_2\lambda_{1\to 2} +  Q_2)} \\&\triangleq C
\end{align*}

\textbf{What about incentive compatible pricing in state 2 to satisfy Lemma~\ref{lem:rangeform2z2}?}
If we only care about that state, we can support any ratio of payments:
\begin{align*}
\text{Let\ \ \ \ } {z_2} &= \left[\frac{ Q_2  T_1 +  Q_1 }{ Q_1 ( Q_2 - \lambda_{2\to 1})}\right]({{m_2 - R_1}}) \triangleq c({{m_2 - R_1}})\\
{R_2} &= \frac{1}{ T_2} \left[m_2 ( T_2 - 1) + z_2 ( Q_2 - \lambda_{2\to 1})\right]\\
\implies \frac{R_2}{R_1} &= \frac{1}{R_1  T_2}\left[{m_2} ( T_2 - 1) + (m_2-R_1)c ( Q_2 - \lambda_{2\to 1})\right]\\
&\to 1 - \frac{1}{ T_2} \leq 1 \text{ as } m_2 \to R_1\\
&\to \infty  \text{ as } m_2 \to \infty
\end{align*}
Thus, we can make the surge state IC for any ratio of payments $\frac{R_2}{R_1} \geq 1$, i.e., $\frac{R_1}{R_2} \leq 1$.

Now, suppose we want to achieve $R_1,R_2$ such that $\frac{R_1}{R_2} \in [0,C)$. From the previous line, we can still set $w_2$ such that every trip in state $2$ is accepted (the derivative with respect to the surge policy is positive everywhere). Then, setting $z_1 = 0$, and $m_1$ to satisfy $R_1$, all trips up to a certain length will be accepted in the non-surge state: By Remark~\ref{rem:basiccontinuityincreasingproperties}, $\paru R(w, \sigma)$ is positive up to a certain value and then negative after that, where $u$ is an upper endpoint of $\sigma_1$. Thus, by Theorem~\ref{thm:surgemodeloptimal_table},  the optimal policy is of the form $\sigma = \{(0, t_1), (0, \infty)\}$. \qed
\subsection{Proofs of auxiliary lemmas}
\label{sec:appdynamiclemmasproofstedious}

\subsubsection{Derivative derivation and comments}

\lemderivatives*
\proof{Proof.}
\begin{align*}
	\mu_i(\{\sigma_j, \sigma_2\}) &= \frac{Q_jT_i}{Q_jT_i + Q_iT_j}\\
	R(w, \sigma) &= \mu_1 (\sigma) R_1(w_1, \sigma_1)  + \mu_2 (\sigma) R_2(w_2, \sigma_2)\\
	&= \left[\frac{1}{ Q_2T_1 + Q_1T_2}\right]\left[Q_2W_1 +  Q_1W_2\right]\\
	{R_i(\sigma_i)} &=  \frac{W_i}{T_i}\\
	\paru Q_i &= \paru \left[\lambda_{i\to j} + \lambda_i\int_{\tau \in \sigma_i}q_{i\to j}(\tau) dF_i(\tau)\right]= \lambda_i q_{i\to j}(u) f_i(u)\\
	\paru W_i &= \paru \left[\lambda_i\int_{\tau \in \sigma_i} w_i(\tau) dF_i(\tau) \right] = \lambda_i w_i(u) f_i(u)\\
	\paru T_i &= \lambda_i f_i(u) u
\end{align*}
\begin{align*}
	\paru R(w, \sigma)&=\left[\frac{\lambda_if_i(u)}{ Q_iT_j + Q_jT_i}\right]\left[\left[q_{i\to j}(u)  W_j + Q_j w_i(u)\right] - R(w, \sigma) (uQ_j + q_{i\to j}(u)T_j) \right]  \\
	&\propto \left[q_{i\to j}(u)  W_j + Q_j w_i(u)\right] - R(w, \sigma) (uQ_j + q_{i\to j}(u)T_j)\\
	&\propto \left[q_{i\to j}(u)  W_j + Q_j w_i(u)\right](Q_iT_j \\&{}\,\,\,+ Q_jT_i) - (Q_iW_j +  Q_jW_i)(uQ_j + q_{i\to j}(u)T_j)\\
	&= q_{i\to j}(u)W_j(Q_iT_j + Q_jT_i) + Q_j w_i(u)(Q_iT_j + Q_jT_i)
	\\&\,\,\,\,- uQ_j(Q_iW_j +  Q_jW_i) - q_{i\to j}(u)T_j(Q_iW_j +  Q_jW_i)\\
	&\propto q_{i\to j}(u)W_jT_i + w_i(u)(Q_iT_j + Q_jT_i) - u(Q_iW_j +  Q_jW_i) - q_{i\to j}(u)T_jW_i\\
	&= q_{i\to j}(u)\left[W_jT_i - T_jW_i\right] + w_i(u)(Q_iT_j + Q_jT_i) - u(Q_iW_j +  Q_jW_i)\\
	&= uT_iT_j \left[\frac{q_{i\to j}(u)}{u}\left(R_j - R_i\right) + \frac{w_i(u)}{u}\left(\frac{Q_i}{T_i} + \frac{Q_j}{T_j}\right) - \left(\frac{Q_i}{T_i}R_j + \frac{Q_j}{T_j}R_i\right) \right]\\
	&\propto \frac{q_{i\to j}(u)}{u}\Delta_{ji} + \frac{w_i(u)}{u}\left(\frac{Q_i}{T_i} + \frac{Q_j}{T_j}\right) - \left(\frac{Q_i}{T_i}R_j + \frac{Q_j}{T_j}R_i\right) & \Delta_{ji} = {R_j} - R_i\\
	& \triangleq r(u, i, w, \sigma)
\end{align*}
\qed
\subsubsection{Lemmas for driver policy in response to affine pricing}

\lemaffinesurgequasiconvex*
\proof{Proof.}
Recall that by definition of strict quasi-convexity, $r(u, i, w, \sigma)$ is strictly quasi-convex if its derivative is strictly negative up to a point, and then strictly positive above that point $u$, for a fixed $\sigma$.

From Lemma~\ref{lem:derivatives},
\begin{align*}
	r(u, i, w, \sigma)
	&= \frac{c_1 a - c_2 q_{i\to j}(u)}{u} + c_3
\end{align*}
For some $c_1> 0,c_2 \geq 0, c_3$.%
For the case of $c_2 = 0$, the result immediately follows. Otherwise:

$\paru r(u, i, w, \sigma)$
\begin{align*}
	&= \paru \left[\frac{c_1 - c_2 q_{i\to j}(u)}{u} + c_3\right]\\
	&= \frac{1}{u^2}\left[ -uc_2 \paru q_{i\to j}(u) - \left[c_1 - c_2 q_{i\to j}(u)\right] \right]\\
	&= \frac{1}{u^2}\left[ -uc_2 \paru \left[\frac{\alpha}{\alpha+\beta}\left[1 - e^{-(\alpha + \beta)u}\right]\right] - \left[c_1 - c_2 \left[\frac{\alpha}{\alpha+\beta}\left[1 - e^{-(\alpha + \beta)u}\right]\right]\right] \right]\\
	&= \frac{1}{u^2}\left[ -uc_2 \left[\alpha e^{-(\alpha + \beta)u}\right] + c_2 \left[\frac{\alpha}{\alpha+\beta}\left[1 - e^{-(\alpha + \beta)u}\right]\right] - c_1\right] \\
	&=\frac{1}{u^2}\left[ -uc_2 \left[\alpha \left[\sum_{n=0}^\infty\frac{u^n (-1)^n(\alpha + \beta)^n}{n!}\right] \right] + c_2 \left[\frac{\alpha}{\alpha+\beta}\left[1 - \left[\sum_{n=0}^\infty\frac{u^n (-1)^n(\alpha + \beta)^n}{n!}\right]\right]\right] - c_1\right] \\
	&=\frac{1}{u^2}\left[ \frac{c_2\alpha}{\alpha+\beta}\left[\sum_{n=0}^\infty\frac{(-1)^{n+1}u^{n+1} (\alpha + \beta)^{n+1}}{n!} + 1 + \sum_{n=0}^\infty\frac{u^n (-1)^{n+1}(\alpha + \beta)^{n}}{n!}\right]   - c_1\right]\\
	&=\frac{1}{u^2}\left[ \frac{c_2\alpha}{\alpha+\beta}\left[\sum_{n'=1}^\infty\frac{(-1)^{n'}u^{n'} (\alpha + \beta)^{n'}}{(n'-1)!} + \sum_{n=1}^\infty\frac{u^n (-1)^{n+1}(\alpha + \beta)^{n}}{n!}\right]   - c_1\right] & n'= n+1\\
	&=\frac{1}{u^2}\left[ \frac{c_2\alpha}{\alpha+\beta}\left[\sum_{n=2}^\infty(-1)^{n}u^{n} (\alpha + \beta)^{n}\left[\frac{1}{(n-1)!} - \frac{1}{n!}\right]\right]   - c_1\right]
\end{align*}
Where last line follows because first ($n=1$) term of summation is zero.

\noindent It is sufficient for the following to be strictly increasing.
\[\frac{c_2\alpha}{\alpha+\beta}\left[\sum_{n=2}^\infty(-1)^{n}u^{n} (\alpha + \beta)^{n}\left[\frac{1}{(n-1)!} - \frac{1}{n!}\right]\right]   - c_1,\]
which holds:
\begin{align*}
	\paru &\left[\frac{c_2\alpha}{\alpha+\beta}\left[\sum_{n=2}^\infty(-1)^{n}u^{n} (\alpha + \beta)^{n}\left[\frac{1}{(n-1)!} - \frac{1}{n!}\right]\right]   - c_1\right]\\
	&= \frac{c_2\alpha}{\alpha+\beta}\left[\sum_{n=2}^\infty(-1)^{n}u^{n-1} (\alpha + \beta)^{n}\left[\frac{n}{(n-1)!} - \frac{n}{n!}\right]\right]\\
	&= \frac{c_2\alpha}{\alpha+\beta}\left[\sum_{n=2}^\infty(-1)^{n}u^{n-1} (\alpha + \beta)^{n}\frac{1}{(n-2)!}\right] =
	\frac{c_2\alpha}{\alpha+\beta}\left[\sum_{n'=0}^\infty(-1)^{n'+2}u^{n'+1} (\alpha + \beta)^{n'+2}\frac{1}{n'!}\right] & n'=n-2\\
	&=c_2\alpha u (\alpha+\beta)\left[\sum_{n=0}^\infty(-1)^{n}u^{n} (\alpha + \beta)^{n}\frac{1}{n!}\right] = c_2\alpha u (\alpha+\beta) e^{-(\alpha + \beta)u} > 0
\end{align*}
\qed

\lemaffinesurgequasiconcave*
\proof{Proof.}
Corollary of Lemma~\ref{lem:affinesurge_quasiconvex}. $r(u, i, w, \sigma)$ is the negative of the previous case, modulo constants that do not affect quasi-concavity.
\qed
\subsubsection{Lemmas for IC policy}

\remderivativesimplification*
\proof{Proof.}
\begin{align*}
w_i(u) &= mu + z q_{i\to j}(u) &m,z\geq 0\\
W_i &= \lambda_i\int_{\tau\in\sigma_i} w_i(\tau) dF_i(\tau)= \lambda_i\int_{\tau\in\sigma_i} \left[m\tau + z q_{i\to j}(\tau)\right] dF_i(\tau)= m (T_i - 1) + z (Q_i - \lambda_{i\to j})
\end{align*}
Then
\begin{align}
W_jT_i - T_jW_i &= R_j T_jT_i - m T_j(T_i - 1) - z T_j(Q_i - \lambda_{i\to j})\nonumber\\
w_i(u)(Q_iT_j + Q_jT_i) &= (mu + z q_{i\to j}(u))(Q_iT_j + Q_jT_i)\nonumber\\
&= q_{i\to j}(u) (zQ_iT_j + zQ_jT_i) + u(mQ_iT_j + mQ_jT_i)\nonumber\\
\paru R(w, \sigma)&\propto q_{i\to j}(u)\left[W_jT_i - T_jW_i\right] + w_i(u)(Q_iT_j + Q_jT_i) - u(Q_iW_j +  Q_jW_i)\label{eq:paruICpolicyline}\\
&= q_{i\to j}(u)\left[ R_j T_jT_i - m T_j(T_i - 1) - z T_j(Q_i - \lambda_{i\to j}) + zQ_iT_j + zQ_jT_i \right]\nonumber\\
&\ \ \ \ \ \ \ \ + u \left[mQ_iT_j + mQ_jT_i -Q_iR_j T_j -  Q_j(m (T_i - 1) + z (Q_i - \lambda_{i\to j}))\right]\nonumber\\
&= q_{i\to j}(u)\left[ (R_j - m) T_jT_i  + mT_j + zQ_jT_i + z T_j\lambda_{i\to j} \right]\nonumber\\
&\ \ \ \ \ \ \ \ + u \left[Q_iT_j (m-R_j) + Q_j(m -z Q_i + z\lambda_{i\to j})\right]\nonumber
\end{align}
Where Line~\eqref{eq:paruICpolicyline} is shown in the proof of Lemma~\ref{lem:derivatives}.
\qed

\remlimitqij*
\proof{Proof.}
Simple application of L'Hopital's rule.
$$ \lim_{u \to 0} \frac{q_{i\to j}(u)}{u} = \lim_{u \to 0} \paru {q_{i\to j}(u)} =  \lim_{u \to 0}\paru \frac{\lambda_{i\to j}}{\lambda_{i\to j}+\lambda_{j\to i}}\left[1 - e^{-(\lambda_{i\to j} + \lambda_{j\to i})u}\right]  = \lambda_{i\to j}$$

\qed

\remvaluesmaximizedwhenacceptall*
\proof{Proof.}
\begin{align*}
\lambda_{i\to j}T_i - Q_i &= \lambda_{i\to j}\left[1 + \lambda_i \int_{\tau\in\sigma_i} \tau dF_i(\tau)\right] - \lambda_{i\to j} - \lambda_i\int_{\sigma_i} q_{i\to j}(\tau) dF_i(\tau)\\
&= \lambda_i \int_{\tau\in\sigma_i} \left[\lambda_{i\to j}\tau - q_{i\to j}(\tau)\right]dF_i(\tau) %
\end{align*}

$\lambda_{i\to j}\tau - q_{i\to j}(\tau)$ is increasing in $\tau$:
\begin{align*}
\frac{\partial }{\partial \tau} \left[\lambda_{i\to j}\tau - q_{i\to j}(\tau)\right] &= \lambda_{i\to j} - \left[\lambda_{i\to j} e^{-(\lambda_{i\to j} + \lambda_{j\to i})\tau}\right] \geq 0
\end{align*}

and $\lambda_{i\to j}\times0 - q_{i\to j}(0) = 0$. Thus, the function being integrated is positive, and so $\lambda_{i\to j}T_i - Q_i \geq 0$ and maximized when $\sigma_i = (0,\infty)$.	Near identical proof holds for $Q_i$.

\qed

\lemrangeformstatetwo*

\proof{Proof.}

Suppose we have $w_2(u) = mu + z q_{2\to 1}(u)$, for some $m> R_1,z\geq 0$.

From Remark~\ref{rem:derivativesimplification},

\begin{align*}\paru R(w, \sigma)\propto&{\ \ \ \ } u \left[\frac{q_{2\to 1}(u)}{u}\left[ (R_1 - m) T_1T_2  + mT_1 + zQ_1T_2 + z T_1\lambda_{2\to 1} \right]\right] \\&+ u\left[Q_2T_1 (m-R_1) + Q_1(m -z Q_2 + z\lambda_{2\to 1})\right]\end{align*}

$T_2, Q_2$ are functions of $\sigma_2$.

As $u \to \infty$, the term in brackets in the first term goes to 0, and thus the first necessary condition is to have the second term always greater than 0.

If the second term is always positive, then the first term may be negative as long as it has a smaller absolute value than the second term. As $u \to 0$, the ratio between (absolute value of) the  first and second terms is maximized. Thus, the second necessary (and sufficient) condition is to have the entire value positive when we
take the limit of $\frac{q_{2\to 1}(u)}{u}$ as $u \to 0$.

These two conditions are sufficient for  $\paru R(w, \sigma) > 0$, for all $u, \sigma_2$.

From the first condition, we need $m,z$ such that:
\begin{align*}
&Q_2T_1 (m-R_1) + Q_1(m -z Q_2 + z\lambda_{2\to 1}) > 0 & \forall T_1,Q_1,Q_2,R_1\\
\iff & \frac{z}{m-R_1} < \frac{Q_2 T_1 + \frac{m}{m-R_1}Q_1 }{Q_1 (Q_2 - \lambda_{2\to 1})}
\end{align*}

From the second condition, and using Remark~\ref{rem:limitqij} we need:
\begin{align*}
&\lambda_{2\to 1}\left[ (R_1 - m) T_1T_2  + mT_1 + zQ_1T_2 + z T_1\lambda_{2\to 1} \right] + \left[Q_2T_1 (m-R_1) + Q_1(m -z Q_2 + z\lambda_{2\to 1})\right] > 0\\
\iff & (m-R_1)T_1(Q_2 - \lambda_{2\to 1}T_2) + m(Q_1 + \lambda_{2\to 1}T_1) + zQ_1(\lambda_{2\to 1}T_2 - Q_2 + \lambda_{2\to 1} ) + zT_1\lambda_{2\to 1}^2 > 0\\
\iff & z \left(Q_1(\lambda_{2\to 1}T_2 - Q_2) + \lambda_{2\to 1}(Q_1 + T_1\lambda_{2\to 1}) \right) > (m-R_1)T_1(\lambda_{2\to 1}T_2 - Q_2) - m(Q_1 + T_1\lambda_{2\to 1})\\
\iff & \frac{z}{m-R_1} > \frac{T_1(\lambda_{2\to 1}T_2 - Q_2) - \frac{m}{m-R_1}\left(Q_1 + T_1\lambda_{2\to 1}\right) }{\left(Q_1(\lambda_{2\to 1}T_2 - Q_2) + \lambda_{2\to 1}(Q_1 + T_1\lambda_{2\to 1}) \right)}
\end{align*}

Putting the conditions together, we need, for all $T_i,Q_i,R_i$:
\begin{align*}
\frac{T_1(\lambda_{2\to 1}T_2 - Q_2) - \frac{m}{m-R_1}\left(Q_1 + T_1\lambda_{2\to 1}\right) }{\left(Q_1(\lambda_{2\to 1}T_2 - Q_2) + \lambda_{2\to 1}(Q_1 + T_1\lambda_{2\to 1}) \right)}< & \frac{z}{m-R_1} < \frac{Q_2 T_1 + \frac{m}{m-R_1}Q_1 }{Q_1 (Q_2 - \lambda_{2\to 1})}
\end{align*}

$m>R_1$ by supposition, and so $\frac{m}{m-R_1}>1$. Thus, the following is sufficient as the constraints become tighter:
\begin{align*}
\frac{T_1(\lambda_{2\to 1}T_2 - Q_2) - \left(Q_1 + T_1\lambda_{2\to 1}\right) }{\left(Q_1(\lambda_{2\to 1}T_2 - Q_2) + \lambda_{2\to 1}(Q_1 + T_1\lambda_{2\to 1}) \right)}<  &\frac{z}{m-R_1} < \frac{Q_2 T_1 + Q_1 }{Q_1 (Q_2 - \lambda_{2\to 1})}\\
\iff\frac{T_1 - \frac{Q_1 + T_1\lambda_{2\to 1}}{(\lambda_{2\to 1}T_2 - Q_2)}}{Q_1 + \frac{\lambda_{2\to 1}(Q_1 + T_1\lambda_{2\to 1})}{(\lambda_{2\to 1}T_2 - Q_2)} }<  &\frac{z}{m-R_1} < \frac{T_1 + \frac{Q_1}{Q_2} }{Q_1 \left(1 - \frac{\lambda_{2\to 1}}{Q_2}\right)}
\end{align*}

It turns out that both constraints are tightest when $\sigma_2 = (0,\infty)$. In the left constraint, the numerator is increasing and the denominator is decreasing with $\lambda_{2\to 1}T_2 - Q_2$, and so the constraint becomes tighter as $\lambda_{2\to 1}T_2 - Q_2$ increases. By Remark~\ref{rem:valuesmaximizedwhenacceptall}, $\lambda_{2\to 1}T_2 - Q_2$ is always positive, and maximized when $\sigma_2 = (0, \infty)$. Similarly, in the right constraint, the numerator decreases and the denominator increases with $Q_2$.

Thus, it is sufficient for the two constraints to be feasible for $\sigma_2 = (0,\infty)$. Then, they are satisfied for all $\sigma'_2$. For feasibility, we need
\begin{align*}
\frac{T_1(\lambda_{2\to 1}T_2 - Q_2) - \left(Q_1 + T_1\lambda_{2\to 1}\right) }{\left(Q_1(\lambda_{2\to 1}T_2 - Q_2) + \lambda_{2\to 1}(Q_1 + T_1\lambda_{2\to 1}) \right)} < \frac{Q_2 T_1 + Q_1 }{Q_1 (Q_2 - \lambda_{2\to 1})}\\
\iff ({T_1(\lambda_{2\to 1}T_2 - Q_2) - \left(Q_1 + T_1\lambda_{2\to 1}\right) }){Q_1 (Q_2 - \lambda_{2\to 1})} \\< ({Q_2 T_1 + Q_1 }){\left(Q_1(\lambda_{2\to 1}T_2 - Q_2) + \lambda_{2\to 1}(Q_1 + T_1\lambda_{2\to 1}) \right)}\\
\iff Q_1 Q_2({T_1(\lambda_{2\to 1}T_2 - Q_2) - \left(Q_1 + T_1\lambda_{2\to 1}\right) }) - Q_1 \lambda_{2\to 1}({T_1(\lambda_{2\to 1}T_2 - Q_2) - \left(Q_1 + T_1\lambda_{2\to 1}\right) })
\\< Q_2 T_1 {\left(Q_1(\lambda_{2\to 1}T_2 - Q_2) + \lambda_{2\to 1}(Q_1 + T_1\lambda_{2\to 1}) \right)} + Q_1{\left(Q_1(\lambda_{2\to 1}T_2 - Q_2) + \lambda_{2\to 1}(Q_1 + T_1\lambda_{2\to 1}) \right)}\\
\iff Q_1 Q_2({ - \left(Q_1 + T_1\lambda_{2\to 1}\right) }) - Q_1 \lambda_{2\to 1}({T_1(\lambda_{2\to 1}T_2 - Q_2) })\\< Q_2 T_1 {\left( \lambda_{2\to 1}(Q_1 + T_1\lambda_{2\to 1}) \right)} + Q_1{\left(Q_1(\lambda_{2\to 1}T_2 - Q_2)  \right)}
\end{align*}

For any valid $Q_i,T_i$, the left hand side of the final line is always non-positive, and the right hand side is always positive, and thus there exist feasible ratios $\frac{z}{m-R_1}$.

\qed

\lemrangeformstateone*
\proof{Proof.}
Similar to previous proof.
Suppose we have $w_1(u) = mu + z q_{1\to 2}(u)$, for some $m= R_2,z\leq 0$.

From Remark~\ref{rem:derivativesimplification},
\begin{align*}\paru R(w, \sigma)=& u \left[\frac{q_{1\to 2}(u)}{u}\left[ (R_2 - m) T_1T_2  + mT_2 + zQ_2T_1 + z T_2\lambda_{1\to 2} \right]\right] \\&+ u\left[Q_1T_2 (m-R_2) + Q_2(m -z Q_1 + z\lambda_{1\to 2})\right]\\
=& u \left[\frac{q_{1\to 2}(u)}{u}\left[ R_2T_2 + zQ_2T_1 + z T_2\lambda_{1\to 2} \right] + \left[Q_2(R_2 -z Q_1 + z\lambda_{1\to 2})\right]\right]
\end{align*}

As before, we have two necessary and sufficient conditions for $\paru R(w, \sigma) > 0$, for all $u, \sigma_1$.

From the first condition, we need $m,z$ such that:
\begin{align*}
&Q_2(R_2 -z (Q_1 - \lambda_{1\to 2})) > 0 & \forall T_2,Q_2,Q_1,R_2\\
\iff & \frac{z}{R_2} < \frac{1 }{(Q_1 - \lambda_{1\to 2})}
\end{align*}

Similarly, the second condition becomes
\begin{align*}
\lambda_{1\to 2}\left[ R_2T_2 + zQ_2T_1 + z T_2\lambda_{1\to 2} \right] &+ \left[Q_2(R_2 -z Q_1 + z\lambda_{1\to 2})\right] > 0\\
\iff 	\lambda_{1\to 2}R_2T_2+ Q_2R_2 &> -z Q_2 (\lambda_{1\to 2}T_1 - Q_1) - z\lambda_{1\to 2}(T_2\lambda_{1\to 2} + Q_2)\\
\iff	\frac{z}{R_2}&> -\frac{(T_2\lambda_{1\to 2} + Q_2) }{Q_2 (\lambda_{1\to 2}T_1 - Q_1) + \lambda_{1\to 2}(T_2\lambda_{1\to 2} + Q_2)}
\end{align*}

Both constraints are tightest when $\sigma_1 = (0,\infty)$. By Remark~\ref{rem:valuesmaximizedwhenacceptall}, $\lambda_{1\to 2}T_1 - Q_1$ is always positive, and maximized when $\sigma_1 = (0, \infty)$, and so the right hand side is always negative.

The constraints are thus feasible when
\[-\frac{(T_2\lambda_{1\to 2} + Q_2) }{Q_2 (\lambda_{1\to 2}T_1 - Q_1) + \lambda_{1\to 2}(T_2\lambda_{1\to 2} + Q_2)} < \frac{1 }{(Q_1 - \lambda_{1\to 2})}\]

which trivially holds as the right hand side is positive and the left hand side is negative.

\qed

\end{document}